\documentclass{aa}  

\usepackage{graphicx}
\usepackage{derivative}
\usepackage{xcolor}
\usepackage{float}
\usepackage{rotating}
\usepackage{booktabs}
\usepackage{mathtools}
\usepackage{placeins}
\usepackage{bm}
\usepackage{longtable}
\usepackage{adjustbox}
\usepackage{supertabular}
\usepackage{multirow}
\usepackage{tabularx}
\usepackage{xurl}
\usepackage{esdiff}
\usepackage{amsmath}
\usepackage{siunitx}
\usepackage[switch]{lineno}

\usepackage{hyperref}
\hypersetup{
  colorlinks=true,
    linkcolor=blue,
    citecolor=blue,
    urlcolor=magenta}

\usepackage{natbib}

\usepackage{txfonts}

\usepackage{color}

\begin{document}

   \title{Measuring stellar surface rotation and activity \\ with the PLATO mission}
   \subtitle{I. Strategy and application to simulated light curves}
    
   \titlerunning{}

   \author{
           S.N.~Breton\inst{1}
           \and
           A.F~Lanza\inst{1}
           \and 
           S.~Messina\inst{1}
           \and
           I. Pagano\inst{1}
           \and
           L.~Bugnet\inst{2}
           \and
           E.~Corsaro\inst{1}
           \and
           R.A.~García\inst{3}
           \and
           S.~Mathur\inst{4,5}
           \and
           A.R.G~Santos\inst{6}
           \and
           S.~Aigrain\inst{7}
           \and 
           L.~Amard\inst{3}
           \and
           A.S.~Brun\inst{3}
           \and
           L.~Degott\inst{8}
           \and
           Q.~Noraz\inst{9,10}
           \and
           D.B.~Palakkatharappil\inst{3}
           \and
           E.~Panetier\inst{11}
           \and 
           A.~Strugarek\inst{3}
           \and
           K.~Belkacem\inst{12}
           \and 
           M.-J~Goupil\inst{12}
           \and
           R.M.~Ouazzani\inst{12}
           \and
           J.~Philidet\inst{12}
           \and
           C.~Renié\inst{12}
           \and
           O.~Roth\inst{12}
          }
    \institute{
    INAF – Osservatorio Astrofisico di Catania, Via S. Sofia, 78, 95123 Catania, Italy \\
    \email{sylvain.breton@inaf.it}
    \and
    Institute of Science and Technology Austria (IST Austria), Am Campus 1, Klosterneuburg, Austria
    \and
    Universit\'e Paris-Saclay, Universit\'e Paris Cit\'e, CEA, CNRS, AIM, 91191, Gif-sur-Yvette, France
    \and
    Instituto de Astrof\'{\i}sica de Canarias, La Laguna, Tenerife, Spain
    \and 
    Departamento de Astrof\'{\i}sica, Universidad de La Laguna, La Laguna, Tenerife, Spain
    \and
    Instituto de Astrof\'isica e Ci\^encias do Espa\c{c}o, Universidade do Porto, CAUP, Rua das Estrelas, PT4150-762 Porto, Portugal
    \and
    Department of Physics, University of Oxford, Oxford, United Kingdom
    \and
    Universit\'e Paris-Sud, CNRS, Institut d’Astrophysique Spatiale, UMR 8617, F-91405, Orsay Cedex, France
    \and
    Rosseland Centre for Solar Physics, University of Oslo, P.O. Box 1029 Blindern, Oslo, NO-0315, Norway
    \and 
    Institute of Theoretical Astrophysics, University of Oslo, P.O. Box 1029 Blindern, Oslo, NO-0315, Norway
    \and
    Universit\'e Paris Cit\'e, Universit\'e Paris-Saclay, CEA, CNRS, AIM, 91191, Gif-sur-Yvette, France
    \and
    LESIA, Observatoire de Paris, CNRS, Universit\'e PSL, Sorbonne Universit\'e, Universit\'e Paris Cit\'e, 5 place Jules Janssen, 92195 Meudon, France
    }

   \date{}

 \abstract{ 
 The Planetary Transits and Oscillations of stars mission (PLATO) will allow us to measure surface rotation and monitor photometric activity of tens of thousands of main sequence solar-type and subgiant stars.
 This paper is the first of a series dedicated to the preparation of the analysis of stellar surface rotation and photospheric activity with the near-future PLATO data.
 We describe in this work the strategy that will be implemented in the PLATO pipeline to measure stellar surface rotation, photometric activity, and long-term modulations.
 The algorithms are applied on both noise-free and noisy simulations of solar-type stars, which include activity cycles, latitudinal differential rotation, and spot evolution. PLATO simulated systematics are included in the noisy light curves.
 We show that surface rotation periods can be recovered with confidence for most of the stars with only six months of observations and that the {recovery rate} of the analysis significantly improves as additional observations are collected. This means that the first PLATO  data release will already provide a substantial set of measurements for this quantity, with a significant refinement on their quality as the instrument obtains longer light curves. 
 Measuring the Schwabe-like magnetic activity cycle during the mission will require that the same field be  observed over a significant timescale (more than four years). Nevertheless, PLATO will provide a vast and robust sample of solar-type stars with constraints on the activity-cycle length. Such a sample is lacking from previous missions dedicated to space photometry.
 }


   \maketitle

\section{Introduction \label{section:introduction}}


An important aspect of the upcoming Planetary Transits and Oscillations of Stars mission \citep[PLATO,][]{Rauer2014} will be its ability to investigate stellar surface rotation and photometric activity modulations of a large sample of stars with convective outer layers, both on the main sequence and at the subgiant stage of evolution. Indeed, as active regions travel over the observed stellar disc, the apparent brightness of the star experiences modulations with periodicities directly related to photospheric surface rotation. Over a longer time frame, the amplitude of these modulations varies due to stellar activity cycles, as observed in the case of the 11 year solar Schwabe cycle \citep[e.g.][]{Salabert2017}. Surface rotation is a crucial proxy with which to estimate stellar ages through gyrochronology \citep[e.g.][]{Skumanich1972,Barnes2003,Barnes2007}, while rotation--activity relations are directly linked to dynamo processes in the convective envelope \citep[see e.g.][and references therein]{Brun2017b}.

Space missions dedicated to long-baseline photometry or including such aspects, such as the Convection, Rotation and planetary Transits satellite \citep[CoRoT,][]{Baglin2006}, \textit{Kepler} \citep{Borucki2010}, K2 \citep{Howell2014}, the Transiting Exoplanet Survey Satellite \citep[TESS,][]{Ricker2015}, or the \textit{Gaia} mission \citep{Gaia2016}, have allowed measurements of stellar surface rotation for tens of thousands of low-mass stars, from the main sequence to evolved stages \citep[][]{Nielsen2013,Reinhold2013a,Garcia2014b,McQuillan2014,Lanzafame2018,Santos2019,Santos2021,Reinhold2020,Gordon2021,Holcomb2022,Distefano2023,Reinhold2023,Claytor2024}.
The availability of such a large amount of data allowed studies of the connection between stellar rotation, age, and activity \citep[e.g.][]{Masuda2022b,Masuda2022a,Santos2023,Mathur2023}, between stellar rotation and metallicity \citep[e.g.][]{Amard2020,See2021,See2023},{} dynamo processes \citep[e.g.][]{Cao2023}, the behaviour of magnetic braking with stellar age \citep[e.g.][]{vanSaders2016,vanSaders2019,Hall2021}, the transition between different fast-rotating regimes \citep[e.g.][]{Lanzafame2019}, and the relation between rotation and stellar flares \citep[e.g.][]{Yang2019,Notsu2019,Raetz2020}.
{Additional connections with structural and evolutionary aspects have also been explored by characterising} the convection--rotation regimes in the envelope {via the estimation of the Rossby number} \citep[e.g.][]{Corsaro2021,Noraz2022} and the role of core--envelope decoupling in rotational evolution \citep[e.g.][]{Spada2020,Angus2020,Lu2022}. 
{Studies related to exoplanetary science could also be performed by analysing} the influence of rotation over the orbital architecture of exoplanets \citep[e.g.][]{McQuillan2013,Messias2022,Garcia2023} and the potential connection between the presence of short-periodic planets and fast stellar rotation with respect to stellar age \citep{Ahuir2021,Deeg2023}. 
{Such measurements are also useful for asteroseismology in order to investigate} the connection between rotation, activity, and pulsations \citep[e.g.][]{Mathur2019}. {Finally, they can be used to perform physically motivated} starspot modelling and characterisation of active features \citep[e.g.][]{Lanza2009,Lanza2019,Ozavci2018,deFreitas2021,Breton2024}.  

In this paper, we describe the algorithms that will be implemented in the PLATO pipeline module dedicated to analysing the surface rotation and activity of the main sequence and subgiant stars observed by the mission. 
To demonstrate the performances of the algorithms, we applied them to a realistic set of simulated light curves and assessed how the inclusion of noise and PLATO systematics impacts the recovery of observables.
The layout of the paper is as follows. In Sect.~\ref{sec:module_description}, we describe the algorithms that will be implemented in the module and we provide the corresponding physical motivations.
In Sect.~\ref{sec:simulation}, we present a set of noise-free and noisy PLATO simulated light curves that were generated for this work. In particular, in the case of noisy light curves, we describe how we handle the correction of the PLATO systematics included in the simulated data.
In Sect.~\ref{sec:application}, we show the results we obtain when applying the algorithms to the simulated light curves. Conclusions and perspectives for future developments are provided in Sect.~\ref{section:conclusion}.

\section{Descriptions of the algorithms  \label{sec:module_description}}

\begin{figure*}[t!]
    \centering
    \includegraphics[width=\textwidth]{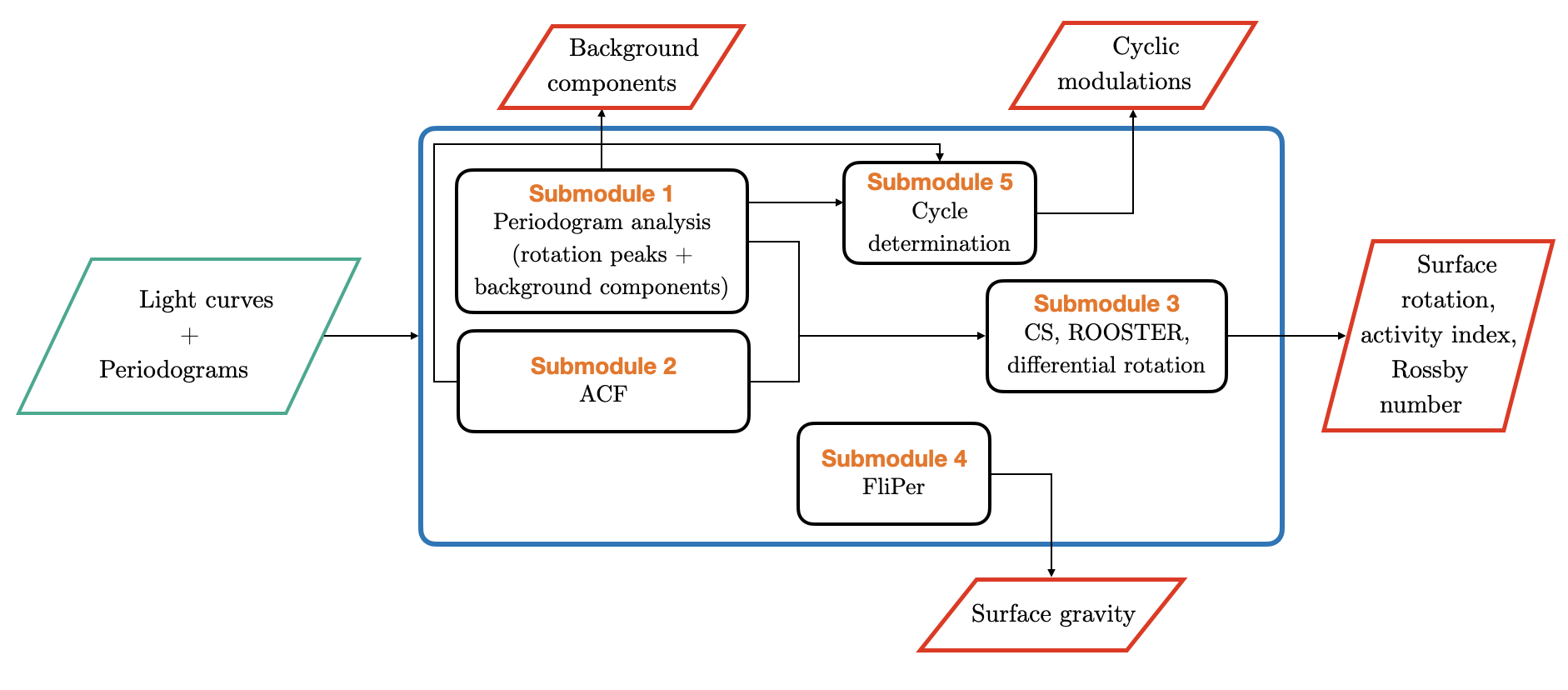}
    \caption{Flow diagram for the module. The main inputs for the module, the light curves, and the periodograms are shown on the left. 
    The different module outputs are encircled in red.
    }
    \label{fig:simplified_flow_diagram}
\end{figure*}

The flow diagram of the module is shown in Fig.~\ref{fig:simplified_flow_diagram} and the corresponding elements are described in the following subsections. 
The diagram is simplified in the sense that, for sake of clarity and concision, we indicate only the physical meaning of the module inputs and outputs, without describing the corresponding data structures  in detail. 
The aim of this work is to show the physical motivations of the implemented algorithms.
In what follows, we therefore describe the abilities of the module thematically, referring each time to the corresponding submodule of the flow diagram.

All analyses in this work are performed using the unofficial open-source prototype module \texttt{star-privateer},\footnote{The code source for the demonstrator is accessible at \url{https://gitlab.com/sybreton/star_privateer} and the corresponding documentation is hosted at \url{https://star-privateer.readthedocs.io/en/latest}. The analysis presented in this paper was performed with {\texttt{v1.1.2}} of the module, which can be downloaded and installed from \texttt{PyPi}: \url{https://pypi.org/project/star-privateer}.} which will be used as a basis for the algorithm implementation in the PLATO pipeline. 

\subsection{Inputs \label{sec:inputs}}

For each target star, two light curves obtained from the same raw data ---but with distinct calibrations--- are provided to the module. The first light curve, {LC1}, preserves low-frequency variability (periods beyond 60 days) and uses a two-hour sampling. 
This sampling is well suited to looking for rotational modulations while reducing the computing time of some of the algorithms compared to analysing light curves with the 25 second sampling that PLATO will also be able to provide. 
The second light curve, {LC2, follows the same calibration procedure as LC1} but an additional high-pass filter with a cutoff of  60~days is applied.
{The effect of the filtering on the variability included in the light curve is illustrated in Appendix~\ref{appendix:filtering}.}
The generalised Lomb-Scargle periodograms \citep[GLS,][]{Lomb1976,Scargle1982,Zechmeister09} of the two light curves are also provided to the module. These periodograms have a frequency binning chosen to ensure independence from one frequency bin to another (therefore with noise distribution following a $\chi^2$ with two degrees of freedom), which corresponds to the {periodogram resolution obtained when applying a discrete Fourier transform algorithm to the light curve: consecutive frequency bins are separated by an interval of $1 / t_{\rm obs}$, where $_{\rm obs}$ is the length of the light curve.} 
It should be noted that the periodogram of the 25s light curve is also provided to the module but is only used for the background fitting step of Submodule~1 (see Sect.~\ref{sec:surface_gravity_background}).

\subsection{Stellar rotation}

\subsubsection{Average rotation period \label{sec:average_rotation_period}}

{Firstly, rotational modulations are searched for in the periodogram of LC2 (Submodule~1). 
In order to obtain a proper estimation of the false-alarm probability of the rotation peaks, a quick non-parametric estimation of the mean background power level in the periodogram is computed with the following methodology: a logarithmically spaced frequency vector of $n=10$ bins is constructed, and the corresponding power values $b$ are computed as
\begin{equation}
    b (\nu_k) = (9/8)^3 \times \mathrm{median} \, (\{x_i\}) \; ,
\end{equation}
where $\{x_i\}$ is the set of power values for which the frequency $\nu_k$ in the resampled vector is the closest to $\nu_i$, the frequency corresponding to $x_i$.
The non-parametric background power level is then computed by interpolating $b$ for each frequency bin of the input periodogram. The use of a corrected median as the metric is motivated by the fact that we aim to estimate the mean power of the periodogram without including the power excess of the rotation peaks.
Dividing the periodogram by this background estimation, we obtain a signal-to-noise-ratio (S/N) periodogram and we can now compute false-alarm probabilities assuming that the likelihood follows a $\chi^2$ with two degrees of freedom \citep{Woodard1984} with a mean of  unity.
The false-alarm probability corresponding to the probability that a given peak of height $z$ is due to a noise fluctuation is computed for each peak as
\begin{equation}
\label{eq:pfa}
    p_{\rm FA} = \mathrm{e}^{-z} \; ,
\end{equation}
and only peaks with a false-alarm probability of smaller than \num{e-6} are considered.

The largest peak in the periodogram is then fitted using a Gaussian profile. The fitted profile is subtracted from the periodogram and the procedure is repeated until there remains no peak above the false-alarm probability threshold.} 

As we look for signal in the 60 day filtered LC2 periodogram for stellar rotation, we consider only peaks with period below 60~days. Given the position of the photometric detection edge at slow rotation encountered in the \textit{Kepler} mission \citep[see e.g.][]{Santos2021,Masuda2022a}, this cutoff should allow us to recover the rotation period for the vast majority of stars where rotational modulations are detectable. 
{The period of the largest fitted peak is kept as a candidate for the average rotation period, $P_{\rm GLS}$, while the other fitted peaks are retained to be analysed as possible differential rotation signatures.} 
The corresponding uncertainties {for each peak} are taken as {half width at half maximum (HWHM)} of the fitted Gaussian profile.
{The spread of power (both in terms of peak width and harmonic pattern) from rotational modulation in the periodogram is due to the combined effect of active region lifetime, differential rotation, and stellar inclination. The method described above allows us to capture the fine structure of this power distribution, especially for slower rotators where the ratio between spot lifetime and rotation period is smaller. Nevertheless, this means that the uncertainty on the average rotation period might be underestimated. To overcome this issue, we smooth the periodogram with a triangular window with a width of one-tenth of the period of the peak of largest amplitude. We then fit a Gaussian profile centred on this period, and we use its HWHM as the uncertainty over $P_{\rm GLS}$.}

Secondly, rotational modulations are searched for {in LC2} by computing its auto-correlation function \citep[ACF,][]{McQuillan2013b} in Submodule~2. Selecting the period corresponding to the highest peak in the Lomb-Scargle of the ACF, a Gaussian window with width equal to one-tenth of this period is applied in order to smooth the ACF. The location of the first local maximum of the ACF is then selected as a candidate value for the average rotation period, {$P_{\rm ACF}$}. 
The ACF value at this period is referred to as $G_\mathrm{ACF}$ \citep[it should be noted that this is a slightly different definition from what is used in][]{Ceillier2017}. In addition, the peak height, $H_\mathrm{ACF}$, defined as the mean difference between $G_\mathrm{ACF}$ and the two surrounding local minima, is also extracted. 
Given the difficulty in estimating a meaningful uncertainty on the periods obtained with the ACF, in its current state the module does not provide an uncertainty value for the ACF measurements. 

In order to increase the amplitude of signals exhibited both by the periodogram and the ACF while decreasing those of signals appearing in only one method, the composite spectrum \citep[CS;][]{Ceillier2016,Ceillier2017} is computed in Submodule~3 by multiplying the ACF by the interpolated LC2 periodogram, normalised with its maximal value. Assuming a normal distribution, the largest peak of the CS is fitted with a Gaussian profile and is selected as a candidate value for the average rotation period, {$P_{\rm CS}$}.

Considering these three candidate values and their uncertainties, the stellar rotation period $P_\mathrm{rot}$ is selected in Submodule~3 using an updated version of the Random fOrest Over STEllar Rotation methodology \citep[ROOSTER,][]{Breton2021}. The ROOSTER implementation in the PLATO pipeline uses two random forest classifiers \citep{Breiman1984,Breiman2001}, {which correspond to the \textit{RotClass} and \textit{PeriodSel} classifiers described in \citet{Breton2021}. We remind the reader that the methodology from \citet{Breton2021} included a third classifier, \textit{PollFlag,} in order to flag possible contaminants in the rotating sample. Nevertheless such an identification role is not devoted to the PLATO rotation and activity module, and this third classifier is left out of the analysis presented here.} 
{\textit{RotClass}} is used to assess the robustness of the detection of rotational modulation in the light curve; it yields a rotation score ranging from 0 to 1, corresponding to the fraction of decision trees in the forest that provided a detection for rotation. This means that the classifier favours a signal detection related to rotation when this score is larger than 0.5, and that it favours an instrumental or a noise fluctuation origin otherwise. 
Considering the three candidate values provided by the periodogram fit, the ACF analysis, and the CS computation, {\textit{PeriodSel} then} selects the stellar rotation period and its corresponding uncertainties.
It should therefore be noted that the uncertainty on $P_\mathrm{rot}$ will be missing if $P_{\rm ACF}$ is selected.
The two classifiers are trained with the same input parameters and hyperparameters. These input parameters include the candidate period values, the corresponding activity indexes (these are described in more detail in Sect.~\ref{sec:activity_index}), and other control parameters such as $H_\mathrm{ACF}$ and $G_\mathrm{ACF}$. It should be noted that, for completeness, the module will provide a rotation period for every target, along with the corresponding rotation score. 

{We briefly discuss some of the differences between the updated ROOSTER methodology used in this work and the original version presented in \citet{Breton2021}. We note that, in the original version, ROOSTER had to select between nine candidate periods. Indeed three calibrated light curves were considered for each target, differing by the choice of cutoff period that was made for the high-pass filter applied in order to mitigate \textit{Kepler} instrumental effects: 20, 55, and 80 days. From each of these light curves, three candidate periods were extracted, and, in the end, the nine possible values were considered by ROOSTER. For PLATO, considering the fact that the Level-0/Level-1 light-curve calibration performed upstream from the stellar analysis system should allow significant mitigation of the effects of PLATO instrumental systematic errors, the choice was made to consider a unique light curve filtered at 60 days, hence reducing the number of candidate periods from nine to three. We also point out that the number of training parameters is significantly reduced with respect to the methodology presented in \citet{Breton2021}. A large number of training parameters were included by these authors in order to study which of them had the greatest influence on the classification results. Building on the results of this previous analysis, we decided here to retain only those parameters that were found by \citet{Breton2021} to have the most significant affect. In what follows, we show how the analysis of the simulated light curves presented in Sect.~\ref{sec:simulation} allows us to validate this choice.} 

{Finally,} the stellar Rossby number, $Ro$, {which is connected to the differential rotation regime in the envelope \citep[see e.g.][]{Brun2017a}}, is estimated from $P_\mathrm{rot}$ and the effective temperature, $T_\mathrm{eff}$, using the prescription from \citet{Noraz2022}. In order to avoid dealing with the uncertainties related to the exact value of the solar Rossby number, $\rm Ro_\odot$, the $Ro$ value provided by the module is normalised by $\rm Ro_\odot$
\begin{equation}
    \frac{Ro}{\mathrm{Ro_\odot}} = \frac{P_\mathrm{rot}}{\rm P_{rot,\odot}} \left( \frac{T_\mathrm{eff}}{\rm T_{eff,\odot}} \right)^{3.29} \; .
\end{equation}

\subsubsection{Differential rotation \label{sec:method_diffrot}}

{
One of the goals of the PLATO rotation and activity analysis module is to directly detect and analyse signatures related to latitudinal differential rotation. 
These measurements are crucial in order to validate magnetohydrodynamic simulation assumptions about the connection between $Ro$ and the differential rotation regime, and provide observational constraints for stellar dynamo models \citep[e.g.][]{Jouve2008}.
However, hare-and-hounds exercises performed in the past \citep[see e.g.][]{Aigrain2015} demonstrated that, because of important degeneracies between differential rotation, stellar inclination, spot latitude, and lifetime \citep{Santos2017,Basri2020}, techniques developed to measure differential rotation directly from the properties of peaks in the periodogram lack robustness and fail in a large proportion of cases. 
As the PLATO tool to measure differential rotation is still under development, we decided to defer this topic to future work.
}


\subsection{Long-term modulations}

\subsubsection{Activity index \label{sec:activity_index}}

The average photometric activity index, $\left< S_\mathrm{ph} \right>$ \citep{Garcia2014b,Mathur2014b}, can be computed once $P_\mathrm{rot}$ is known (Submodule~3).
First, the $S_\mathrm{ph}$ time series is computed as the standard deviation of light-curve segments of length $5 \times P_\mathrm{rot}$. Here, $\left< S_\mathrm{ph} \right>$ is taken as the mean of this time series, {corrected by subtracting the photon noise}. This method provides a reliable proxy for unveiling signatures of stellar activity \citep{Salabert2016,Salabert2017}.  
{We remind the reader that, due to the combined impact that stellar inclination and latitude of active regions have on observed variability, this proxy has to be considered as a lower limit of the real level of photospheric activity. For the same level of starspot coverage, it should also be remembered that the observed stellar variability will be impacted by the spot-to-faculae coverage ratio and the spot distribution on the photosphere \citep[see e.g.][]{Basri2018,Luger2021}.}

\subsubsection{Cyclic modulations \label{sec:cyclic_modulations}}


Submodule~5 is designed to detect cyclic modulations in the light curves of stars observed by PLATO. These modulations can be related to year-long Schwabe-like magnetic activity cycles or to shorter periodicities analogous to the Rieger modulations  \citep[e.g.][]{Rieger1984,Gurgenashvili2022}
{or the quasi-biennial oscillations \citep[QBO, e.g.][]{Mehta2022}, both observed in the Sun. In principle, as shown by the solar cases, signatures from these shorter cycles are expected to be low-amplitude perturbations superimposed on the stronger Schwabe-like cycle.}

In principle, amplitude modulations of the envelope of the light curve might be related to beating phenomena arising from differential rotation and close periodicities in the observable rotational signal \citep[see][]{Mathur2014}. Nevertheless, these beating periodicities affect the envelope and are not visible in the Fourier decomposition of the signal, and therefore cross-validation of the detection by comparing the Fourier analysis and the ACF analysis should allow us to minimise the occurrence of such signatures in the reported measurements. 

Similarly to the search for rotational modulations, significant peaks are fitted in the {LC1} periodogram beyond 100~days (Submodule~1) and the false-alarm probability is computed, keeping only signals where this probability is below \num{e-6}. 
{As the frequency region explored in this step has a very small number of frequency bins, the method described in Sect.~\ref{sec:average_rotation_period} to estimate the background will most likely filter out the relevant signal. The S/N of the peaks,  $z$ , required to compute the false-alarm probability through Eq.~(\ref{eq:pfa}) is therefore estimated by comparing the height of the peaks with the mean power level in the spectrum.}
The ACF of {LC1} is then computed (Submodule~2). Before this step, in order to remove fluctuations that shorter-period rotational modulation will create, we smooth the light curve with a 60 day Gaussian window. 
Comparing periods extracted with the ACF to the ones fitted in the periodogram, we keep only signals agreeing within {$30~\%$; this somewhat high threshold is chosen to ensure that there is an excess of power in the LC1 periodogram in the vicinity of periodicities detected by the ACF.} 
To minimise the number of false positives, we also consider only signals above a certain threshold in terms of $G_\mathrm{ACF}$ and $H_\mathrm{ACF}$, depending on the length of the light curve. Our main aim here is to minimise the number of false positives in the validated periods. To limit the rate of false positives for the shortest temporal baselines, we put a stringent threshold on the length of the accepted light curves, retaining only those shorter than four years, and imposing the criteria $G_\mathrm{ACF} > 0.5$ and $H_\mathrm{ACF} > 0.7$ in order to validate a signal. 
For light curves with a length equal to four years, we relax this constraint to $G_\mathrm{ACF} > 0.2$ and $H_\mathrm{ACF} > 0.5$.
For longer light curves, we impose $G_\mathrm{ACF} > 0.2$ and $H_\mathrm{ACF} > 0.35$. Comparing the results for the different temporal baselines we use in what follows, we show that this choice allows us to efficiently discard false positives while validating a significant amount of cyclic modulation recovery. We underline that these thresholds are optimised using the simulated light curves we analyse in this work. They will therefore be reassessed when dealing with real data in order to optimise the yield while limiting the number of false positives.
{In addition to this, complementary indicators from spectroscopic ground observations will be available for some PLATO stars. The existence of such a sample will be useful to refine the design of the algorithm dedicated to validating the existence of cyclic modulations in targets for which only space photometry is available to constrain long-term activity.}


Additionally, the ACF of the $S_\mathrm{ph}$ time series is computed and its periodogram is searched for significant peaks (Submodule~3). If their values are compatible with the periodicities measured with the periodogram and the light curve ACF, they shall be provided as additional outputs of the module. 

\subsection{Surface gravity and background \label{sec:surface_gravity_background}}


In addition to the algorithms for surface rotation and stellar activity presented in detail in this paper, the PLATO pipeline module will be used to measure the level of convective granulation by fitting the {25 second LC1} periodogram background profile (Submodule~1) with the high-DImensional And multi-MOdal NesteD Sampling legacy code \citep[DIAMONDS,][]{Corsaro2014,Corsaro2015}. Finally, using the Flicker in Power \citep[FliPer,][]{Bastien2016,Bugnet2018} random forest methodology, the module will be able to provide a model-independent estimate of the logarithm of the stellar surface gravity parameter ($\log g$, Submodule~4).

\section{A dataset of simulated light curves \label{sec:simulation}}


\subsection{Light-curve simulation procedure \label{sec:simulation_procedure}}

\begin{figure}[ht!]
    \centering
    \includegraphics[width=0.49\textwidth]{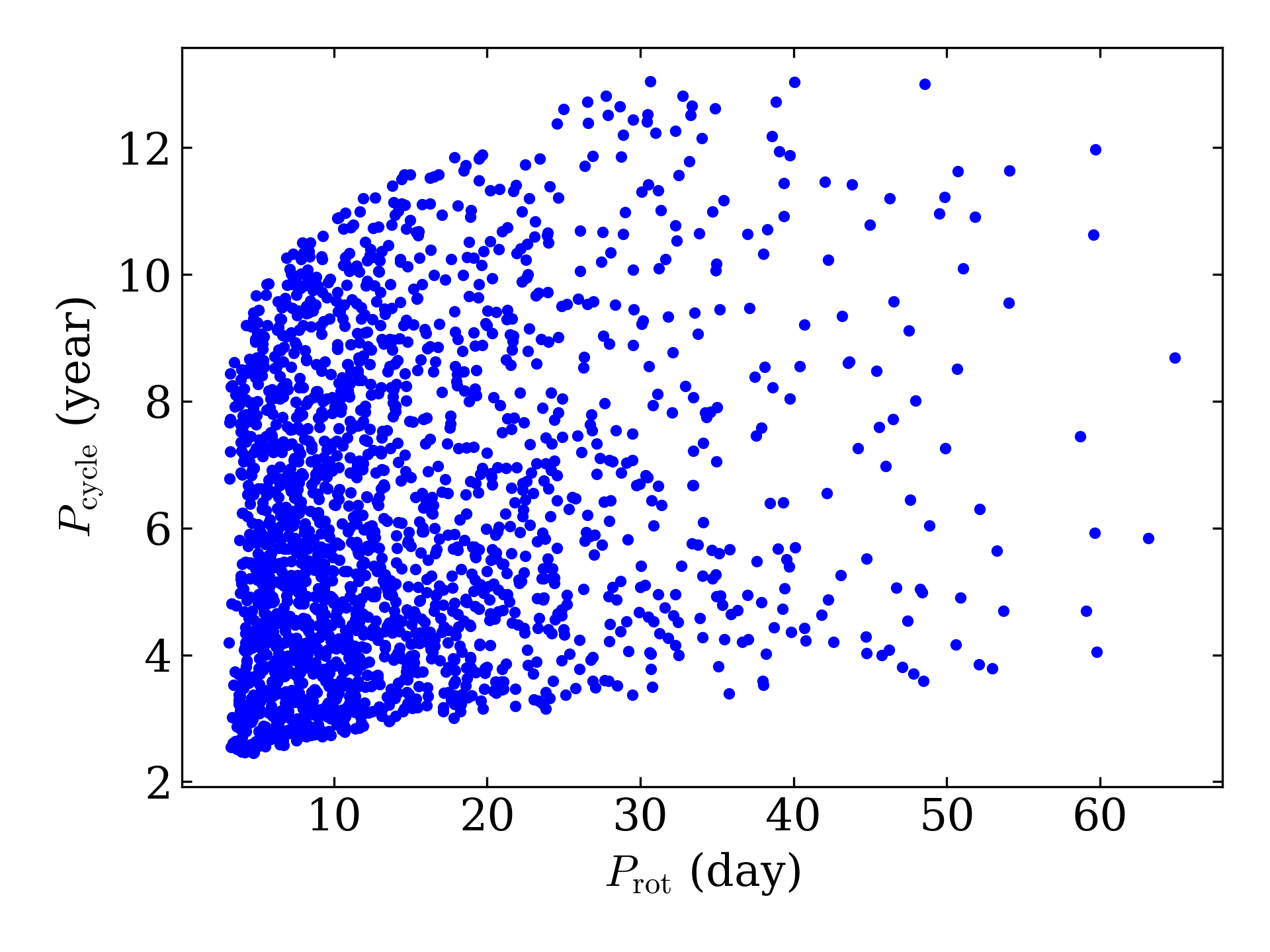}
    \caption{Rotation period $P_\mathrm{rot}$ and cyclic periodicity $P_\mathrm{cycle}$ distribution of the simulated sample.}
    \label{fig:prot_pcycle_distribution}
\end{figure}

\begin{figure}[ht!]
    \centering
    \includegraphics[width=0.49\textwidth]{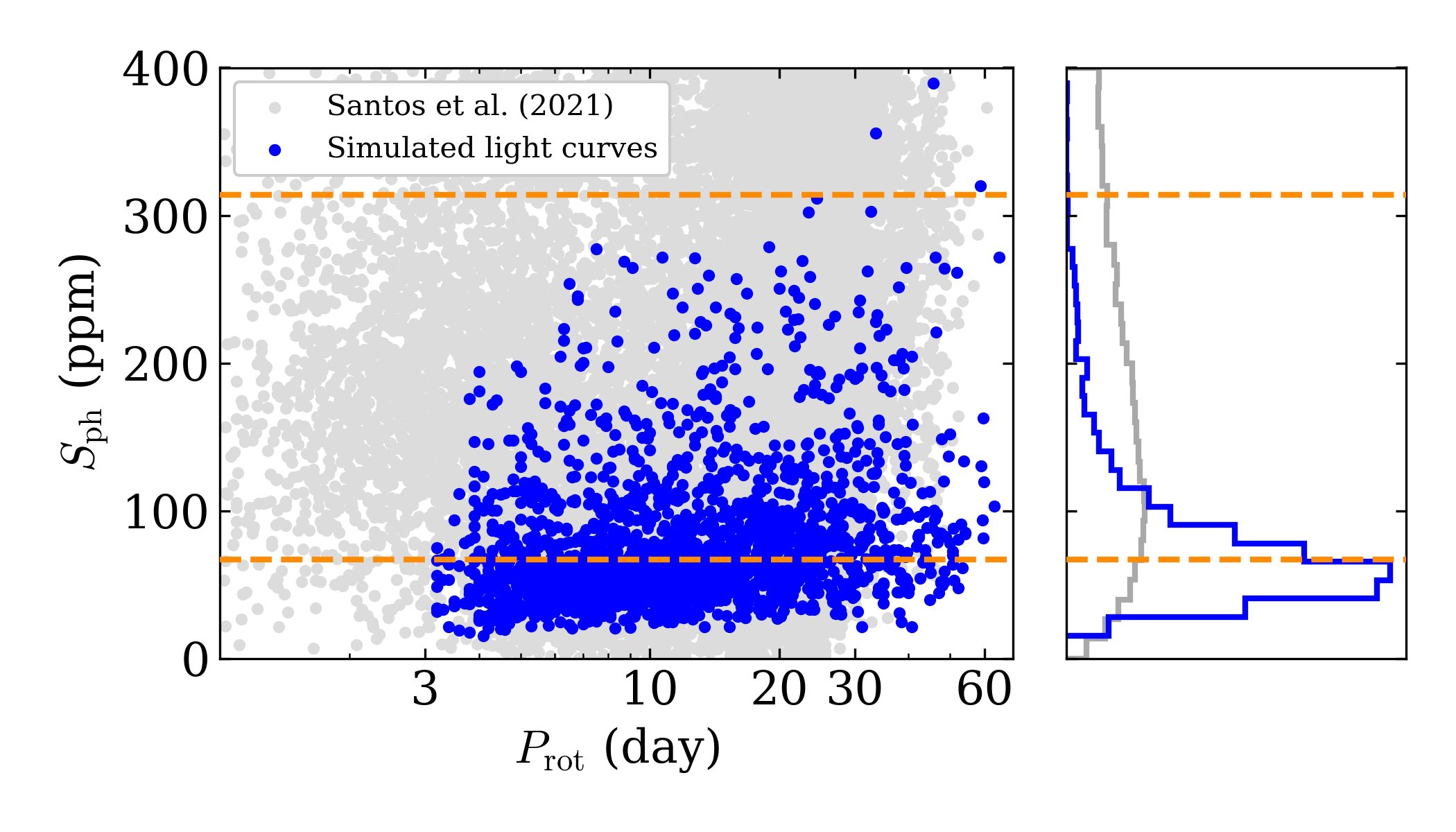}
    \caption{Rotation period $P_\mathrm{rot}$ and activity index $S_\mathrm{ph}$ distribution of the simulated sample (blue) compared with the sample from \citet[][grey]{Santos2021}. The right panel shows the projected $S_{\rm ph}$ density distribution for the two populations. The projected density distribution for the \citet{Santos2021} targets is computed considering only stars with $S_\mathrm{ph}<400$~ppm. The dashed horizontal orange lines correspond to the solar minimal and maximal values measured by \citet{Salabert2016}.}
    \label{fig:sph_distribution}
\end{figure}

\begin{figure*}[ht!]
    \centering
    \includegraphics[width=0.98\textwidth]{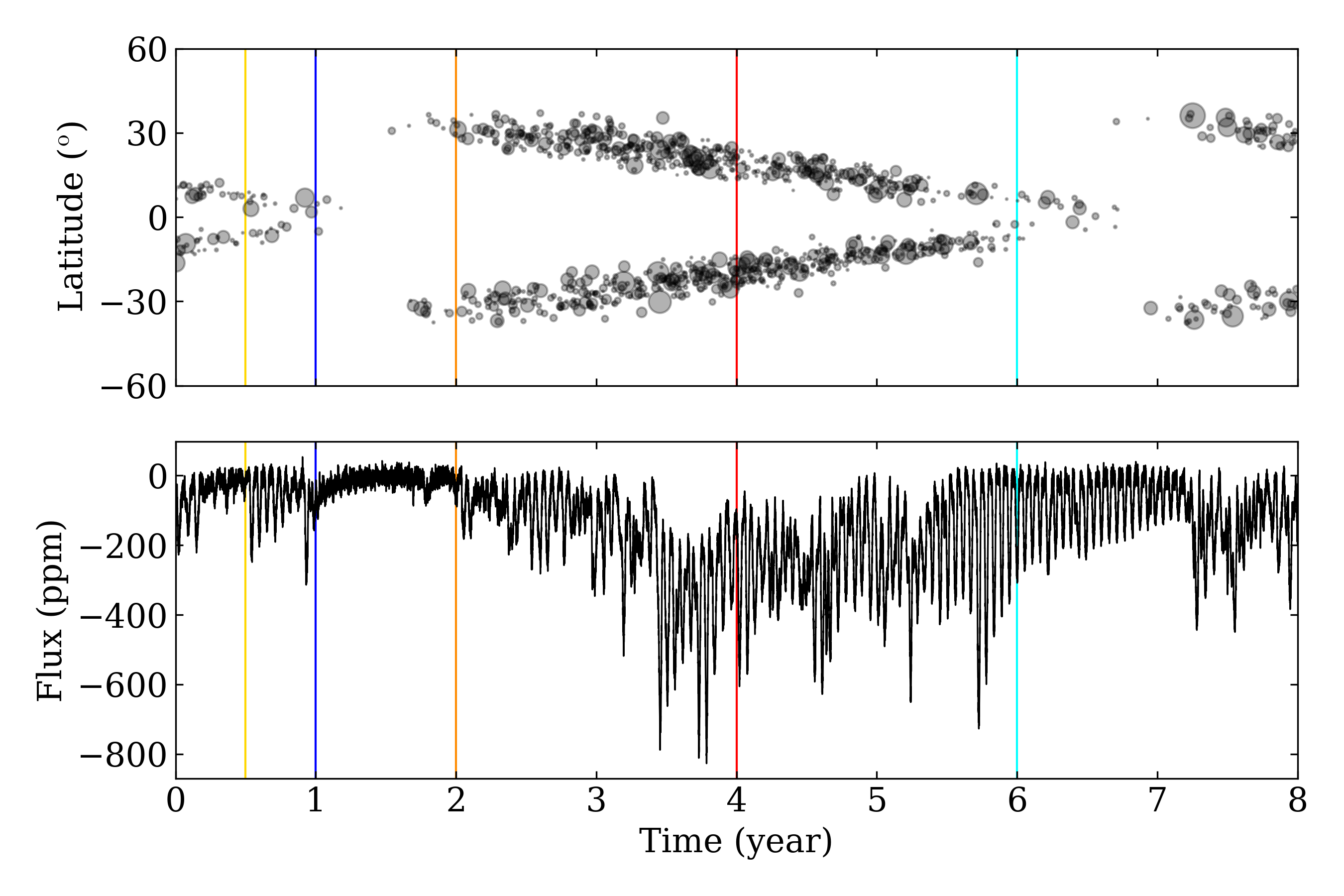}
    \caption{Example of spot activity included in the simulations (\textit{top}) and the corresponding 8 year light curve, including convective granulation (\textit{bottom}). The simulation has $P_\mathrm{rot} = 21$~days, $P_\mathrm{cycle} = 5.4$~years, $\delta_\mathrm{cycle} = 0.45$~years, and $i = 83^{\rm o}$. The vertical coloured lines highlight the extent of the light-curve segments considered in this work in addition to the full 8 year light curves: 6 months (yellow), 1 year (blue), 2 years (orange), 4 years (red), and 6 years (cyan).}
    \label{fig:example_simulation}
\end{figure*}

The rotational modulations are simulated with the \texttt{pyspot} code following the procedure described in \citet{Aigrain2015}. 
{The \texttt{pyspot} code allows simulation of the brightness variations induced by stellar dark spots while omitting the facular contribution and neglecting limb-darkening effects.}
Simulated spots emerge close to active latitudes and decay with time, with a decay rate following a log-normal distribution \citep{MartinezPillet1993}. 
Each simulated time series has an activity cycle period, $P_\mathrm{cycle}$. During the course of the cycle, active latitudes, spot number, and area vary. 
As in the solar case, active centres migrate from high to low latitudes during the course of the activity cycle, allowing the reconstruction of a butterfly diagram for each simulation. 
Additionally, the overlap between cycles is included \citep[e.g.][]{Wilson1988}. 
Differential rotation is also taken into account as the rotation period of the simulated spots is a function of their latitude of emergence. {The differential rotation law used in this work is solar-like for all simulations ---that is the rotation period increases with latitude. No active nesting is included in the simulations ---in other words, there is no preferred active longitude for the emergence of the spots.} 
The contribution of convective granulation is included in the simulations as Harvey models \citep{Harvey1985}, considering the scaling laws inferred by \citet{Kallinger2014}.
{More details are provided on this aspect in Appendix~\ref{appendix:granulation_level}.}

Given that they encompass only spot rotational modulation ({no faculae}), spot evolution, granulation, and cyclic activity, the set of simulations we generate here is ideal. When using them to test the algorithms described in Sect.~\ref{sec:module_description}, we assume that we are dealing with single stars with light curves where planetary transits and flares have been completely removed.

In practice, using the PLATO Input Catalog (PIC) from \citet{Montalto2021}, we select 2000 stars with parameters meeting the apparent magnitude $V$ and effective temperature $T_\mathrm{eff}$ pre-requisites necessary to be included in the {PLATO P1 or P2 sample. The P1 sample will consist of F5 to K7 dwarf and subgiant stars with magnitudes of $8.5 \leq V \leq 11$, while the P2 sample will consist of F5 to K7 dwarf and subgiant stars with magnitudes of $V \leq 8.5$.}
The main parameters of each simulation are the stellar spin inclination angle $i$, the average rotation period $P_\mathrm{rot}$ (used as the reference value for our period recovery exercise), the minimum and maximum rotation periods, $P_\mathrm{min}$ and $P_\mathrm{max}$, the maximum latitude of activity $\lambda_\mathrm{max}$, the activity cycle length $P_\mathrm{cycle}$, and the overlap between consecutive cycles $\delta_\mathrm{cycle}$. These are generated drawing from a distribution that depends on the effective temperature $T_\mathrm{eff}$, using the prescriptions from \citet{Meunier2019}. The method used to draw these parameters is described in detail in Appendix~\ref{appendix:draw_rot}. 
{In particular, it should be noted that $P_\mathrm{rot}$, $P_\mathrm{min}$, and $P_\mathrm{max}$ are chosen such that $P_\mathrm{rot}$ is the average between $P_\mathrm{min}$ and $P_\mathrm{max}$, as can be seen from Eq.~(\ref{eq:differential_rotation}).} 
We also account for stellar inclination in the simulations: each star has an inclination $i$ drawn considering a uniform $\sin i$ distribution to ensure that the orientation of the stellar rotation axis is isotropic.

The rotational and activity modulations in the light curves were generated with the \texttt{pyspot} code. We used the PLATO Solar-Like Simulator\footnote{The PSLS documentation is accessible at \url{https://sites.lesia.obspm.fr/psls}. We used \texttt{v1.5} of the PSLS in this work.} \citep[PSLS,][]{Samadi2019} in order to add PLATO instrumental systematics, camera random noise, and convective granulation. 
We generated 2000 light curves with rotational modulations. With the same input parameters, we generated 2000 other light curves that include only the granulation signal. 
Here, these light curves without rotational modulation were used to train the ROOSTER algorithms to distinguish light curves with  a rotation signature from light curves without.

The $P_\mathrm{rot}$ versus $P_\mathrm{cycle}$ distribution of the simulated sample is shown in Fig.~\ref{fig:prot_pcycle_distribution}. 
The $P_\mathrm{rot}$ values range from a few days to a few tens of days, while $P_\mathrm{cycle}$ are comprised within 2 and 14 years. For this reason, in this work, only light curves longer than 2 years encompass complete activity cycles for some of the simulated targets.
Although, in our simulations, stars with short rotation periods are more likely to have shorter activity cycles \citep[see e.g. Fig.~6 from][]{Meunier2019}, there is no strong correlation between $P_\mathrm{rot}$ and $P_\mathrm{cycle}$ in the simulated sample (see Eq.~(\ref{eq:prot_pcycle}) for the relation between $P_\mathrm{rot}$ and $P_\mathrm{cycle}$).
The actual connection between $P_\mathrm{rot}$ and $P_\mathrm{cycle}$ is probably more complex, and both the form of the correlation and its origin are debated \citep[see e.g.][]{Bonanno2022}

In order to compare the variability amplitude of our simulations to an observational sample, we computed the $S_\mathrm{ph}$ distribution of the simulated light curves. We represent this distribution in Fig.~\ref{fig:sph_distribution}, and we compare the $P_\mathrm{rot}$ versus $S_\mathrm{ph}$ of our simulated sample with the \textit{Kepler} targets from \citet{Santos2021}. We also show the maximum and minimum level of photometric solar variability as measured by \citet{Salabert2016}. It appears that, in terms of $\left< S_\mathrm{ph} \right>$, our simulated light curves coincide with the moderate-to-low-activity \textit{Kepler} targets from \citet{Santos2021}. The vast majority of them have $\left< S_\mathrm{ph} \right>$ below the maximal solar level, and about half of them are below the minimal solar level of activity. We underline that it is important to focus on these types of targets, as these are the ones for which measurement of surface rotation is likely to prove challenging because of their low amplitude of variability. Moreover, these stars are also the ones for which the detection of acoustic oscillations (p modes) is expected, as high levels of magnetic activity inhibit stochastic pulsations in solar-type stars \citep[see][]{Chaplin2011b,Mathur2019} 


Each light curve is generated on a 8 year baseline, which is the maximum time frame over which a field may be observed by the PLATO mission\footnote{In this scenario, PLATO would observe only one of the two selected long-pointing fields \citep{Nascimbeni2022}. While the first PLATO long-pointing field was announced by the PLATO Science Working Team on 11 June 2023, the total amount of time that PLATO will remain in this field is conditioned on a mission performance assessment after one year of flight. See the PLATO first field announcement: \url{https://platomission.com/2023/07/11/first-plato-long-duration-observation-phase-lop-field-selected/}.}. 
{It should be noted that, in order to save computing time and storage space, after having validated that this did not introduce any bias in the photon noise level of the noisy simulations (see Appendix~\ref{appendix:photon_noise}), we directly generated the light curves with the two-hour sampling that will be used by the rotation and activity module.}

In order to provide a picture of the ability of the module at the different stages of the mission, from the first data release to legacy catalogues, we performed this analysis on simulations for several different baselines: 6 months, and 1, 2, 4, 6, and 8 years.  
We show example simulated light curves in Fig.~\ref{fig:example_simulation}, with $P_\mathrm{rot} = 21$~days, $P_\mathrm{cycle} = 5.4$~years, $\delta_\mathrm{cycle} = 0.45$~years, and $i=83^\mathrm{o}$. 
The vertical coloured lines highlight the extent of the segments corresponding to each temporal baseline.  
At the beginning of the light curve, the simulated star is in the descending phase of its activity cycle, and the latitude of spot formation gradually comes closer to the equator. The appearance of spots at a higher latitude marks the beginning of a new cycle. It is interesting to note that, considering the different baseline we are using in this work, we are going to analyse the light curves at different stages of the activity cycle. In the example of Fig.~\ref{fig:example_simulation}, for light curves shorter than two years, only the minimum of activity of the star can be observed, and thus the $\left< S_\mathrm{ph} \right>$ we would be inferring shall be biased towards small values. A four-year baseline enables us to additionally monitor the rising phase of the activity cycle \citep[e.g. in the case of the \textit{Kepler} light curve of the bright solar analogue HD 173701; see e.g.][]{Karoff2018}. A 6 year baseline is longer than the stellar activity cycle but the descending phase of the current cycle is missing. Finally, the 8 year baseline allows observation of  the entire cycle and also includes the first year of a new cycle.

\subsection{PLATO systematics and random noise}

The PLATO systematics included in the light curves were simulated using the PSLS abilities described in \citet{Samadi2019}. For each simulation, we consider the case where the star is observed together by the four groups of six cameras of PLATO \citep[see e.g.][]{Nascimbeni2022}. 
The random noise in each simulation is computed with the \texttt{PLATO\_SIMU} option, considering the target input $V$ magnitude and the reference PLATO simulations included in PSLS, with a medium drift level. 
Our set of simulations therefore represents what we can expect from good quality targets located at the centre of the PLATO field of view. 

\subsection{Correction of systematics}

In order to analyse the noisy light curves with the algorithms described in Sect.~\ref{sec:module_description}, we implement a correction strategy, with the aim being to produce corrected light curves that are similar ---in terms of their properties--- to those that will be provided to the rotation and activity PLATO module. 
The first step is to apply a low-pass filter with a 400 day cutoff in order to isolate long-term variability that could be related to stellar cyclic activity. The extracted signal is subtracted from the light curves, which are then folded on a quarter-period of 90 days. At this step, the quarter modulation trend is extracted as the median value at each phase of the folding. The trend is then subtracted quarter-wise from the light curve and the filtered low trend is added again.  

\begin{figure}[ht!]
    \centering
    \includegraphics[width=0.49\textwidth]{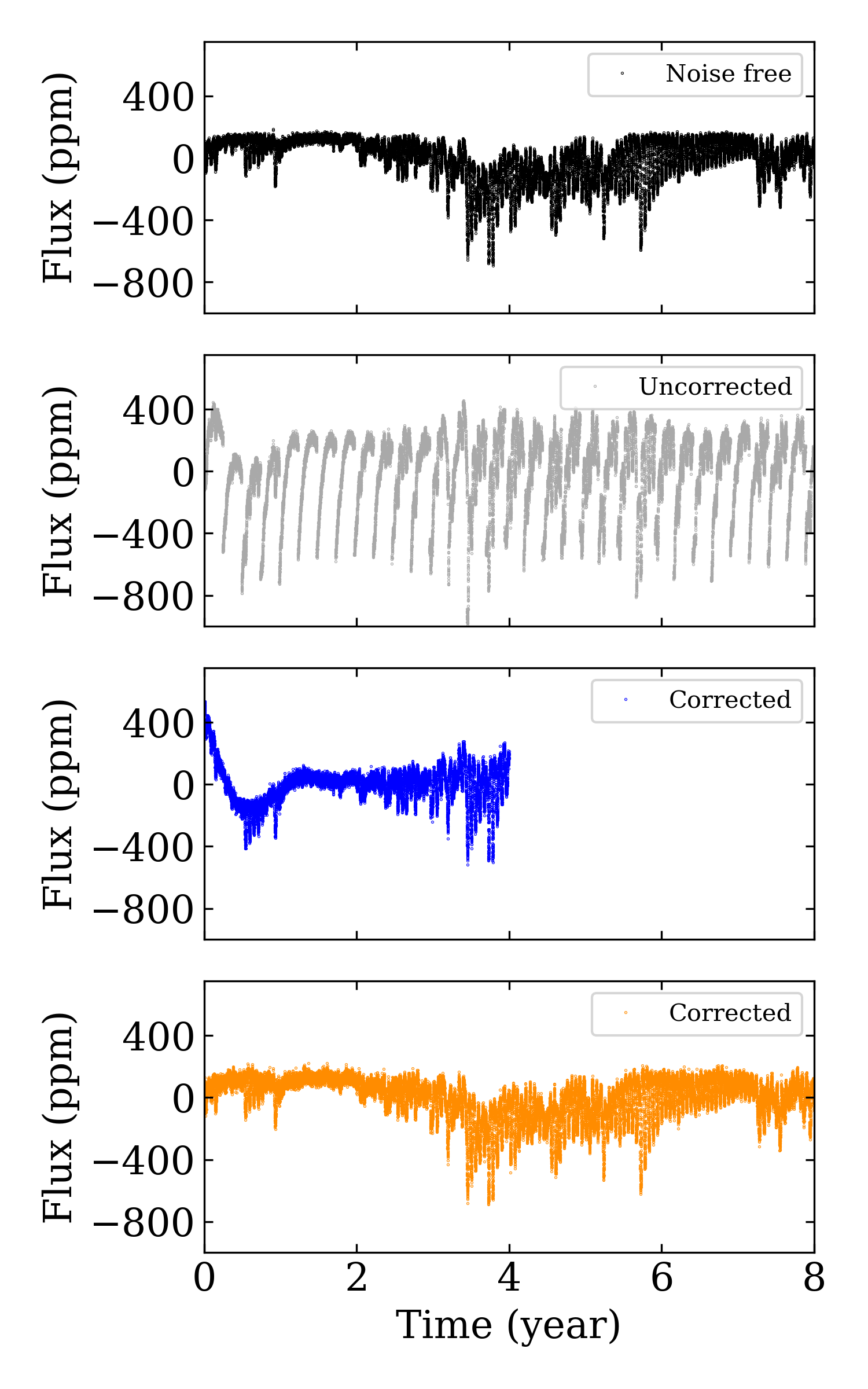}
    \caption{Comparison between the 8 year noise-free light curve (\textit{black}), the uncorrected 8 year noisy light curve (\textit{grey}), the corrected 4 year light curve (\textit{blue}), and the corrected 8 year light curve (\textit{orange}).}
    \label{fig:correction_example}
\end{figure}

The two upper panels of Fig.~\ref{fig:correction_example} show how inclusion of the PLATO systematics affects the light curve presented in Fig.~\ref{fig:example_simulation}. 
The effect of the drift during each 90 day quarter is apparent. 
In order to illustrate how a given segment of the corrected light curve can vary depending on the total observation length considered for the correction, the two lower panels show the light curve obtained by applying our correction procedure on the first half of the light curve and on its full extent, respectively. We note that a remnant of the quarter modulation is still clearly visible at the beginning of the 4 year corrected light curve.

\begin{figure}[ht!]
    \centering
    \includegraphics[width=0.49\textwidth]{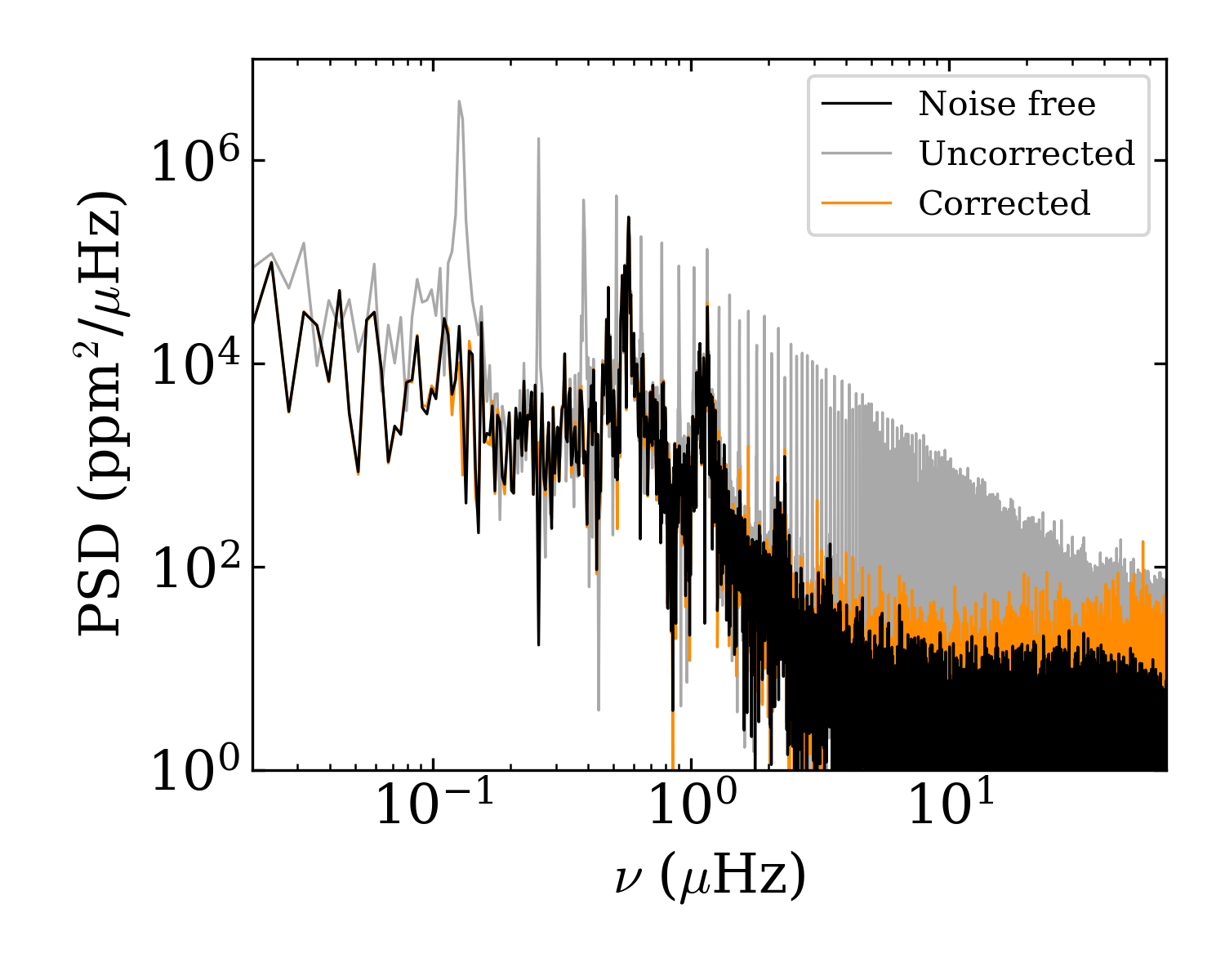}
    \caption{Comparison between the PSD obtained from the noise-free light curve (\textit{black}), the uncorrected noisy light curve (\textit{grey}), and the corrected noisy light curve (\textit{orange}).}
    \label{fig:correction_example_psd}
\end{figure}

 Figure~\ref{fig:correction_example_psd} shows a comparison of the power spectral density (PSD) of the noise-free light curve, of the uncorrected noisy light curve, and of the corrected noisy light curve. We clearly see that the PSD of the uncorrected noisy light curves is dominated by the harmonics of the 90 day quarter modulation. For the vast majority of the simulated noisy light curves, our correction procedure allows removal of the power contribution of this modulation almost entirely, as illustrated here. The impact of the random noise in the corrected light curves is visible in the PSD beyond a few $\mu$Hz as a higher flat noise level in the corrected PSD than in the noise-free PSD. 

\section{Performances on simulated light curves \label{sec:application}}

\subsection{Recovery of the rotation period \label{sec:rotation_period_recovery}}

\begin{figure}[ht!]
    \centering
    \includegraphics[width=0.49\textwidth]{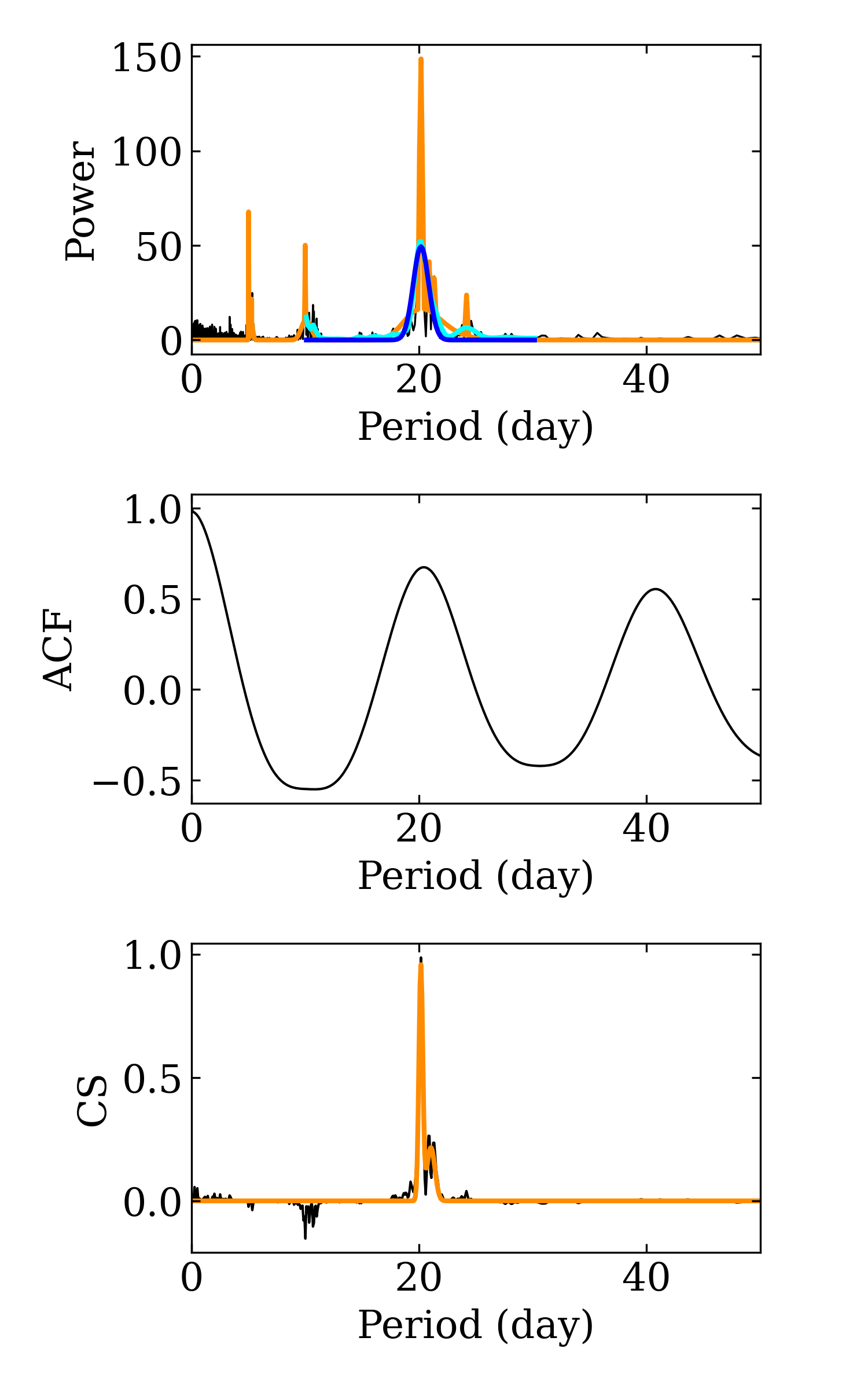}
    \caption{Example of GLS periodogram (\textit{top}), ACF (\textit{middle}), and CS (\textit{bottom}) analyses performed to recover the rotational modulation of the light curve shown in Fig.~\ref{fig:example_simulation}, using the {full} time series. 
    {The GLS periodogram was normalised to account for the background contribution according to the method described in Sect.~\ref{sec:average_rotation_period}.}
    The {set of Gaussian profiles fitted on the GLS periodogram and the CS} are shown in orange.
    {The smoothed periodogram is shown in cyan, while the Gaussian profile fitted on it and centred on $P_{\rm LS}$ is shown in blue.}
    }
    \label{fig:example_analysis_rotation}
\end{figure}

The 8000 simulated light curves (2000 with rotation and 2000 without rotation for both the noise-free and the corrected noisy cases) are analysed homogeneously, considering the periodogram, the ACF, and the CS (see Sect.~\ref{sec:average_rotation_period}). In Fig.~\ref{fig:example_analysis_rotation}, we show an example of this analysis. The light curve used for the example is the noise free light curve shown in Fig.~\ref{fig:example_simulation}, and in this case, the analysis is performed on a {full time series} after applying a high pass filter with 60 day cutoff {to obtain the LC2 input} as explained in Sect.~\ref{sec:inputs} {(see also Appendix~\ref{appendix:filtering})}. 
The main rotation peak around 21~days is fitted both in the periodogram and in the CS. 
{As seen in the periodogram, additional peaks are also fitted, namely the second and third harmonics of the average rotation period, close to 10 and 5~days, respectively.}
The ACF exhibits a very clear periodicity related to the rotational modulation. Because of the decreased signal coherence, the envelope amplitude of the ACF decays as the shift in period increases \citep{Santos2021_ACF}.

\begin{table}[t!]
    \centering
    \caption{Summary of the results obtained for the ROOSTER \textit{RotClass} classifier dedicated to assess the presence of a rotational modulation.}
    \begin{tabular}{cccc}
    \hline\hline
              & Noise free & Noisy \\
    Baseline  &  Accuracy (\%) & Accuracy (\%) \\
    \hline
    6 month   & 99.2 & 98.5 \\
    1 year    & 99.4 & 98.7 \\
    2 years   & 99.9 & 98.9 \\
    4 years   & 100 & 99.0 \\
    6 years   & 100 & 99.5 \\
    8 years   & 100 & 99.5 \\
    \end{tabular}
    \label{tab:rooster_summary_0}
\end{table}

\begin{figure*}[ht!]
    \centering
    \includegraphics[width=0.49\textwidth]{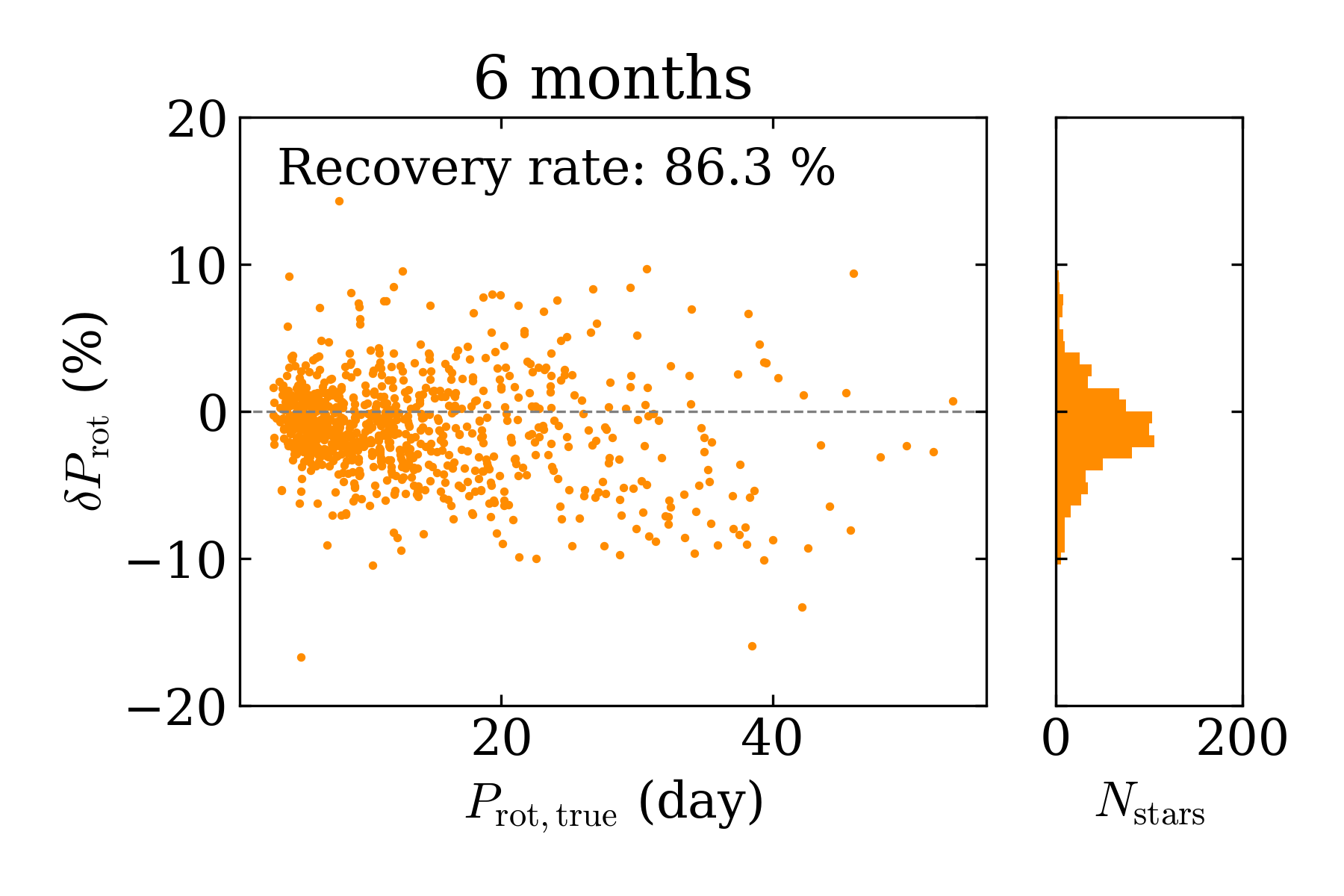}
    \includegraphics[width=0.49\textwidth]{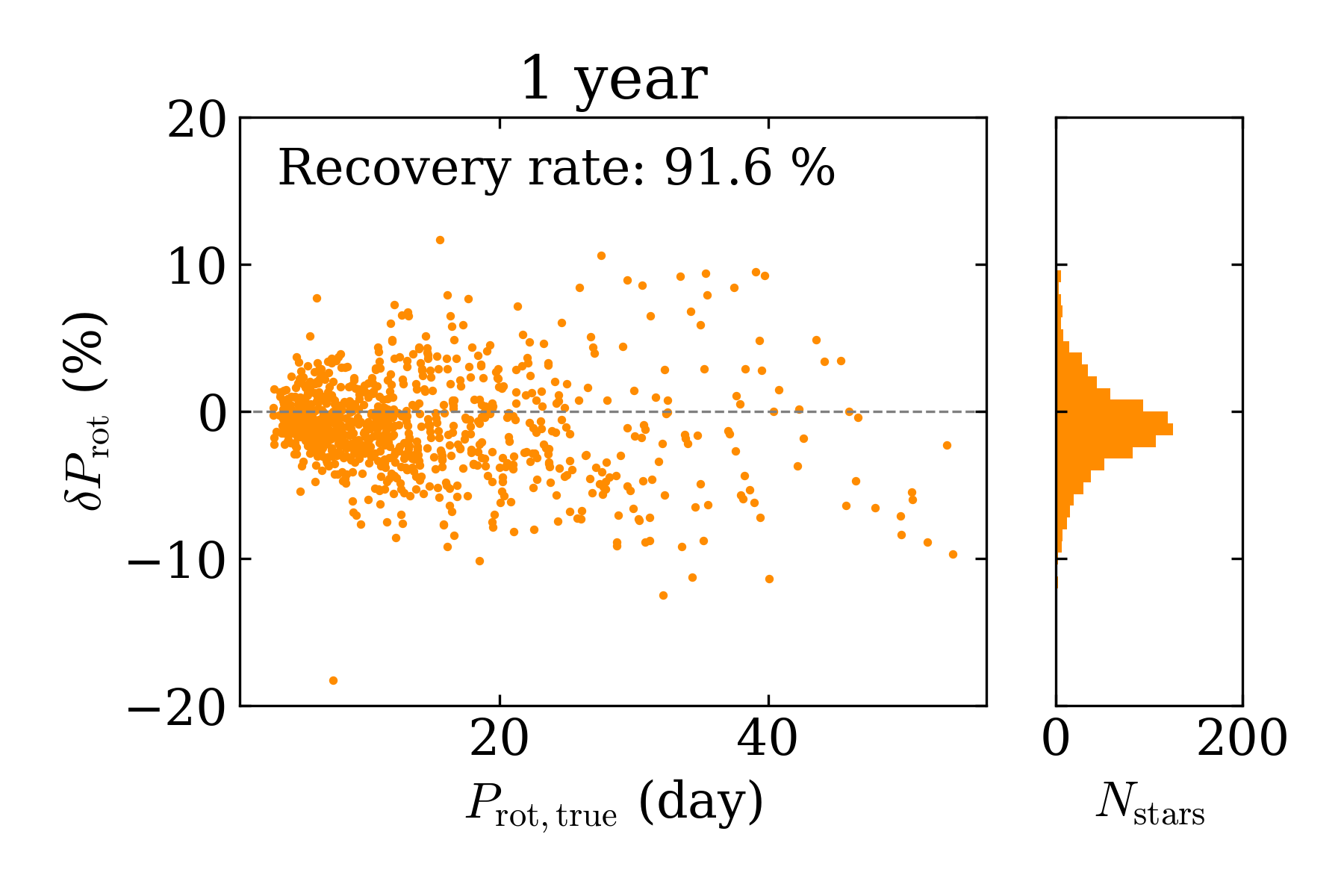}
    \includegraphics[width=0.49\textwidth]{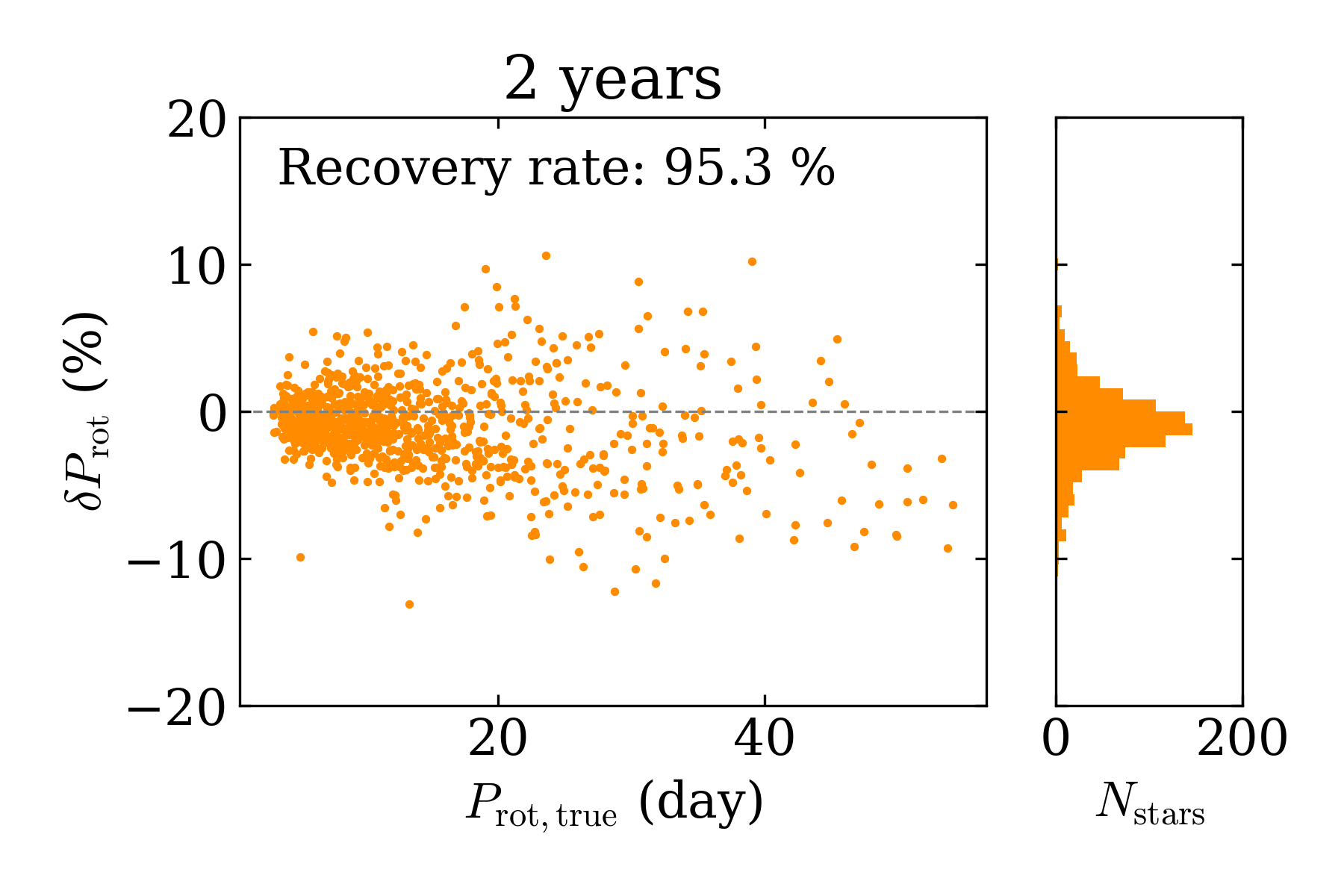}
    \includegraphics[width=0.49\textwidth]{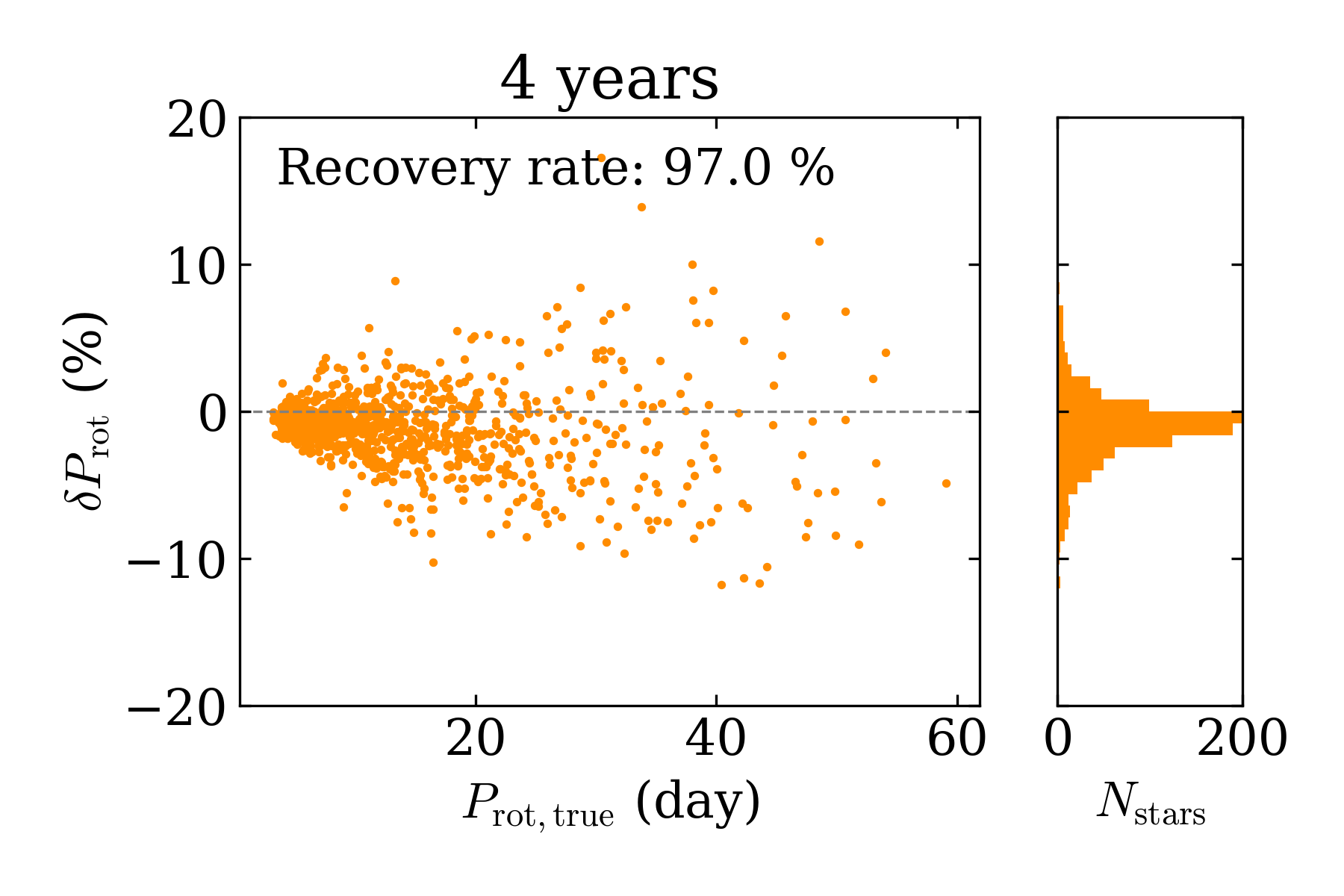}
    \includegraphics[width=0.49\textwidth]{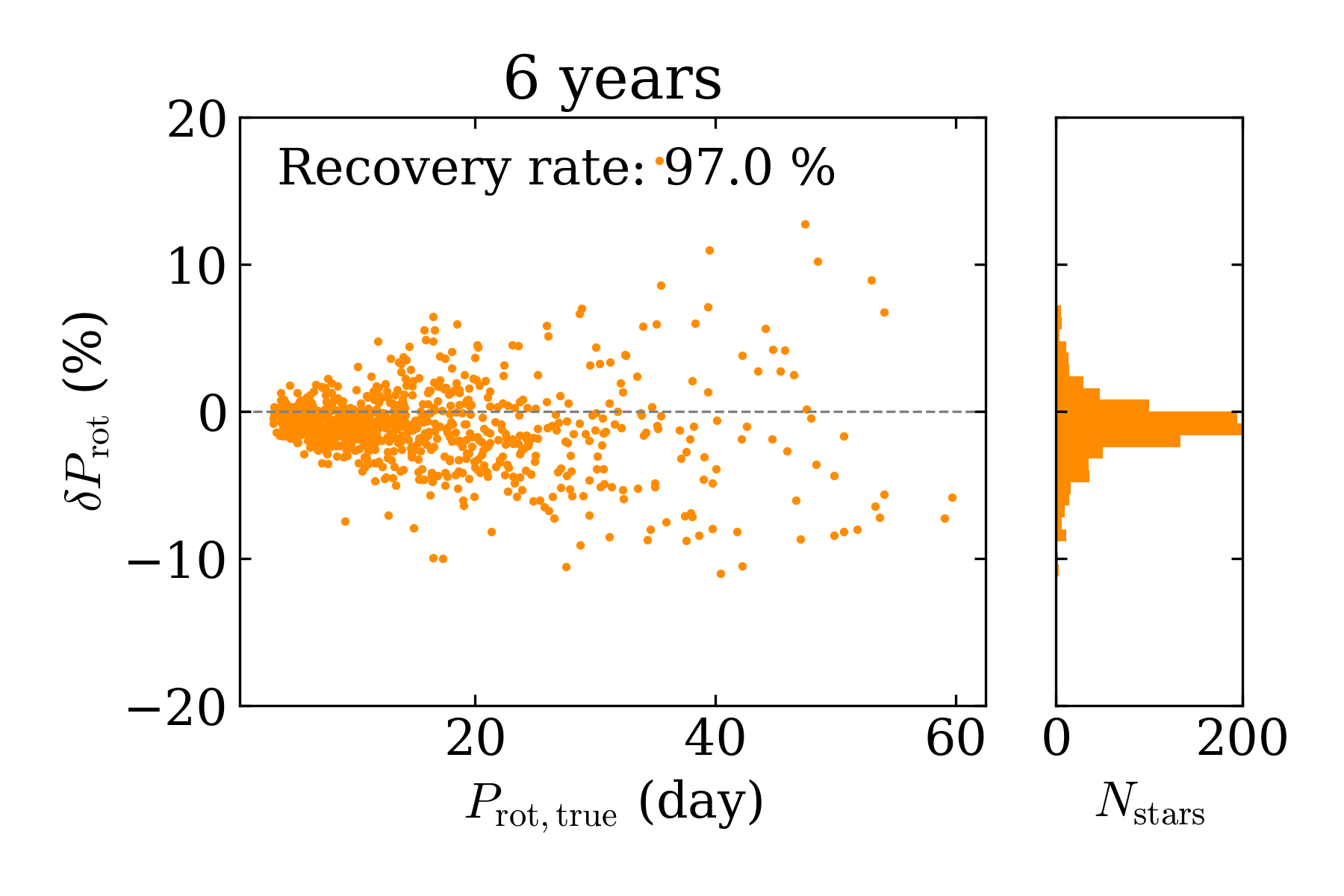}
    \includegraphics[width=0.49\textwidth]{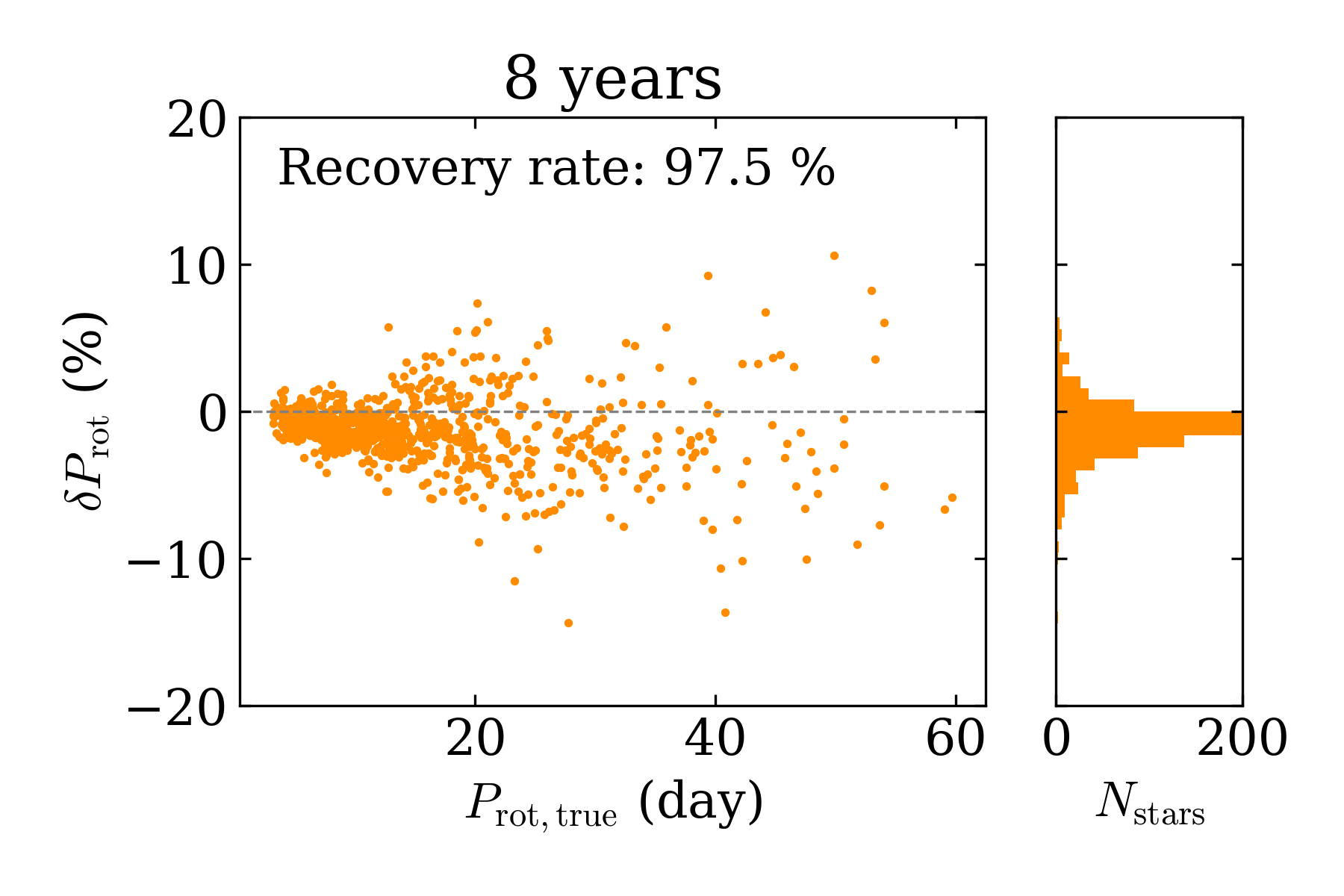}
    \caption{Rotation period recovery for the different temporal baselines considered in this work, in the case of noise-free light curves: 6 months (\textit{top left}), 1 year (\textit{top right}), 2 years (\textit{middle left}), 4 years (\textit{middle right}), 6 years (\textit{bottom left}), and 8 years (\textit{bottom right}). The histogram to the right of each panels shows the $\delta P_\mathrm{rot}$ distribution for the complete tested sample.}
    \label{fig:rotation_period_recovery}
\end{figure*}

\begin{figure*}[ht!]
    \centering
    \includegraphics[width=0.49\textwidth]{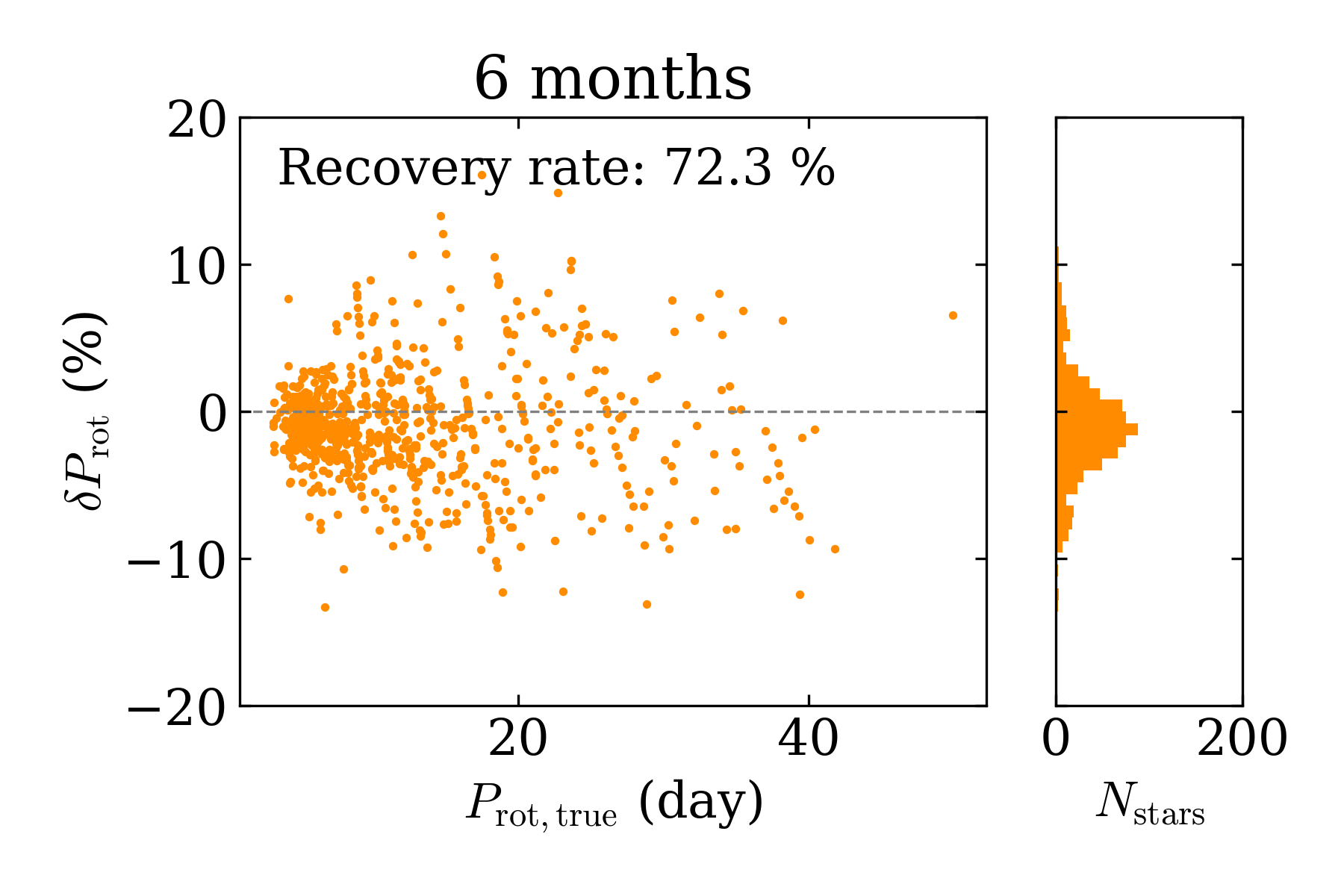}
    \includegraphics[width=0.49\textwidth]{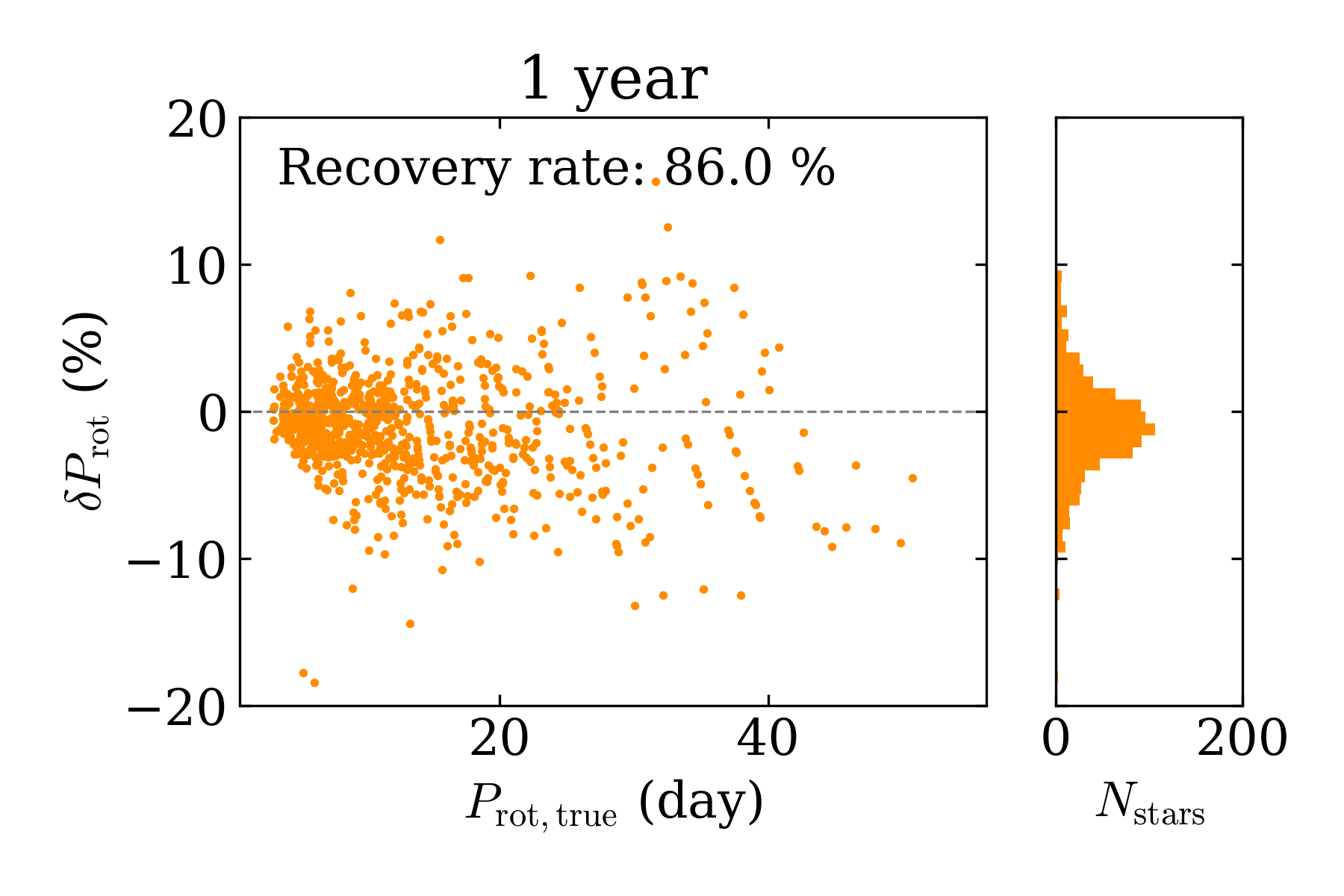}
    \includegraphics[width=0.49\textwidth]{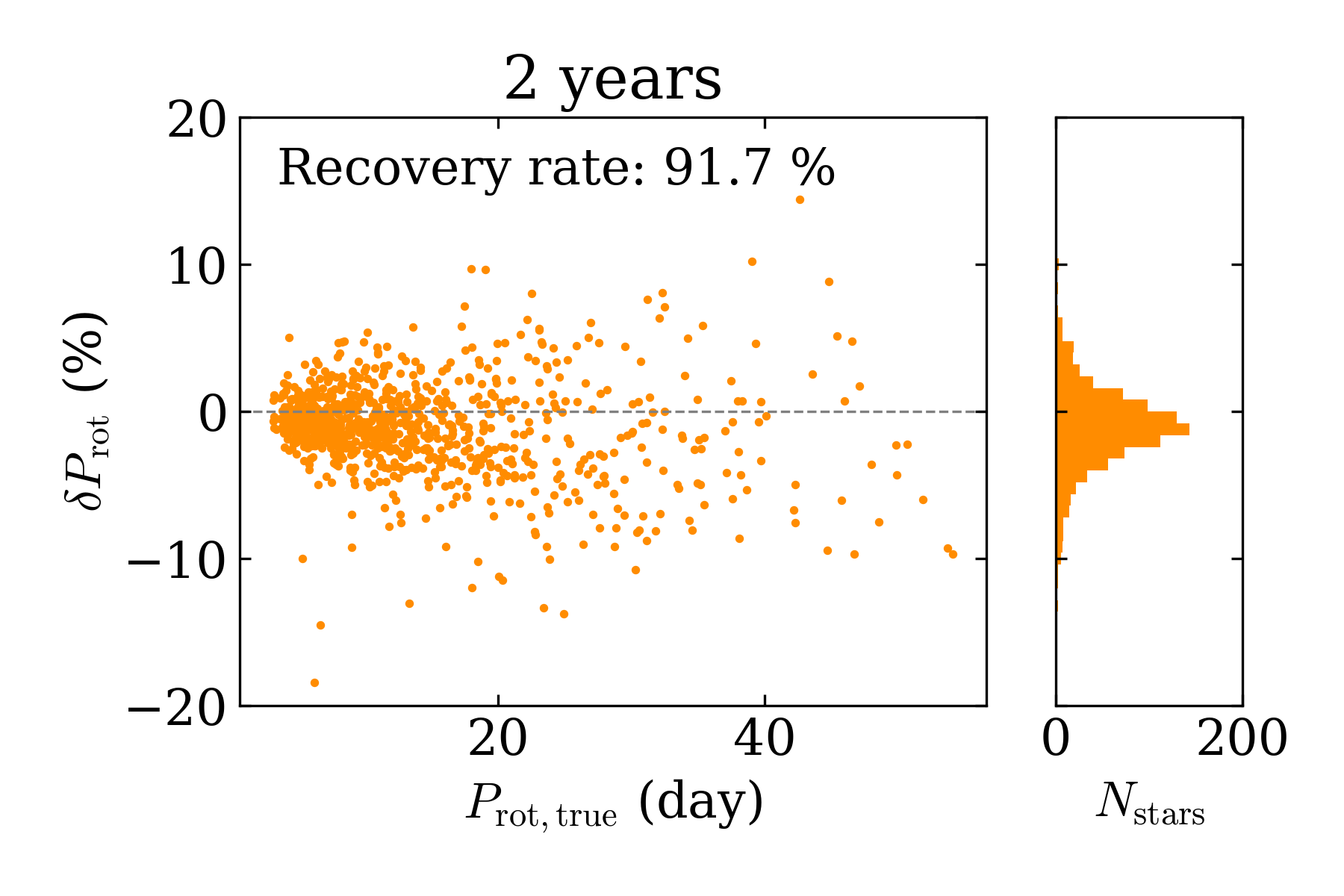}
    \includegraphics[width=0.49\textwidth]{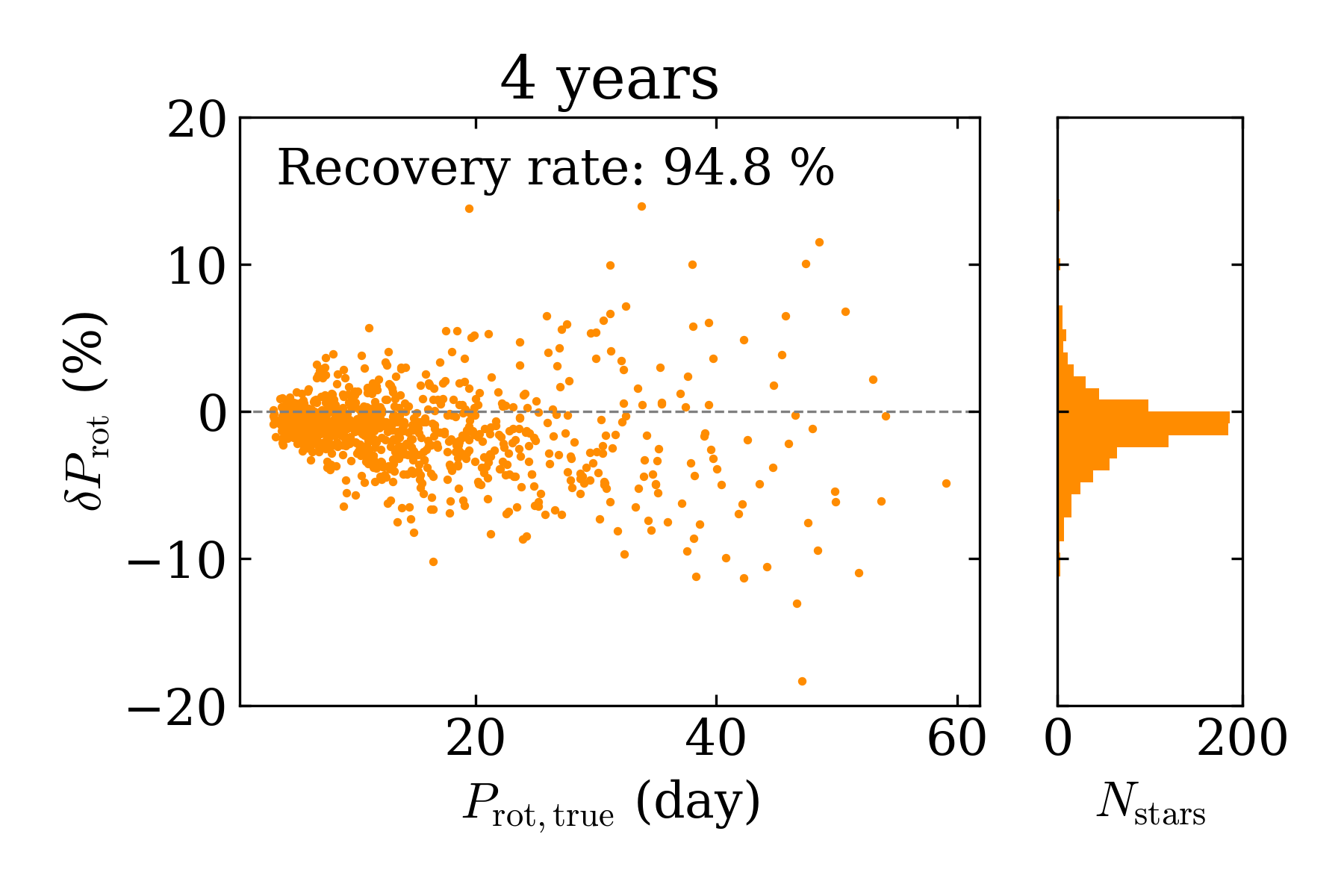}
    \includegraphics[width=0.49\textwidth]{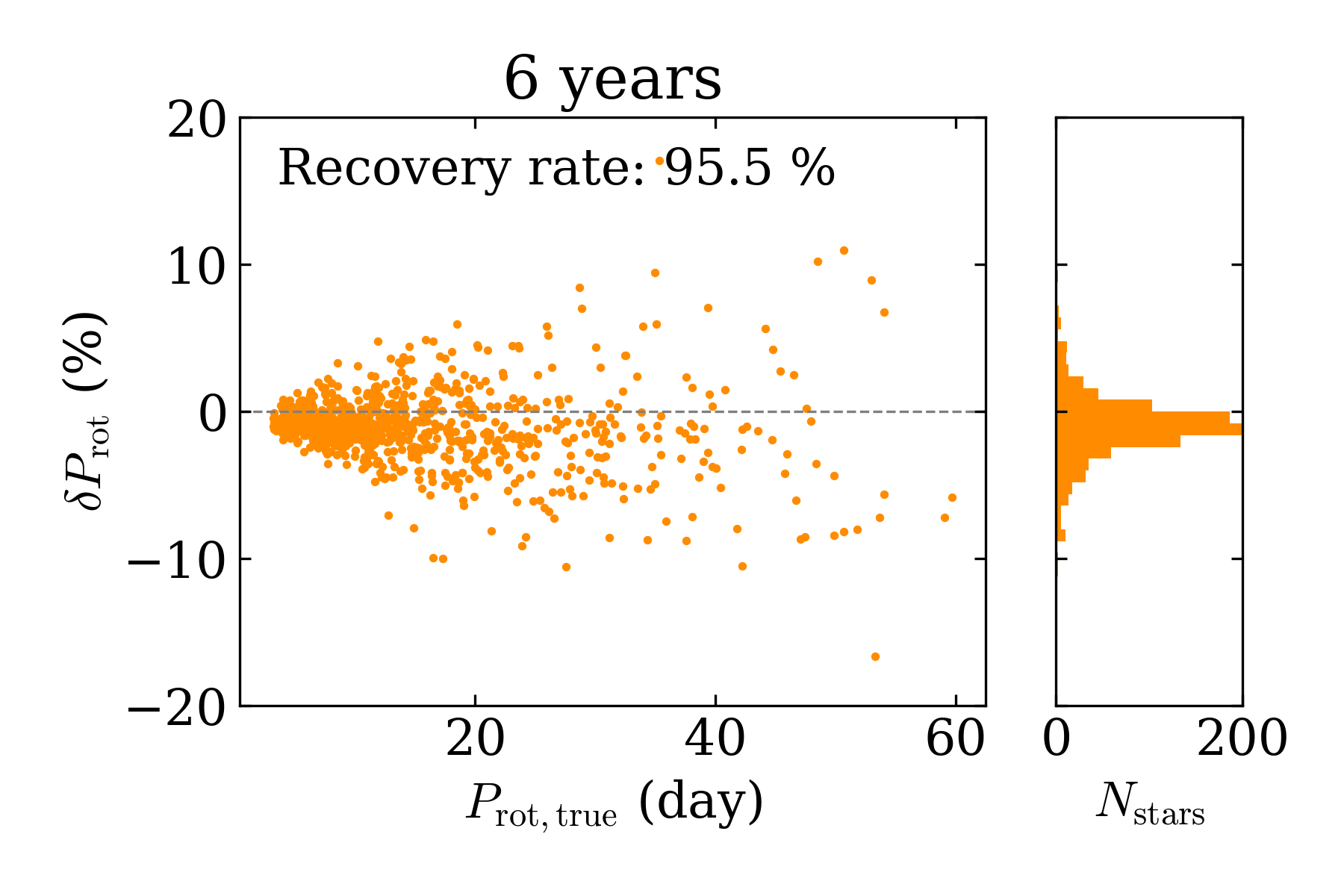}
    \includegraphics[width=0.49\textwidth]{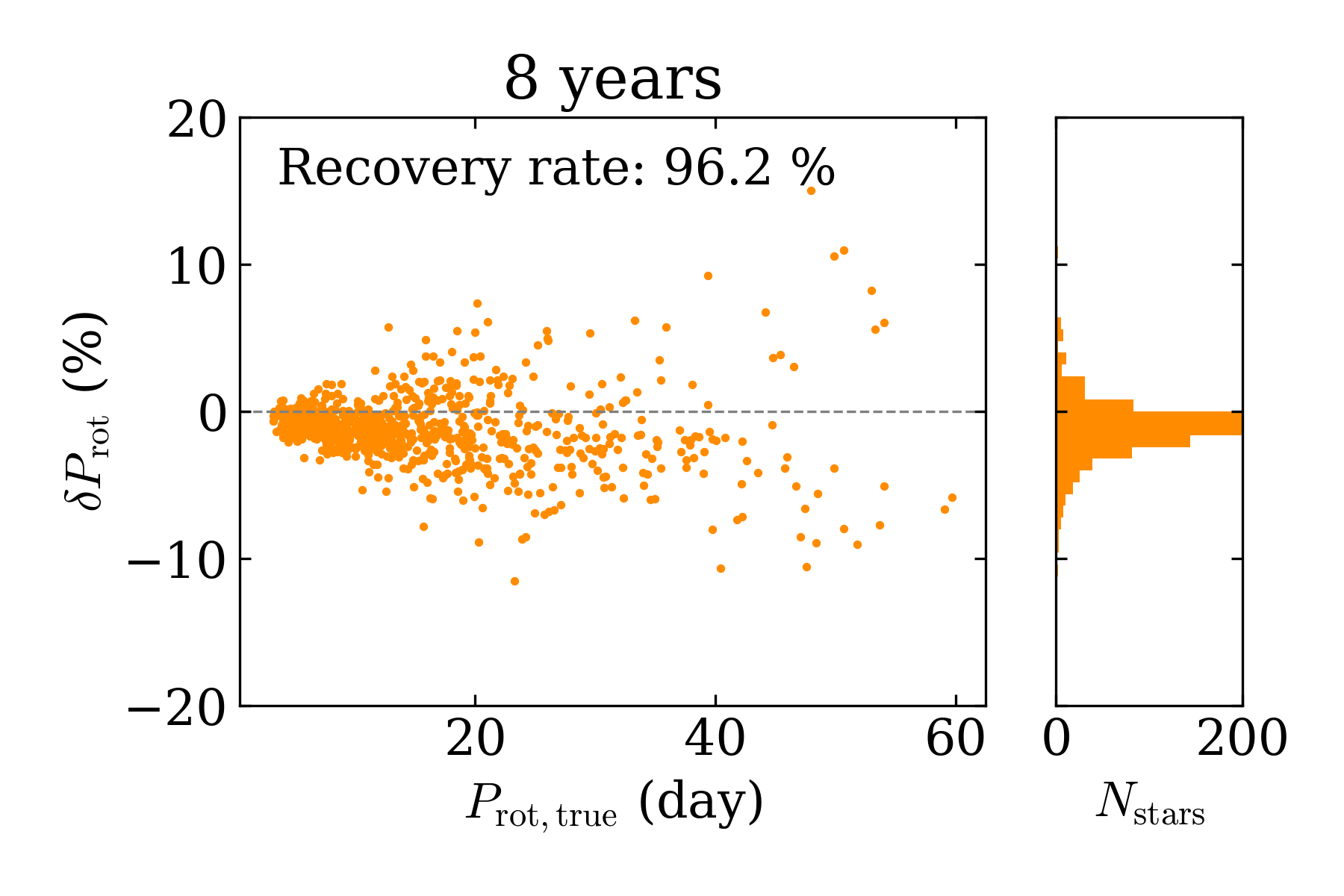}
    \caption{Same as Fig.~\ref{fig:rotation_period_recovery} but for the case of corrected noisy light curves.}
    \label{fig:rotation_period_recovery_corr}
\end{figure*}

The ROOSTER methodology is applied after this first step. The random forest classifiers have to be trained beforehand. To this purpose, we split the light curves between a training set and a test set. 
We consider noise-free and corrected noisy light curves  separately.
Initially, each set includes 1000 light curves with rotational modulations and 1000 stars without rotation. However, for light curves with rotational modulations, we compare the results of the previous analysis with the true rotation period. If no candidate value coincides with the true rotation period with a 10\% tolerance, the light curve is removed from the set. This is done in the training set because we flag the candidate period closest to the true rotation period as the one that should be selected by the random forest classifier. If this happens in the test set, we consider that the methodology was not able to correctly measure the rotation period and, taking this into account, we correct the global accuracy score of the classifier dedicated to select the rotation periods.
{This corrected score corresponds to the recovery rate of our method, which, among 1000 light curves, is the fraction for which \textit{PeriodSel} provides a rotation period that is consistent with the reference value to within 10\%. This 10\% threshold is obviously arbitrary, but chosen to be reasonably commensurable to the typical uncertainties measured on $P_{\rm rot}$ (see e.g. Fig.~\ref{fig:additional_recovery_diagnostics} and \ref{fig:additional_recovery_diagnostics_corr}).}
For all considered temporal baselines, we use the same split between training and test set. We note that the analysis performed by \citet{Breton2021} demonstrated that the performances of the methodology were not dependent on the exact composition of the training set.
{We also stress that the period recovery exercise performed here is designed in such a way that the training set and the test set both incorporate the same physical and instrumental ingredients, with ground truth known in both cases. This will obviously not be the case with actual PLATO observations, which could in turn affect the algorithm performance. The composition of the training set that will be used in the PLATO pipeline has yet to be determined and describing what will be its exact composition is out of the scope of this paper. To ensure that the best performance is achieved, the light curves included in the training set will be carefully selected, with the possibility to incorporate relevant light curves from previous space missions (\textit{Kepler}/K2, TESS) together with PLATO simulations. It is planned that the module performance will be evaluated within the PLATO consortium after launch, once the first observations are available, and the possibility to include a subset of real PLATO targets with rotation measurements validated independently from ROOSTER is under consideration.}

In what follows, we report the rotation period recovery obtained for the test set.
First, {using the \textit{RotClass} classifier}, we evaluate the accuracy with which ROOSTER is able to distinguish between light curves with rotational modulations and light curves without a rotational signal. 
We note that the main issue that \citet{Breton2021} encountered at this step was related to training ROOSTER to properly distinguish genuine stellar signals from instrumental modulations from \textit{Kepler}.
{
We define the classifier accuracy as the fraction of light curves correctly classified. As summarised in Table~\ref{tab:rooster_summary_0}, for the noise-free light curves, the accuracy of \textit{RotClass} is therefore 99.2\%, 99.4\%, and 99.9\% for the 6 month, 1 year, and 2 year light curves, respectively, and is 100~\% for all other time-series lengths. 
This accuracy slightly decreases for the corrected noisy light curves because of the occasional remnant presence of periodic quarter modulations that can be confused with rotational modulation. In this case, we obtain an accuracy of 98.5\%, 98.7\%, 98.9\%, 99.0\%, 99.5\%, and 99.5\% for the 6 month, 1 year, 2 year, 4 year, 6 year, and 8 year light curves, respectively.}
When inspecting the misclassified targets, it appears that they correspond to light curves that include rotational modulations but with an apparent rotational signal of very low amplitude comparable to or below the level of remnant instrumental noise, generally because the simulated star is in a minimum of activity or because the stellar inclination is low (almost pole-on view). Nevertheless, the overall accuracy of this step remains very high even for the noisy data.

We then assess the quality of the period retrieval by ROOSTER \textit{PeriodSel}.
The summary of this analysis is shown in Fig.~\ref{fig:rotation_period_recovery} for the noise-free light curves, and in Fig.~\ref{fig:rotation_period_recovery_corr} for the corrected noisy light curves. These two figures show comparisons of the {rotation period} recovery for the different temporal baselines considered for the exercise.
We show the relative error $\delta P_\mathrm{rot} = (P_\mathrm{rot,recovered} - P_\mathrm{rot,true}) / P_\mathrm{rot,true}$ against $P_\mathrm{rot,true}$.
For light curves shorter than 2 years, it should be noted that we clearly see resolution effects from the periodogram binning on the recovered distribution. 
We also note that the spread in relative error is larger for longer periods.
For every range of periods, the spread in $\delta P_\mathrm{rot}$ decreases for longer temporal baselines and the accuracy of the ROOSTER significantly improves. 
{
For short temporal baselines, we note that some of the recovered periods are located close to ridges related to the frequency binning of the GLS periodogram (this is especially visible in the two top panels of Fig.~\ref{fig:rotation_period_recovery_corr}).

In the case of noise-free light curves, it goes from 86.3~\% for the 6 month temporal baseline to 97.5~\% for the 8 year temporal baseline. Nevertheless, the accuracy of the rotation period recovery is already satisfying after only six months of observations, as the majority of the recovered periods are  within 10~\% of the true period. 
It should be noted that, as we increase the temporal baseline, the number of light curves removed from the test set ---because we did not manage to properly measure the rotation period--- substantially decreases, from 114 for the 6 month baseline to only 3 for the 8 year baseline.  
These results are summarised in Table~\ref{tab:rooster_summary}.
Considering the corrected noisy light curves, the accuracy scores are significantly affected with respect to the noise-free case. In particular, in the case of 6 month light curves, the accuracy score goes down to 72.3\%.
Nevertheless, we see again that considering longer time series allows recovery of the rotation periods with a significantly larger accuracy, obtaining a 86.0\% accuracy for 1 year light curves, 91.7\% for 2 year light curves, 94.8\% for 4 year light curves, 95.5\% for 6 year light curves, and 96.2\% for 8 year light curves.
For stars for which a valid period candidate value was obtained from either GLS, ACF, or CS, we note that it can occur that the first overtone (second harmonic) of the rotation period is selected rather than the actual $P_{\rm rot}$, similarly to what was discussed in \citet{Breton2021}. In order to illustrate this last aspect, we provide additional recovery diagnostics in Appendix~\ref{appendix:additional_recovery_diagnostics}. 

The performance we obtain when analysing both the noise-free and noisy simulated samples validates that, as discussed in Sect.~\ref{sec:average_rotation_period}, the modifications made to the ROOSTER methodology with respect to the \citet{Breton2021} version are justified, even if some aspects are still susceptible to evolution as the performance of the algorithms are assessed with real PLATO observations.}

\begin{table*}[t!]
    \centering
    \caption{Summary of the results obtained for the ROOSTER \textit{PeriodSel} classifier dedicated to selecting the rotation period.}
    \begin{tabular}{ccccc}
    \hline\hline
    & \multicolumn{2}{c}{Noise free} & \multicolumn{2}{c}{Noisy} \\
    Baseline  & $N_\mathrm{star}$ & Recovery rate (\%) & $N_\mathrm{star}$ & Recovery rate (\%) \\
    \hline
    6 month   & 886/1000 & 86.3 & 762/1000 & 72.3 \\
    1 year    & 944/1000 & 91.6 & 905/1000 & 86.0 \\
    2 years   & 975/1000 & 95.3 & 956/1000 & 91.7 \\
    4 years   & 990/1000 & 97.0 & 978/1000 & 94.8 \\
    6 years   & 992/1000 & 97.0 & 978/1000 & 95.5 \\
    8 years   & 997/1000 & 97.5 & 983/1000 & 96.2 \\
    \end{tabular}
    \tablefoot{$N_\mathrm{star}$ is the number of targets to which the ROOSTER classifier was actually applied. The accuracy score is corrected to account for the light curves for which the periodogram, ACF, and CS analysis methods were not able to provide a value corresponding to the actual $P_\mathrm{rot}$ of the simulation, as explained in the main body of the paper.  
    }
    \label{tab:rooster_summary}
\end{table*}

\subsection{Long-term modulation recovery}

\begin{figure}[ht!]
    \centering
    \includegraphics[width=0.49\textwidth]{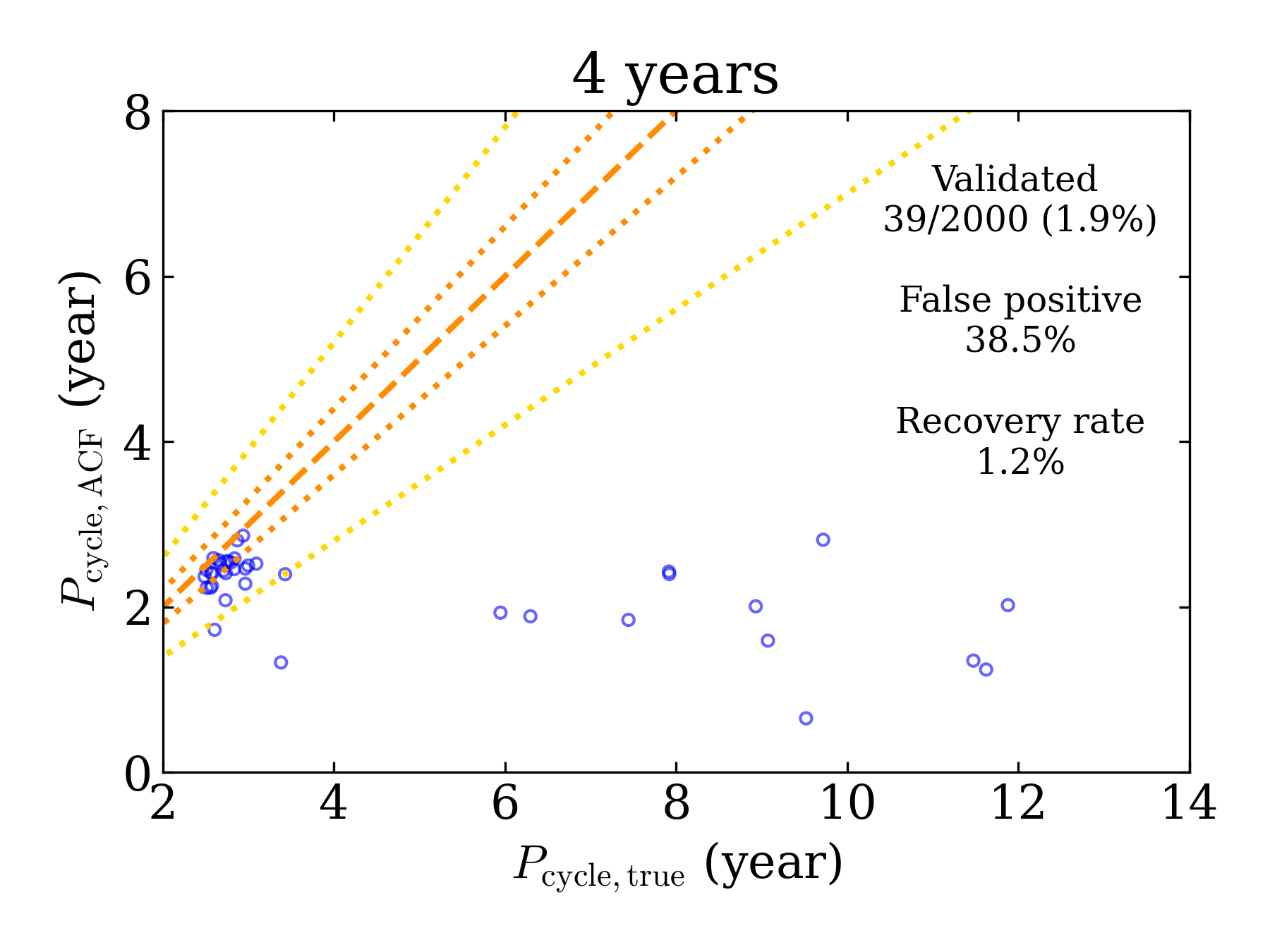}
    \includegraphics[width=0.49\textwidth]{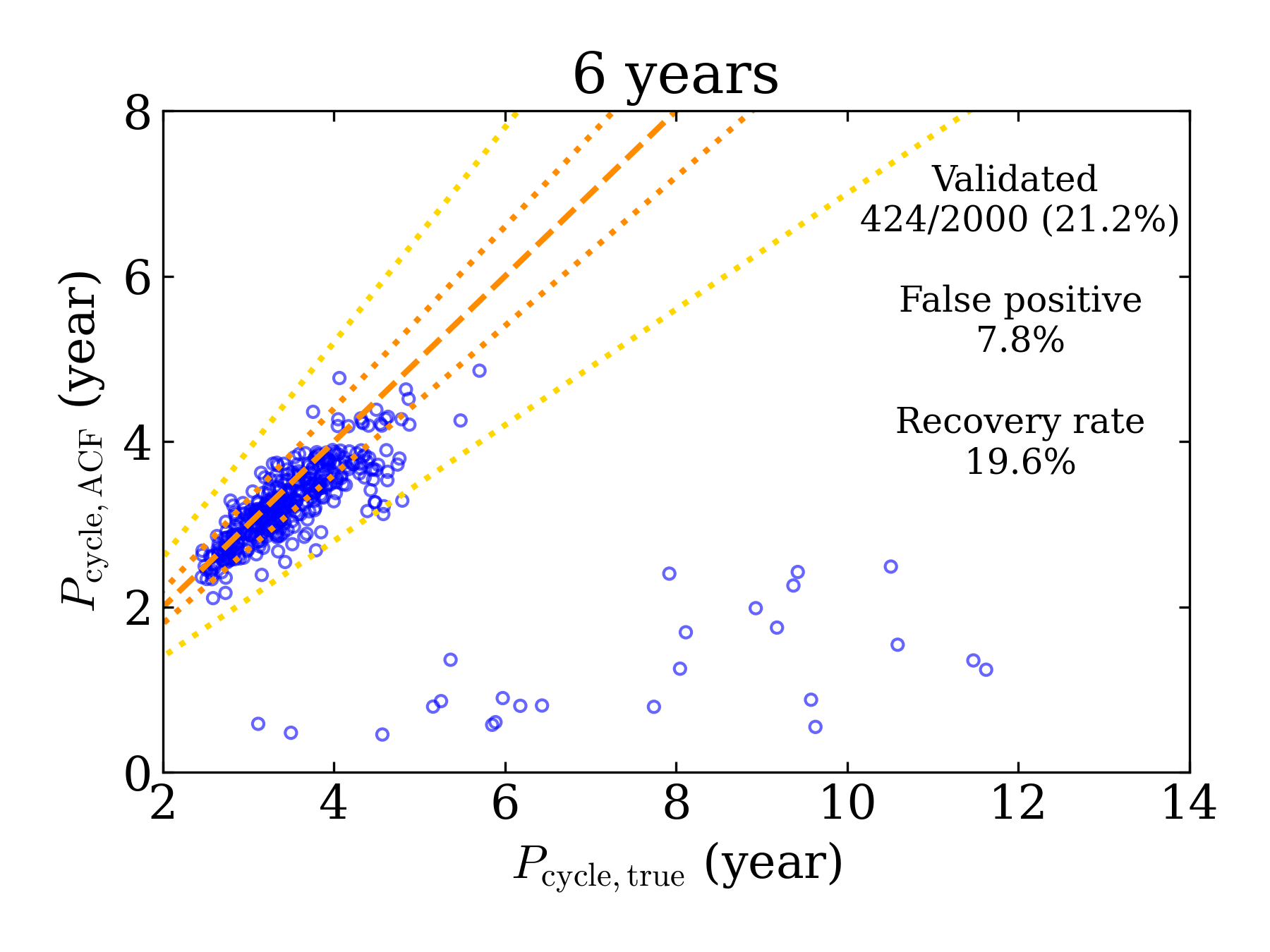}
    \includegraphics[width=0.49\textwidth]{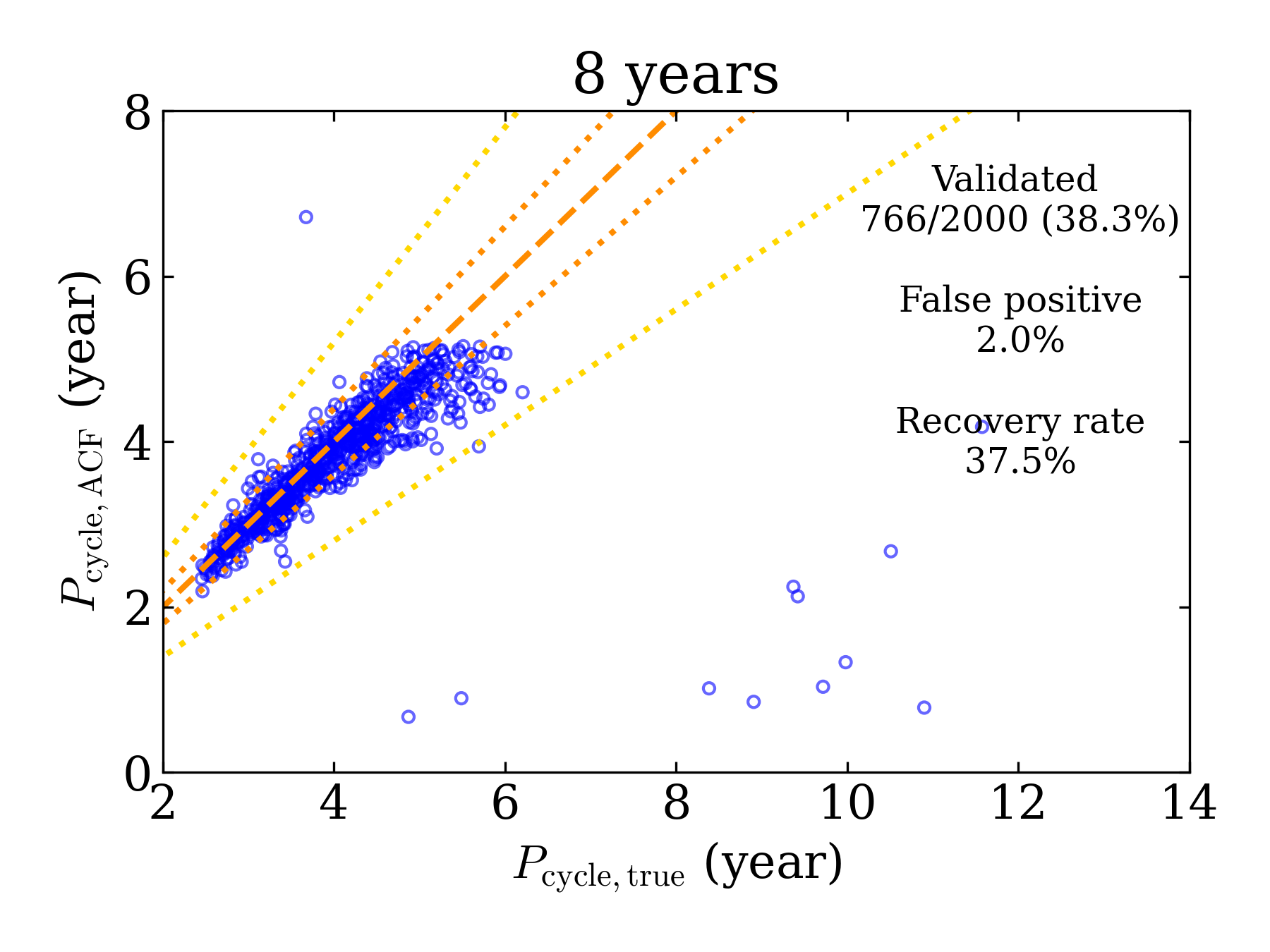}
    \caption{Measured long-term modulations (\textit{blue}) compared to the cycle length included in the simulations in the case of noise-free light curves for the temporal baselines considered in this work, where we retrieve modulations: 4 years (\textit{top}), 6 years (\textit{middle}), and 8 years (\textit{bottom}). The dashed orange line corresponds to the 1:1 line, while the upper dotted orange line and the lower dotted orange line are the 1:1.1 and 1:0.9 lines, respectively. The upper dotted yellow line and the lower dotted yellow line are the 1:1.3 and 1:0.7 lines, respectively}
    \label{fig:cycle_recovery}
\end{figure}

\begin{figure}[ht!]
    \centering
    \includegraphics[width=0.49\textwidth]{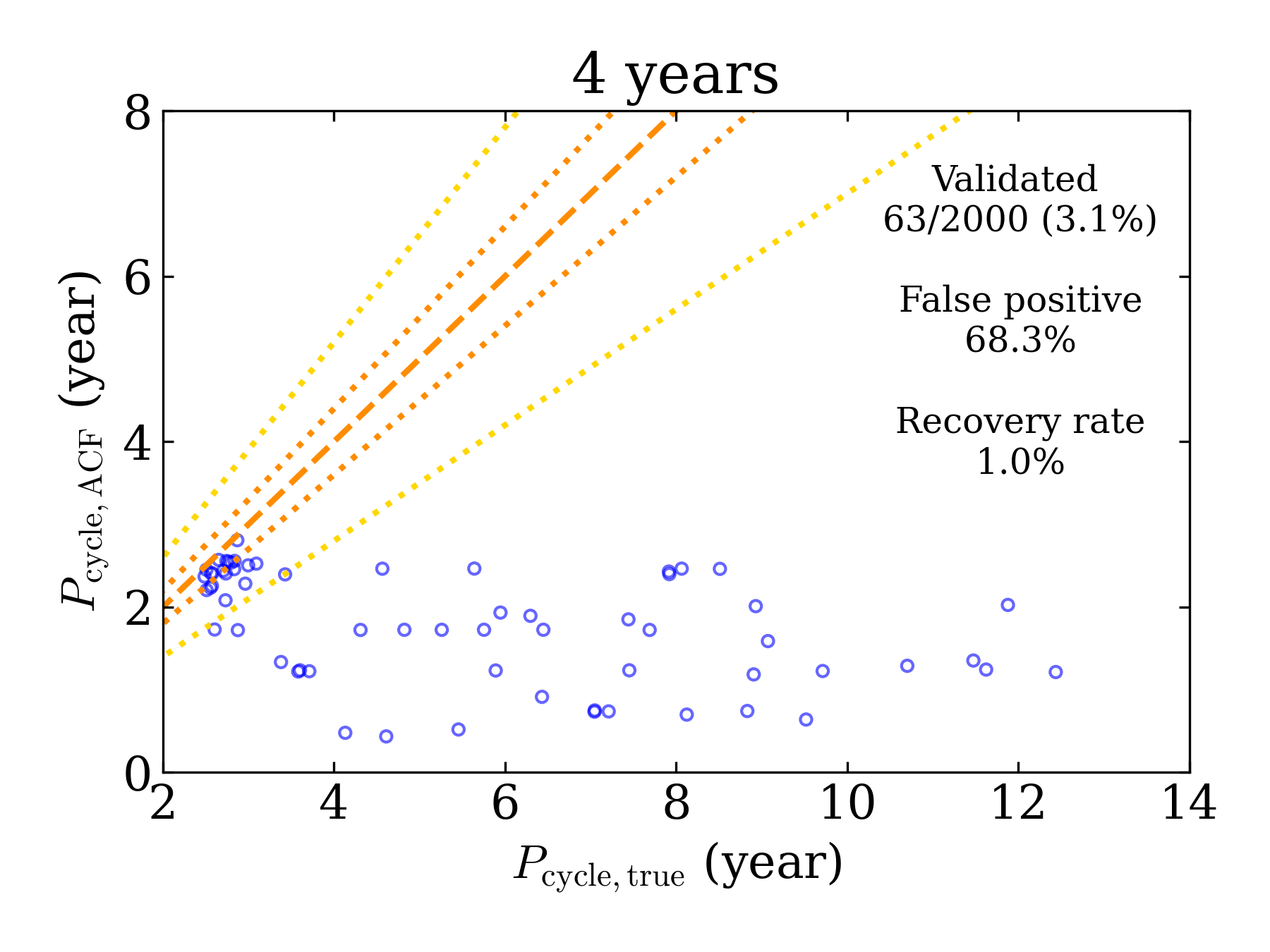}
    \includegraphics[width=0.49\textwidth]{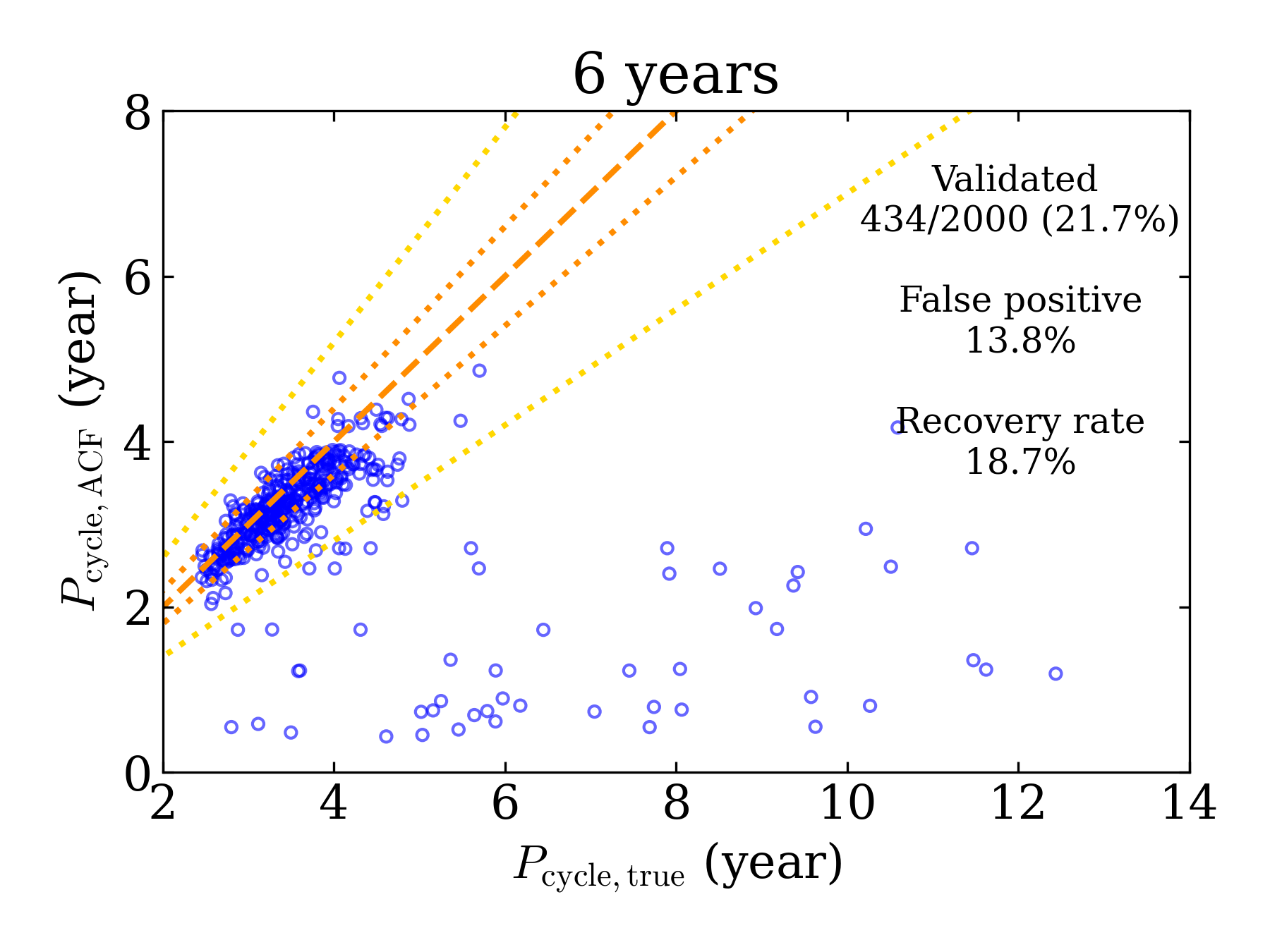}
    \includegraphics[width=0.49\textwidth]{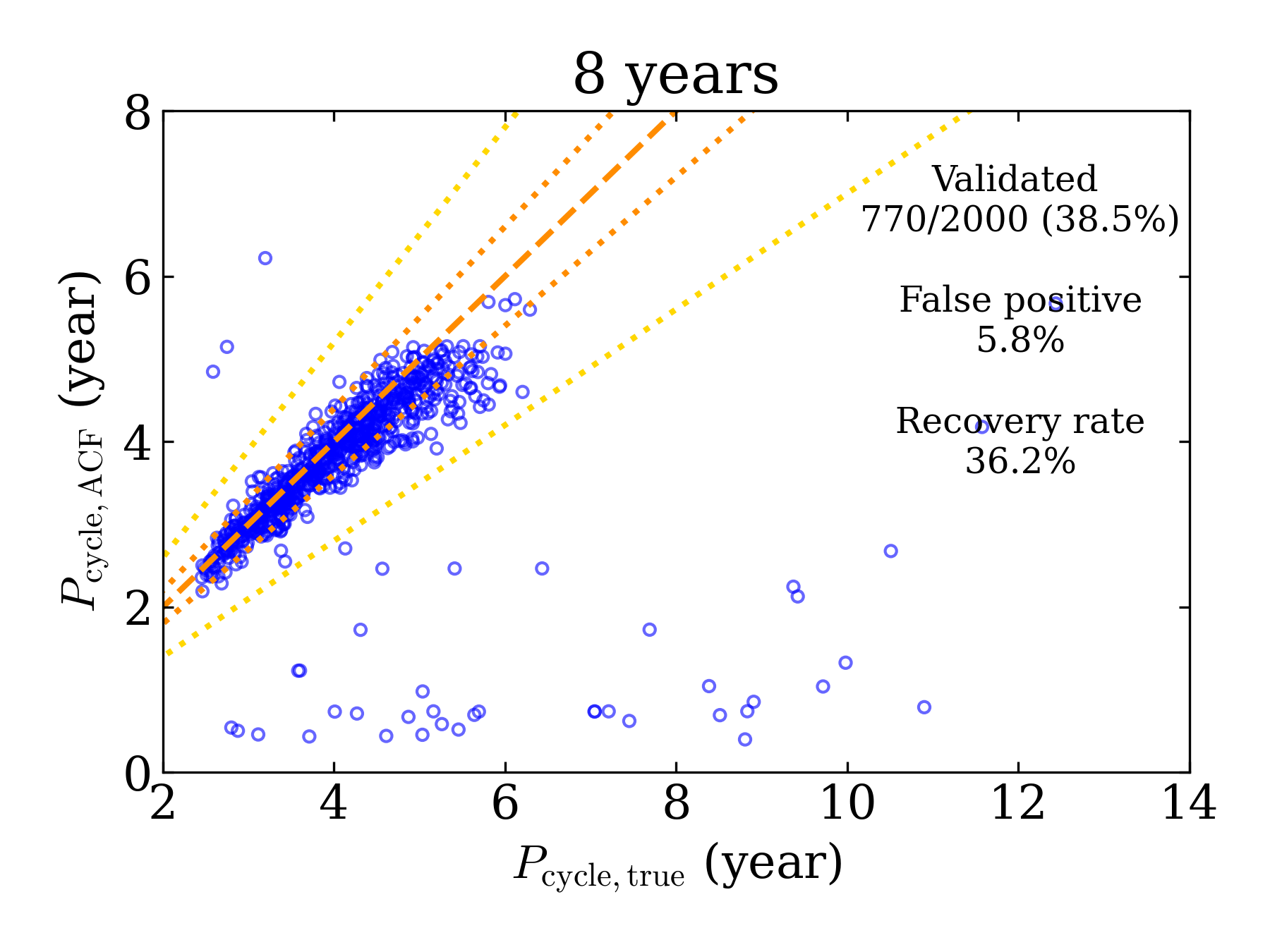}
    \caption{Same as Fig.~\ref{fig:cycle_recovery} but in the case of corrected noisy light curves.}
    \label{fig:cycle_recovery_corr}
\end{figure}

Most of the $P_\mathrm{cycle}$ included in the simulated light curves are comparable to or larger than the observing time (see Fig.~\ref{fig:prot_pcycle_distribution}). This means that the corresponding excess of power will not be resolved in the periodogram and that the corresponding period measurement will be biased by the frequency bin distribution of the periodogram {(in other words, the values of the frequency bins will act as attractors in the distribution of $P_\mathrm{cycle}$ measured in the GLS periodogram)}. This effect can already be seen in the case of rotation periods, where, for observing times of shorter than two years, the recovered values at long periods are distributed along ridges related to the frequency binning of the periodogram. Checking whether or not these excesses of power exist is nevertheless useful in order to validate the modulations detected in the ACF. Indeed, the rotation frequency beating signature will be measured by the ACF but will not produce peaks in the periodogram as they are not fundamental frequencies in the Fourier decomposition of the time series. 

In Fig.~\ref{fig:cycle_recovery}, we show the periods that we recover when computing the ACF of the LC1 noise-free light curves compared to the actual $P_\mathrm{cycle}$ values of the simulated light curves. 
{We remind the reader that periods for which the $H_{\rm ACF}$ and $G_{\rm ACF}$ do not meet the validation thresholds defined in Sect.~\ref{sec:cyclic_modulations} are discarded. As stated in Sect.~\ref{sec:cyclic_modulations}, to be validated, periods measured in the ACF must also be consistent within 30~\% with one of the periods measured in the GLS. In what follows, for each light curve, in cases where we validate several periods, we consider the recovered $P_{\rm cycle}$ to be the validated period with the largest $H_{\rm ACF}$. We then define measurements that are not consistent within 30~\% with the reference $P_{\rm cycle}$  as false positives. Given the low-frequency of the modulation we are trying to recover in this exercise, this criterion is deliberately chosen to be more permissive than the acceptance range we set (10~\%) for rotation period recovery.}
We show the results of the same analysis applied to the LC1 noisy light curves in Fig.~\ref{fig:cycle_recovery_corr}.
As expected, the activity cycle modulations are too long to be properly recovered with light curves of shorter than 4 years. The $G_\mathrm{ACF}$ and $H_\mathrm{ACF}$ thresholds are tuned to discard all false positives in this case, and therefore we do not show diagrams corresponding to the 6 month, 1 year, and 2 year light curves. 
For longer temporal baselines, it is important to note that, even if we are able to discard most of the modulations that are not actually related to a cycle, the selection methodology still yields a small number of false positives.
Nevertheless, with the 4 year light curves, we start recovering periods in the ACF that are related to the activity cycle of some stars with $P_\mathrm{cycle} \leq 4$~year. However, we note that the longest periods measured with the ACF are underestimated compared to the true value. 
We also have a significant fraction of false positives: in the noise-free sample, we validate activity modulations for 1.9\% of the simulated light curves (39 stars) but 38.5\% of these validations are actually false positives. Things improve significantly when considering 6 year and 8 year baselines, although a significant portion of the recovered periods are still underestimated. Nevertheless, with a 6 year baseline, we validate a cyclic modulation for 21.2\% of the sample (424 stars), with only 7.8\% false positives, while with an 8 year baseline, we improve upon this score to obtain a cycle detection for 38.3~\% of the sample (766 stars), with only 2.0\% false positives. We note that  we obtain a greater  number of false positives in the case of corrected noisy light curves. For the 4 year temporal baseline, we validate 3.1\% of the sample (63 stars), but 68.3\% of them are actually false positives. For the 6 year temporal baseline, we validate a cyclic modulation in 21.7\% of the sample (434 stars), with 13.8\% being false positives. For the 8 year temporal baseline, we validate a cyclic modulation in 38.5\% of the sample (770 stars), with 5.8\% false positives. Obtaining a similar yield in actual PLATO observations would provide vast and robust samples of solar-type stars with strong constraints on their magnetic activity cycle length, a catalogue which is still lacking from previous space missions dedicated to photometry.  

We finally reiterate that, in real PLATO observations, variability modulations with both smaller amplitude and shorter period than the simulated activity cycles, such as Rieger- or QBO-like cycles, will also be searched for in the light curves. Due to the uncertainties related to the physical behaviour and the manifestation of these types of cycles in the light curve, they were not included in the simulations used in the present work. If detectable, properly
constraining these modulations will require shorter temporal baselines.

\section{Conclusion \label{section:conclusion}}

This paper is the first in a series of publications dedicated to discussing the possibilities offered by PLATO to monitor the surface rotation and magnetic activity of solar-type stars. Here, we present the algorithms that will be implemented for this purpose in the standard analysis pipeline of the mission. 

The algorithms were applied to a set of simulated light curves specifically computed for this work, encompassing convective granulation, spot-induced flux rotational modulation, spot emergence and decay, magnetic activity cycle, latitudinal differential rotation, and migration of active latitudes with time. 
To produce these simulations, we interfaced the PSLS with the \texttt{pyspot} code in order to include PLATO instrumental systematics and camera random noise in the light curves. To assess how instrumental effects could affect the analysis of stellar signal, we compared the performances of the rotation and activity algorithms when considering noise-free and noisy light curves.

Regarding the algorithmic performances, we demonstrate that the time of observation that PLATO is able to dedicate to each field will significantly affect the rotation period recovery rate of the observed targets, especially with slow rotators. 
{
Nevertheless, we show that an acceptable recovery rate (above 72~\% for the noisy set of light curves) for the recovery of average rotation periods should already be reached with only 6 months of observations. This rate will be significantly improved as more data are collected, reaching a 86.0\% score for 1 year time series, and 91.7\% for 2 year time series.}
Longer observations will allow improvement of the quality of the rotation measurements, especially for slow rotators. In the noisy set of light curves, we are indeed able to obtain a recovery rate of almost 95\% for the recovery of average rotation period for a 4 year temporal baseline (which is the length of the nominal PLATO mission).
{
PLATO instrumental systematics as predicted by PSLS do not significantly limit rotation measurements, either by sensibly reducing the fraction of stars for  which we detect rotational modulation or by biasing the recovered period. In addition to the core mission objectives of detecting exoplanets and characterising their host stars through asteroseismology, photometric measurements of stellar surface rotation will therefore be an important added value of the PLATO mission.

Under the assumptions made to include activity cycles in the simulated light curves}, we are also able to show that recovering Schwabe-like magnetic activity cycles will require observation of the same field for at least four years, and that staying in the same field for six or eight years will significantly improve the mission yield in regard to this observable. Indeed, considering the case of noisy four-year  light curves, we are able to detect an activity cycle for {3.1\% of the simulated sample, with 68.3\% false positives, while extending the observing time to eight years allows us to constrain the activity cycle for 38.5\% of the sample and reduce the false positives to 5.8~\% of the validated periods.} Such measurements of variability due to long-term activity will be of primary importance for improving our understanding of the dependencies of the dynamo mechanisms of  solar-type stars on age, structure, and dynamics (e.g. mean rotation rate, convective flow velocity, and density gradient). 

The data products yielded by the module are designed to be of direct use to the community.   
Until the launch and during flying operations, the algorithms presented in this work will undergo continuous improvement within the PLATO consortium in order to provide the community with the most accurate measurements of rotation and magnetic activity upon each data release. 

\begin{acknowledgements}
This work presents results from the European Space Agency (ESA) space mission PLATO. The PLATO payload, the PLATO Ground Segment and PLATO data processing are joint developments of ESA and the PLATO Mission Consortium (PMC). Funding for the PMC is provided at national levels, in particular by countries participating in the PLATO Multilateral Agreement (Austria, Belgium, Czech Republic, Denmark, France, Germany, Italy, Netherlands, Portugal, Spain, Sweden, Switzerland, Norway, and United Kingdom) and institutions from Brazil. Members of the PLATO Consortium can be found at \url{https://platomission.com}. The ESA PLATO mission website is \url{https://www.cosmos.esa.int/plato}. The authors thank the teams working for PLATO for all their work. 
They acknowledge the critical reading and the constructive comments from the anonymous referee that significantly allowed improving the original version of this paper.
They finally thank R.~Samadi for helpful advice and suggestions concerning the PSLS abilities.
S.N.B, A.F.L, S.Me, I.P and E.C acknowledge support from PLATO ASI-INAF agreement no. 2022-28-HH.0 "PLATO Fase D".
S.N.B, L.A, A.S.B, Q.N, and A.S acknowledge financial support by ERC Whole Sun Synergy grant \#810218. 
S.N.B, R.A.G, L.A, A.S.B, Q.N., D.B.P, E.P, and A.S acknowledge the support from PLATO CNES grant.  
R.A.G, D.B.P, and E.P acknowledge the support from SoHO/GOLF CNES grant.
A.S.B, Q.N, and A.S acknowledge the support from INSU/PNST grant and Solar Orbiter CNES grant.
A.S acknowledges funding from  from the European Union’s Horizon-2020 research and innovation program (grant agreement no. 776403 ExoplANETS-A) and the Programme National de Plan\'etologie (PNP). 
A.R.G.S acknowledges the support from the FCT through national funds and FEDER through COMPETE2020 (UIDB/04434/2020, UIDP/04434/2020, 2022.03993.PTDC) and the support from the FCT through the work contract No. 2020.02480.CEECIND/CP1631/CT0001. S.Ma~acknowledges support from the Spanish Ministry of Science and Innovation (MICINN) with the Ram\'on y Cajal fellowship no.~RYC-2015-17697 and through AEI under the Severo Ochoa Centres of Excellence Programme 2020--2023 (CEX2019-000920-S). 
S.Ma acknowledges support from the  Spanish Ministry of Science and Innovation (MICINN) with the grant no.~PID2019-107187GB-I00. 
M.J.G., K.B., R.M.O, J.P, O.R., C.R. acknowledge support from CNES. The computations were performed with the IRFU/CEA Saclay server facilities, funded by ERC Synergy grant WholeSun No.810218, the P2IO Labex emergence project FlarePredict, and CNES PLATO funds.
\\
\textit{Software:} 
\texttt{star-privateer} (this work),
\texttt{pyspot} \citep{Aigrain2015}, PSLS \citep{Samadi2019},
\texttt{numpy} \citep{harris2020array}, \texttt{matplotlib} \citep{Hunter:2007}, \texttt{scipy} \citep{2020SciPy-NMeth}, 
\texttt{astropy} \citep{astropy:2022},
\texttt{pandas} \citep{mckinney-proc-scipy-2010,reback2020pandas},
\texttt{scikit-learn} \citep{scikit-learn}.
\end{acknowledgements}

\bibliographystyle{aa} 
\bibliography{biblio} 

\begin{thebibliography}{105}
\expandafter\ifx\csname natexlab\endcsname\relax\def\natexlab#1{#1}\fi

\bibitem[{{Ahuir} {et~al.}(2021){Ahuir}, {Strugarek}, {Brun}, \& {Mathis}}]{Ahuir2021}
{Ahuir}, J., {Strugarek}, A., {Brun}, A.~S., \& {Mathis}, S. 2021, \aap, 650, A126

\bibitem[{{Aigrain} {et~al.}(2015){Aigrain}, {Llama}, {Ceillier}, {Chagas}, {Davenport}, {Garc{\'\i}a}, {Hay}, {Lanza}, {McQuillan}, {Mazeh}, {de Medeiros}, {Nielsen}, \& {Reinhold}}]{Aigrain2015}
{Aigrain}, S., {Llama}, J., {Ceillier}, T., {et~al.} 2015, \mnras, 450, 3211

\bibitem[{{Amard} {et~al.}(2020){Amard}, {Roquette}, \& {Matt}}]{Amard2020}
{Amard}, L., {Roquette}, J., \& {Matt}, S.~P. 2020, \mnras, 499, 3481

\bibitem[{{Angus} {et~al.}(2020){Angus}, {Beane}, {Price-Whelan}, {Newton}, {Curtis}, {Berger}, {van Saders}, {Kiman}, {Foreman-Mackey}, {Lu}, {Anderson}, \& {Faherty}}]{Angus2020}
{Angus}, R., {Beane}, A., {Price-Whelan}, A.~M., {et~al.} 2020, \aj, 160, 90

\bibitem[{{Astropy Collaboration} {et~al.}(2022){Astropy Collaboration}, {Price-Whelan}, {Lim}, {Earl}, {Starkman}, {Bradley}, {Shupe}, {Patil}, {Corrales}, {Brasseur}, {N{"o}the}, {Donath}, {Tollerud}, {Morris}, {Ginsburg}, {Vaher}, {Weaver}, {Tocknell}, {Jamieson}, {van Kerkwijk}, {Robitaille}, {Merry}, {Bachetti}, {G{"u}nther}, {Aldcroft}, {Alvarado-Montes}, {Archibald}, {B{'o}di}, {Bapat}, {Barentsen}, {Baz{'a}n}, {Biswas}, {Boquien}, {Burke}, {Cara}, {Cara}, {Conroy}, {Conseil}, {Craig}, {Cross}, {Cruz}, {D'Eugenio}, {Dencheva}, {Devillepoix}, {Dietrich}, {Eigenbrot}, {Erben}, {Ferreira}, {Foreman-Mackey}, {Fox}, {Freij}, {Garg}, {Geda}, {Glattly}, {Gondhalekar}, {Gordon}, {Grant}, {Greenfield}, {Groener}, {Guest}, {Gurovich}, {Handberg}, {Hart}, {Hatfield-Dodds}, {Homeier}, {Hosseinzadeh}, {Jenness}, {Jones}, {Joseph}, {Kalmbach}, {Karamehmetoglu}, {Ka{l}uszy{'n}ski}, {Kelley}, {Kern}, {Kerzendorf}, {Koch}, {Kulumani}, {Lee}, {Ly}, {Ma}, {MacBride}, {Maljaars}, {Muna}, {Murphy}, {Norman}, {O'Steen},
  {Oman}, {Pacifici}, {Pascual}, {Pascual-Granado}, {Patil}, {Perren}, {Pickering}, {Rastogi}, {Roulston}, {Ryan}, {Rykoff}, {Sabater}, {Sakurikar}, {Salgado}, {Sanghi}, {Saunders}, {Savchenko}, {Schwardt}, {Seifert-Eckert}, {Shih}, {Jain}, {Shukla}, {Sick}, {Simpson}, {Singanamalla}, {Singer}, {Singhal}, {Sinha}, {Sip{H{o}}cz}, {Spitler}, {Stansby}, {Streicher}, {{{S}}umak}, {Swinbank}, {Taranu}, {Tewary}, {Tremblay}, {Val-Borro}, {Van Kooten}, {Vasovi{'c}}, {Verma}, {de Miranda Cardoso}, {Williams}, {Wilson}, {Winkel}, {Wood-Vasey}, {Xue}, {Yoachim}, {Zhang}, {Zonca}, \& {Astropy Project Contributors}}]{astropy:2022}
{Astropy Collaboration}, {Price-Whelan}, A.~M., {Lim}, P.~L., {et~al.} 2022, apj, 935, 167

\bibitem[{{Baglin} {et~al.}(2006){Baglin}, {Auvergne}, {Boisnard}, {Lam-Trong}, {Barge}, {Catala}, {Deleuil}, {Michel}, \& {Weiss}}]{Baglin2006}
{Baglin}, A., {Auvergne}, M., {Boisnard}, L., {et~al.} 2006, in 36th COSPAR Scientific Assembly, Vol.~36, 3749

\bibitem[{{Barnes}(2003)}]{Barnes2003}
{Barnes}, S.~A. 2003, \apj, 586, 464

\bibitem[{Barnes(2007)}]{Barnes2007}
Barnes, S.~A. 2007, ApJ, 669, 1167

\bibitem[{{Basri}(2018)}]{Basri2018}
{Basri}, G. 2018, \apj, 865, 142

\bibitem[{{Basri} \& {Shah}(2020)}]{Basri2020}
{Basri}, G. \& {Shah}, R. 2020, \apj, 901, 14

\bibitem[{{Bastien} {et~al.}(2016){Bastien}, {Stassun}, {Basri}, \& {Pepper}}]{Bastien2016}
{Bastien}, F.~A., {Stassun}, K.~G., {Basri}, G., \& {Pepper}, J. 2016, \apj, 818, 43

\bibitem[{{Bonanno} \& {Corsaro}(2022)}]{Bonanno2022}
{Bonanno}, A. \& {Corsaro}, E. 2022, \apjl, 939, L26

\bibitem[{{Borucki} {et~al.}(2010){Borucki}, {Koch}, {Basri}, {Batalha}, {Brown}, {Caldwell}, {Caldwell}, {Christensen-Dalsgaard}, {Cochran}, {DeVore}, {Dunham}, {Dupree}, {Gautier}, {Geary}, {Gilliland}, {Gould}, {Howell}, {Jenkins}, {Kondo}, {Latham}, {Marcy}, {Meibom}, {Kjeldsen}, {Lissauer}, {Monet}, {Morrison}, {Sasselov}, {Tarter}, {Boss}, {Brownlee}, {Owen}, {Buzasi}, {Charbonneau}, {Doyle}, {Fortney}, {Ford}, {Holman}, {Seager}, {Steffen}, {Welsh}, {Rowe}, {Anderson}, {Buchhave}, {Ciardi}, {Walkowicz}, {Sherry}, {Horch}, {Isaacson}, {Everett}, {Fischer}, {Torres}, {Johnson}, {Endl}, {MacQueen}, {Bryson}, {Dotson}, {Haas}, {Kolodziejczak}, {Van Cleve}, {Chandrasekaran}, {Twicken}, {Quintana}, {Clarke}, {Allen}, {Li}, {Wu}, {Tenenbaum}, {Verner}, {Bruhweiler}, {Barnes}, \& {Prsa}}]{Borucki2010}
{Borucki}, W.~J., {Koch}, D., {Basri}, G., {et~al.} 2010, Science, 327, 977

\bibitem[{Breiman(2001)}]{Breiman2001}
Breiman, L. 2001, Machine Learning, 45, 5

\bibitem[{Breiman {et~al.}(1984)Breiman, Friedman, Stone, \& Olshen}]{Breiman1984}
Breiman, L., Friedman, J., Stone, C., \& Olshen, R. 1984, Classification and Regression Trees, The Wadsworth and Brooks-Cole statistics-probability series (Taylor \& Francis)

\bibitem[{{Breton} {et~al.}(2024){Breton}, {Lanza}, \& {Messina}}]{Breton2024}
{Breton}, S.~N., {Lanza}, A.~F., \& {Messina}, S. 2024, \aap, 682, A67

\bibitem[{{Breton} {et~al.}(2021){Breton}, {Santos}, {Bugnet}, {Mathur}, {Garc{\'\i}a}, \& {Pall{\'e}}}]{Breton2021}
{Breton}, S.~N., {Santos}, A.~R.~G., {Bugnet}, L., {et~al.} 2021, \aap, 647, A125

\bibitem[{Brun \& Browning(2017)}]{Brun2017b}
Brun, A.~S. \& Browning, M.~K. 2017, Living Rev. Solar Phys., 14, 4

\bibitem[{{Brun} {et~al.}(2017){Brun}, {Strugarek}, {Varela}, {Matt}, {Augustson}, {Emeriau}, {DoCao}, {Brown}, \& {Toomre}}]{Brun2017a}
{Brun}, A.~S., {Strugarek}, A., {Varela}, J., {et~al.} 2017, \apj, 836, 192

\bibitem[{{Bugnet} {et~al.}(2018){Bugnet}, {Garc{\'\i}a}, {Davies}, {Mathur}, {Corsaro}, {Hall}, \& {Rendle}}]{Bugnet2018}
{Bugnet}, L., {Garc{\'\i}a}, R.~A., {Davies}, G.~R., {et~al.} 2018, \aap, 620, A38

\bibitem[{{Cao} {et~al.}(2023){Cao}, {Pinsonneault}, \& {van Saders}}]{Cao2023}
{Cao}, L., {Pinsonneault}, M.~H., \& {van Saders}, J.~L. 2023, \apjl, 951, L49

\bibitem[{{Ceillier} {et~al.}(2017){Ceillier}, {Tayar}, {Mathur}, {Salabert}, {Garc{\'\i}a}, {Stello}, {Pinsonneault}, {van Saders}, {Beck}, \& {Bloemen}}]{Ceillier2017}
{Ceillier}, T., {Tayar}, J., {Mathur}, S., {et~al.} 2017, \aap, 605, A111

\bibitem[{{Ceillier} {et~al.}(2016){Ceillier}, {van Saders}, {Garc{\'\i}a}, {Metcalfe}, {Creevey}, {Mathis}, {Mathur}, {Pinsonneault}, {Salabert}, \& {Tayar}}]{Ceillier2016}
{Ceillier}, T., {van Saders}, J., {Garc{\'\i}a}, R.~A., {et~al.} 2016, \mnras, 456, 119

\bibitem[{{Chaplin} {et~al.}(2011){Chaplin}, {Bedding}, {Bonanno}, {Broomhall}, {Garc{\'\i}a}, {Hekker}, {Huber}, {Verner}, {Basu}, {Elsworth}, {Houdek}, {Mathur}, {Mosser}, {New}, {Stevens}, {Appourchaux}, {Karoff}, {Metcalfe}, {Molenda-{\.Z}akowicz}, {Monteiro}, {Thompson}, {Christensen-Dalsgaard}, {Gilliland}, {Kawaler}, {Kjeldsen}, {Ballot}, {Benomar}, {Corsaro}, {Campante}, {Gaulme}, {Hale}, {Handberg}, {Jarvis}, {R{\'e}gulo}, {Roxburgh}, {Salabert}, {Stello}, {Mullally}, {Li}, \& {Wohler}}]{Chaplin2011b}
{Chaplin}, W.~J., {Bedding}, T.~R., {Bonanno}, A., {et~al.} 2011, \apjl, 732, L5

\bibitem[{{Claytor} {et~al.}(2024){Claytor}, {van Saders}, {Cao}, {Pinsonneault}, {Teske}, \& {Beaton}}]{Claytor2024}
{Claytor}, Z.~R., {van Saders}, J.~L., {Cao}, L., {et~al.} 2024, \apj, 962, 47

\bibitem[{{Corsaro} {et~al.}(2021){Corsaro}, {Bonanno}, {Mathur}, {Garc{\'\i}a}, {Santos}, {Breton}, \& {Khalatyan}}]{Corsaro2021}
{Corsaro}, E., {Bonanno}, A., {Mathur}, S., {et~al.} 2021, \aap, 652, L2

\bibitem[{{Corsaro} \& {De Ridder}(2014)}]{Corsaro2014}
{Corsaro}, E. \& {De Ridder}, J. 2014, \aap, 571, A71

\bibitem[{{Corsaro} {et~al.}(2015){Corsaro}, {De Ridder}, \& {Garc{\'\i}a}}]{Corsaro2015}
{Corsaro}, E., {De Ridder}, J., \& {Garc{\'\i}a}, R.~A. 2015, \aap, 579, A83

\bibitem[{{de Freitas} {et~al.}(2021){de Freitas}, {Lanza}, {da Silva Gomes}, \& {Das Chagas}}]{deFreitas2021}
{de Freitas}, D.~B., {Lanza}, A.~F., {da Silva Gomes}, F.~O., \& {Das Chagas}, M.~L. 2021, \aap, 650, A40

\bibitem[{{Deeg} {et~al.}(2023){Deeg}, {Georgieva}, {Nowak}, {Persson}, {Cale}, {Murgas}, {Pall{\'e}}, {Godoy-Rivera}, {Dai}, {Ciardi}, {Murphy}, {Beck}, {Burke}, {Cabrera}, {Carleo}, {Cochran}, {Collins}, {Csizmadia}, {El Mufti}, {Fridlund}, {Fukui}, {Gandolfi}, {Garc{\'\i}a}, {Guenther}, {Guerra}, {Grziwa}, {Isaacson}, {Isogai}, {Jenkins}, {K{\'a}bath}, {Korth}, {Lam}, {Latham}, {Luque}, {Lund}, {Livingston}, {Mathis}, {Mathur}, {Narita}, {Orell-Miquel}, {Osborne}, {Parviainen}, {Plavchan}, {Redfield}, {Rodriguez}, {Schwarz}, {Seager}, {Smith}, {Van Eylen}, {Van Zandt}, {Winn}, \& {Ziegler}}]{Deeg2023}
{Deeg}, H.~J., {Georgieva}, I.~Y., {Nowak}, G., {et~al.} 2023, \aap, 677, A12

\bibitem[{{Distefano} {et~al.}(2023){Distefano}, {Lanzafame}, {Brugaletta}, {Holl}, {Lanza}, {Messina}, {Pagano}, {Audard}, {Jevardat de Fombelle}, {Lecoeur-Taibi}, {Mowlavi}, {Nienartowicz}, {Rimoldini}, {Evans}, {Riello}, {Garc{\'\i}a-Lario}, {Gavras}, \& {Eyer}}]{Distefano2023}
{Distefano}, E., {Lanzafame}, A.~C., {Brugaletta}, E., {et~al.} 2023, \aap, 674, A20

\bibitem[{{Gaia Collaboration} {et~al.}(2016){Gaia Collaboration}, {Prusti}, {de Bruijne}, {Brown}, {Vallenari}, {Babusiaux}, {Bailer-Jones}, {Bastian}, {Biermann}, {Evans}, {Eyer}, {Jansen}, {Jordi}, {Klioner}, {Lammers}, {Lindegren}, {Luri}, {Mignard}, {Milligan}, {Panem}, {Poinsignon}, {Pourbaix}, {Randich}, {Sarri}, {Sartoretti}, {Siddiqui}, {Soubiran}, {Valette}, {van Leeuwen}, {Walton}, {Aerts}, {Arenou}, {Cropper}, {Drimmel}, {H{\o}g}, {Katz}, {Lattanzi}, {O'Mullane}, {Grebel}, {Holland}, {Huc}, {Passot}, {Bramante}, {Cacciari}, {Casta{\~n}eda}, {Chaoul}, {Cheek}, {De Angeli}, {Fabricius}, {Guerra}, {Hern{\'a}ndez}, {Jean-Antoine-Piccolo}, {Masana}, {Messineo}, {Mowlavi}, {Nienartowicz}, {Ord{\'o}{\~n}ez-Blanco}, {Panuzzo}, {Portell}, {Richards}, {Riello}, {Seabroke}, {Tanga}, {Th{\'e}venin}, {Torra}, {Els}, {Gracia-Abril}, {Comoretto}, {Garcia-Reinaldos}, {Lock}, {Mercier}, {Altmann}, {Andrae}, {Astraatmadja}, {Bellas-Velidis}, {Benson}, {Berthier}, {Blomme}, {Busso}, {Carry}, {Cellino}, {Clementini},
  {Cowell}, {Creevey}, {Cuypers}, {Davidson}, {De Ridder}, {de Torres}, {Delchambre}, {Dell'Oro}, {Ducourant}, {Fr{\'e}mat}, {Garc{\'\i}a-Torres}, {Gosset}, {Halbwachs}, {Hambly}, {Harrison}, {Hauser}, {Hestroffer}, {Hodgkin}, {Huckle}, {Hutton}, {Jasniewicz}, {Jordan}, {Kontizas}, {Korn}, {Lanzafame}, {Manteiga}, {Moitinho}, {Muinonen}, {Osinde}, {Pancino}, {Pauwels}, {Petit}, {Recio-Blanco}, {Robin}, {Sarro}, {Siopis}, {Smith}, {Smith}, {Sozzetti}, {Thuillot}, {van Reeven}, {Viala}, {Abbas}, {Abreu Aramburu}, {Accart}, {Aguado}, {Allan}, {Allasia}, {Altavilla}, {{\'A}lvarez}, {Alves}, {Anderson}, {Andrei}, {Anglada Varela}, {Antiche}, {Antoja}, {Ant{\'o}n}, {Arcay}, {Atzei}, {Ayache}, {Bach}, {Baker}, {Balaguer-N{\'u}{\~n}ez}, {Barache}, {Barata}, {Barbier}, {Barblan}, {Baroni}, {Barrado y Navascu{\'e}s}, {Barros}, {Barstow}, {Becciani}, {Bellazzini}, {Bellei}, {Bello Garc{\'\i}a}, {Belokurov}, {Bendjoya}, {Berihuete}, {Bianchi}, {Bienaym{\'e}}, {Billebaud}, {Blagorodnova}, {Blanco-Cuaresma}, {Boch},
  {Bombrun}, {Borrachero}, {Bouquillon}, {Bourda}, {Bouy}, {Bragaglia}, {Breddels}, {Brouillet}, {Br{\"u}semeister}, {Bucciarelli}, {Budnik}, {Burgess}, {Burgon}, {Burlacu}, {Busonero}, {Buzzi}, {Caffau}, {Cambras}, {Campbell}, {Cancelliere}, {Cantat-Gaudin}, {Carlucci}, {Carrasco}, {Castellani}, {Charlot}, {Charnas}, {Charvet}, {Chassat}, {Chiavassa}, {Clotet}, {Cocozza}, {Collins}, {Collins}, {Costigan}, {Crifo}, {Cross}, {Crosta}, {Crowley}, {Dafonte}, {Damerdji}, {Dapergolas}, {David}, {David}, {De Cat}, {de Felice}, {de Laverny}, {De Luise}, {De March}, {de Martino}, {de Souza}, {Debosscher}, {del Pozo}, {Delbo}, {Delgado}, {Delgado}, {di Marco}, {Di Matteo}, {Diakite}, {Distefano}, {Dolding}, {Dos Anjos}, {Drazinos}, {Dur{\'a}n}, {Dzigan}, {Ecale}, {Edvardsson}, {Enke}, {Erdmann}, {Escolar}, {Espina}, {Evans}, {Eynard Bontemps}, {Fabre}, {Fabrizio}, {Faigler}, {Falc{\~a}o}, {Farr{\`a}s Casas}, {Faye}, {Federici}, {Fedorets}, {Fern{\'a}ndez-Hern{\'a}ndez}, {Fernique}, {Fienga}, {Figueras}, {Filippi},
  {Findeisen}, {Fonti}, {Fouesneau}, {Fraile}, {Fraser}, {Fuchs}, {Furnell}, {Gai}, {Galleti}, {Galluccio}, {Garabato}, {Garc{\'\i}a-Sedano}, {Gar{\'e}}, {Garofalo}, {Garralda}, {Gavras}, {Gerssen}, {Geyer}, {Gilmore}, {Girona}, {Giuffrida}, {Gomes}, {Gonz{\'a}lez-Marcos}, {Gonz{\'a}lez-N{\'u}{\~n}ez}, {Gonz{\'a}lez-Vidal}, {Granvik}, {Guerrier}, {Guillout}, {Guiraud}, {G{\'u}rpide}, {Guti{\'e}rrez-S{\'a}nchez}, {Guy}, {Haigron}, {Hatzidimitriou}, {Haywood}, {Heiter}, {Helmi}, {Hobbs}, {Hofmann}, {Holl}, {Holland}, {Hunt}, {Hypki}, {Icardi}, {Irwin}, {Jevardat de Fombelle}, {Jofr{\'e}}, {Jonker}, {Jorissen}, {Julbe}, {Karampelas}, {Kochoska}, {Kohley}, {Kolenberg}, {Kontizas}, {Koposov}, {Kordopatis}, {Koubsky}, {Kowalczyk}, {Krone-Martins}, {Kudryashova}, {Kull}, {Bachchan}, {Lacoste-Seris}, {Lanza}, {Lavigne}, {Le Poncin-Lafitte}, {Lebreton}, {Lebzelter}, {Leccia}, {Leclerc}, {Lecoeur-Taibi}, {Lemaitre}, {Lenhardt}, {Leroux}, {Liao}, {Licata}, {Lindstr{\o}m}, {Lister}, {Livanou}, {Lobel}, {L{\"o}ffler},
  {L{\'o}pez}, {Lopez-Lozano}, {Lorenz}, {Loureiro}, {MacDonald}, {Magalh{\~a}es Fernandes}, {Managau}, {Mann}, {Mantelet}, {Marchal}, {Marchant}, {Marconi}, {Marie}, {Marinoni}, {Marrese}, {Marschalk{\'o}}, {Marshall}, {Mart{\'\i}n-Fleitas}, {Martino}, {Mary}, {Matijevi{\v{c}}}, {Mazeh}, {McMillan}, {Messina}, {Mestre}, {Michalik}, {Millar}, {Miranda}, {Molina}, {Molinaro}, {Molinaro}, {Moln{\'a}r}, {Moniez}, {Montegriffo}, {Monteiro}, {Mor}, {Mora}, {Morbidelli}, {Morel}, {Morgenthaler}, {Morley}, {Morris}, {Mulone}, {Muraveva}, {Musella}, {Narbonne}, {Nelemans}, {Nicastro}, {Noval}, {Ord{\'e}novic}, {Ordieres-Mer{\'e}}, {Osborne}, {Pagani}, {Pagano}, {Pailler}, {Palacin}, {Palaversa}, {Parsons}, {Paulsen}, {Pecoraro}, {Pedrosa}, {Pentik{\"a}inen}, {Pereira}, {Pichon}, {Piersimoni}, {Pineau}, {Plachy}, {Plum}, {Poujoulet}, {Pr{\v{s}}a}, {Pulone}, {Ragaini}, {Rago}, {Rambaux}, {Ramos-Lerate}, {Ranalli}, {Rauw}, {Read}, {Regibo}, {Renk}, {Reyl{\'e}}, {Ribeiro}, {Rimoldini}, {Ripepi}, {Riva}, {Rixon},
  {Roelens}, {Romero-G{\'o}mez}, {Rowell}, {Royer}, {Rudolph}, {Ruiz-Dern}, {Sadowski}, {Sagrist{\`a} Sell{\'e}s}, {Sahlmann}, {Salgado}, {Salguero}, {Sarasso}, {Savietto}, {Schnorhk}, {Schultheis}, {Sciacca}, {Segol}, {Segovia}, {Segransan}, {Serpell}, {Shih}, {Smareglia}, {Smart}, {Smith}, {Solano}, {Solitro}, {Sordo}, {Soria Nieto}, {Souchay}, {Spagna}, {Spoto}, {Stampa}, {Steele}, {Steidelm{\"u}ller}, {Stephenson}, {Stoev}, {Suess}, {S{\"u}veges}, {Surdej}, {Szabados}, {Szegedi-Elek}, {Tapiador}, {Taris}, {Tauran}, {Taylor}, {Teixeira}, {Terrett}, {Tingley}, {Trager}, {Turon}, {Ulla}, {Utrilla}, {Valentini}, {van Elteren}, {Van Hemelryck}, {van Leeuwen}, {Varadi}, {Vecchiato}, {Veljanoski}, {Via}, {Vicente}, {Vogt}, {Voss}, {Votruba}, {Voutsinas}, {Walmsley}, {Weiler}, {Weingrill}, {Werner}, {Wevers}, {Whitehead}, {Wyrzykowski}, {Yoldas}, {{\v{Z}}erjal}, {Zucker}, {Zurbach}, {Zwitter}, {Alecu}, {Allen}, {Allende Prieto}, {Amorim}, {Anglada-Escud{\'e}}, {Arsenijevic}, {Azaz}, {Balm}, {Beck}, {Bernstein},
  {Bigot}, {Bijaoui}, {Blasco}, {Bonfigli}, {Bono}, {Boudreault}, {Bressan}, {Brown}, {Brunet}, {Bunclark}, {Buonanno}, {Butkevich}, {Carret}, {Carrion}, {Chemin}, {Ch{\'e}reau}, {Corcione}, {Darmigny}, {de Boer}, {de Teodoro}, {de Zeeuw}, {Delle Luche}, {Domingues}, {Dubath}, {Fodor}, {Fr{\'e}zouls}, {Fries}, {Fustes}, {Fyfe}, {Gallardo}, {Gallegos}, {Gardiol}, {Gebran}, {Gomboc}, {G{\'o}mez}, {Grux}, {Gueguen}, {Heyrovsky}, {Hoar}, {Iannicola}, {Isasi Parache}, {Janotto}, {Joliet}, {Jonckheere}, {Keil}, {Kim}, {Klagyivik}, {Klar}, {Knude}, {Kochukhov}, {Kolka}, {Kos}, {Kutka}, {Lainey}, {LeBouquin}, {Liu}, {Loreggia}, {Makarov}, {Marseille}, {Martayan}, {Martinez-Rubi}, {Massart}, {Meynadier}, {Mignot}, {Munari}, {Nguyen}, {Nordlander}, {Ocvirk}, {O'Flaherty}, {Olias Sanz}, {Ortiz}, {Osorio}, {Oszkiewicz}, {Ouzounis}, {Palmer}, {Park}, {Pasquato}, {Peltzer}, {Peralta}, {P{\'e}turaud}, {Pieniluoma}, {Pigozzi}, {Poels}, {Prat}, {Prod'homme}, {Raison}, {Rebordao}, {Risquez}, {Rocca-Volmerange}, {Rosen},
  {Ruiz-Fuertes}, {Russo}, {Sembay}, {Serraller Vizcaino}, {Short}, {Siebert}, {Silva}, {Sinachopoulos}, {Slezak}, {Soffel}, {Sosnowska}, {Strai{\v{z}}ys}, {ter Linden}, {Terrell}, {Theil}, {Tiede}, {Troisi}, {Tsalmantza}, {Tur}, {Vaccari}, {Vachier}, {Valles}, {Van Hamme}, {Veltz}, {Virtanen}, {Wallut}, {Wichmann}, {Wilkinson}, {Ziaeepour}, \& {Zschocke}}]{Gaia2016}
{Gaia Collaboration}, {Prusti}, T., {de Bruijne}, J.~H.~J., {et~al.} 2016, \aap, 595, A1

\bibitem[{{Garc{\'\i}a} \& {Ballot}(2019)}]{Garcia2019}
{Garc{\'\i}a}, R.~A. \& {Ballot}, J. 2019, Living Reviews in Solar Physics, 16, 4

\bibitem[{{Garc{\'\i}a} {et~al.}(2014){Garc{\'\i}a}, {Ceillier}, {Salabert}, {Mathur}, {van Saders}, {Pinsonneault}, {Ballot}, {Beck}, {Bloemen}, {Campante}, {Davies}, {do Nascimento}, {Mathis}, {Metcalfe}, {Nielsen}, {Su{\'a}rez}, {Chaplin}, {Jim{\'e}nez}, \& {Karoff}}]{Garcia2014b}
{Garc{\'\i}a}, R.~A., {Ceillier}, T., {Salabert}, D., {et~al.} 2014, \aap, 572, A34

\bibitem[{{Garc{\'\i}a} {et~al.}(2023){Garc{\'\i}a}, {Gourv{\`e}s}, {Santos}, {Strugarek}, {Godoy-Rivera}, {Mathur}, {Delsanti}, {Breton}, {Beck}, {Brun}, \& {Mathis}}]{Garcia2023}
{Garc{\'\i}a}, R.~A., {Gourv{\`e}s}, C., {Santos}, A.~R.~G., {et~al.} 2023, \aap, 679, L12

\bibitem[{{Gordon} {et~al.}(2021){Gordon}, {Davenport}, {Angus}, {Foreman-Mackey}, {Agol}, {Covey}, {Ag{\"u}eros}, \& {Kipping}}]{Gordon2021}
{Gordon}, T.~A., {Davenport}, J. R.~A., {Angus}, R., {et~al.} 2021, \apj, 913, 70

\bibitem[{{Gurgenashvili} {et~al.}(2022){Gurgenashvili}, {Zaqarashvili}, {Kukhianidze}, {Reiners}, {Reinhold}, \& {Lanza}}]{Gurgenashvili2022}
{Gurgenashvili}, E., {Zaqarashvili}, T.~V., {Kukhianidze}, V., {et~al.} 2022, \aap, 660, A33

\bibitem[{{Hall} {et~al.}(2021){Hall}, {Davies}, {van Saders}, {Nielsen}, {Lund}, {Chaplin}, {Garc{\'\i}a}, {Amard}, {Breimann}, {Khan}, {See}, \& {Tayar}}]{Hall2021}
{Hall}, O.~J., {Davies}, G.~R., {van Saders}, J., {et~al.} 2021, Nature Astronomy, 5, 707

\bibitem[{Harris {et~al.}(2020)Harris, Millman, van~der Walt, Gommers, Virtanen, Cournapeau, Wieser, Taylor, Berg, Smith, Kern, Picus, Hoyer, van Kerkwijk, Brett, Haldane, del R{\'{i}}o, Wiebe, Peterson, G{\'{e}}rard-Marchant, Sheppard, Reddy, Weckesser, Abbasi, Gohlke, \& Oliphant}]{harris2020array}
Harris, C.~R., Millman, K.~J., van~der Walt, S.~J., {et~al.} 2020, Nature, 585, 357

\bibitem[{{Harvey}(1985)}]{Harvey1985}
{Harvey}, J. 1985, in ESA Special Publication, Vol. 235, Future Missions in Solar, Heliospheric \& Space Plasma Physics, ed. E.~{Rolfe} \& B.~{Battrick}, 199

\bibitem[{{Holcomb} {et~al.}(2022){Holcomb}, {Robertson}, {Hartigan}, {Oelkers}, \& {Robinson}}]{Holcomb2022}
{Holcomb}, R.~J., {Robertson}, P., {Hartigan}, P., {Oelkers}, R.~J., \& {Robinson}, C. 2022, \apj, 936, 138

\bibitem[{{Howell} {et~al.}(2014){Howell}, {Sobeck}, {Haas}, {Still}, {Barclay}, {Mullally}, {Troeltzsch}, {Aigrain}, {Bryson}, {Caldwell}, {Chaplin}, {Cochran}, {Huber}, {Marcy}, {Miglio}, {Najita}, {Smith}, {Twicken}, \& {Fortney}}]{Howell2014}
{Howell}, S.~B., {Sobeck}, C., {Haas}, M., {et~al.} 2014, \pasp, 126, 398

\bibitem[{Hunter(2007)}]{Hunter:2007}
Hunter, J.~D. 2007, Computing in Science \& Engineering, 9, 90

\bibitem[{{Jouve} {et~al.}(2008){Jouve}, {Brun}, {Arlt}, {Brandenburg}, {Dikpati}, {Bonanno}, {K{\"a}pyl{\"a}}, {Moss}, {Rempel}, {Gilman}, {Korpi}, \& {Kosovichev}}]{Jouve2008}
{Jouve}, L., {Brun}, A.~S., {Arlt}, R., {et~al.} 2008, \aap, 483, 949

\bibitem[{{Kallinger} {et~al.}(2014){Kallinger}, {De Ridder}, {Hekker}, {Mathur}, {Mosser}, {Gruberbauer}, {Garc{\'\i}a}, {Karoff}, \& {Ballot}}]{Kallinger2014}
{Kallinger}, T., {De Ridder}, J., {Hekker}, S., {et~al.} 2014, \aap, 570, A41

\bibitem[{{Kallinger} {et~al.}(2016){Kallinger}, {Hekker}, {Garcia}, {Huber}, \& {Matthews}}]{Kallinger2016}
{Kallinger}, T., {Hekker}, S., {Garcia}, R.~A., {Huber}, D., \& {Matthews}, J.~M. 2016, Science Advances, 2, 1500654

\bibitem[{{Karoff} {et~al.}(2018){Karoff}, {Metcalfe}, {Santos}, {Montet}, {Isaacson}, {Witzke}, {Shapiro}, {Mathur}, {Davies}, {Lund}, {Garcia}, {Brun}, {Salabert}, {Avelino}, {van Saders}, {Egeland}, {Cunha}, {Campante}, {Chaplin}, {Krivova}, {Solanki}, {Stritzinger}, \& {Knudsen}}]{Karoff2018}
{Karoff}, C., {Metcalfe}, T.~S., {Santos}, {\^A}. R.~G., {et~al.} 2018, \apj, 852, 46

\bibitem[{{Kjeldsen} \& {Bedding}(1995)}]{Kjeldsen1995}
{Kjeldsen}, H. \& {Bedding}, T.~R. 1995, \aap, 293, 87

\bibitem[{{Lanza} {et~al.}(2019){Lanza}, {Netto}, {Bonomo}, {Parviainen}, {Valio}, \& {Aigrain}}]{Lanza2019}
{Lanza}, A.~F., {Netto}, Y., {Bonomo}, A.~S., {et~al.} 2019, \aap, 626, A38

\bibitem[{{Lanza} {et~al.}(2009){Lanza}, {Pagano}, {Leto}, {Messina}, {Aigrain}, {Alonso}, {Auvergne}, {Baglin}, {Barge}, {Bonomo}, {Boumier}, {Collier Cameron}, {Comparato}, {Cutispoto}, {de Medeiros}, {Foing}, {Kaiser}, {Moutou}, {Parihar}, {Silva-Valio}, \& {Weiss}}]{Lanza2009}
{Lanza}, A.~F., {Pagano}, I., {Leto}, G., {et~al.} 2009, \aap, 493, 193

\bibitem[{{Lanzafame} {et~al.}(2019){Lanzafame}, {Distefano}, {Barnes}, \& {Spada}}]{Lanzafame2019}
{Lanzafame}, A.~C., {Distefano}, E., {Barnes}, S.~A., \& {Spada}, F. 2019, \apj, 877, 157

\bibitem[{{Lanzafame} {et~al.}(2018){Lanzafame}, {Distefano}, {Messina}, {Pagano}, {Lanza}, {Eyer}, {Guy}, {Rimoldini}, {Lecoeur-Taibi}, {Holl}, {Audard}, {de Fombelle}, {Nienartowicz}, {Marchal}, \& {Mowlavi}}]{Lanzafame2018}
{Lanzafame}, A.~C., {Distefano}, E., {Messina}, S., {et~al.} 2018, \aap, 616, A16

\bibitem[{{Lomb}(1976)}]{Lomb1976}
{Lomb}, N.~R. 1976, \apss, 39, 447

\bibitem[{{Lu} {et~al.}(2022){Lu}, {Curtis}, {Angus}, {David}, \& {Hattori}}]{Lu2022}
{Lu}, Y.~L., {Curtis}, J.~L., {Angus}, R., {David}, T.~J., \& {Hattori}, S. 2022, \aj, 164, 251

\bibitem[{{Luger} {et~al.}(2021){Luger}, {Foreman-Mackey}, {Hedges}, \& {Hogg}}]{Luger2021}
{Luger}, R., {Foreman-Mackey}, D., {Hedges}, C., \& {Hogg}, D.~W. 2021, \aj, 162, 123

\bibitem[{{Martinez Pillet} {et~al.}(1993){Martinez Pillet}, {Moreno-Insertis}, \& {Vazquez}}]{MartinezPillet1993}
{Martinez Pillet}, V., {Moreno-Insertis}, F., \& {Vazquez}, M. 1993, \aap, 274, 521

\bibitem[{{Masuda}(2022{\natexlab{a}})}]{Masuda2022b}
{Masuda}, K. 2022{\natexlab{a}}, \apj, 937, 94

\bibitem[{{Masuda}(2022{\natexlab{b}})}]{Masuda2022a}
{Masuda}, K. 2022{\natexlab{b}}, \apj, 933, 195

\bibitem[{{Mathur} {et~al.}(2023){Mathur}, {Claytor}, {Santos}, {Garc{\'\i}a}, {Amard}, {Bugnet}, {Corsaro}, {Bonanno}, {Breton}, {Godoy-Rivera}, {Pinsonneault}, \& {van Saders}}]{Mathur2023}
{Mathur}, S., {Claytor}, Z.~R., {Santos}, {\^A}. R.~G., {et~al.} 2023, \apj, 952, 131

\bibitem[{{Mathur} {et~al.}(2014{\natexlab{a}}){Mathur}, {Garc{\'\i}a}, {Ballot}, {Ceillier}, {Salabert}, {Metcalfe}, {R{\'e}gulo}, {Jim{\'e}nez}, \& {Bloemen}}]{Mathur2014}
{Mathur}, S., {Garc{\'\i}a}, R.~A., {Ballot}, J., {et~al.} 2014{\natexlab{a}}, \aap, 562, A124

\bibitem[{{Mathur} {et~al.}(2019){Mathur}, {Garc{\'\i}a}, {Bugnet}, {Santos}, {Santiago}, \& {Beck}}]{Mathur2019}
{Mathur}, S., {Garc{\'\i}a}, R.~A., {Bugnet}, L., {et~al.} 2019, Frontiers in Astronomy and Space Sciences, 6, 46

\bibitem[{{Mathur} {et~al.}(2014{\natexlab{b}}){Mathur}, {Salabert}, {Garc{\'\i}a}, \& {Ceillier}}]{Mathur2014b}
{Mathur}, S., {Salabert}, D., {Garc{\'\i}a}, R.~A., \& {Ceillier}, T. 2014{\natexlab{b}}, Journal of Space Weather and Space Climate, 4, A15

\bibitem[{{McQuillan} {et~al.}(2013{\natexlab{a}}){McQuillan}, {Aigrain}, \& {Mazeh}}]{McQuillan2013b}
{McQuillan}, A., {Aigrain}, S., \& {Mazeh}, T. 2013{\natexlab{a}}, \mnras, 432, 1203

\bibitem[{{McQuillan} {et~al.}(2013{\natexlab{b}}){McQuillan}, {Mazeh}, \& {Aigrain}}]{McQuillan2013}
{McQuillan}, A., {Mazeh}, T., \& {Aigrain}, S. 2013{\natexlab{b}}, \apjl, 775, L11

\bibitem[{{McQuillan} {et~al.}(2014){McQuillan}, {Mazeh}, \& {Aigrain}}]{McQuillan2014}
{McQuillan}, A., {Mazeh}, T., \& {Aigrain}, S. 2014, \apjs, 211, 24

\bibitem[{{Mehta} {et~al.}(2022){Mehta}, {Jain}, {Tripathy}, {Kiefer}, {Kolotkov}, \& {Broomhall}}]{Mehta2022}
{Mehta}, T., {Jain}, K., {Tripathy}, S.~C., {et~al.} 2022, \mnras, 515, 2415

\bibitem[{{Messias} {et~al.}(2022){Messias}, {de Oliveira}, {Gomes}, {Arruda Gon{\c{c}}alves}, {Canto Martins}, {Le{\~a}o}, \& {De Medeiros}}]{Messias2022}
{Messias}, Y.~S., {de Oliveira}, L.~L.~A., {Gomes}, R.~L., {et~al.} 2022, \apjl, 930, L23

\bibitem[{{Meunier} {et~al.}(2019){Meunier}, {Lagrange}, {Boulet}, \& {Borgniet}}]{Meunier2019}
{Meunier}, N., {Lagrange}, A.~M., {Boulet}, T., \& {Borgniet}, S. 2019, \aap, 627, A56

\bibitem[{{Montalto} {et~al.}(2021){Montalto}, {Piotto}, {Marrese}, {Nascimbeni}, {Prisinzano}, {Granata}, {Marinoni}, {Desidera}, {Ortolani}, {Aerts}, {Alei}, {Altavilla}, {Benatti}, {B{\"o}rner}, {Cabrera}, {Claudi}, {Deleuil}, {Fabrizio}, {Gizon}, {Goupil}, {Heras}, {Magrin}, {Malavolta}, {Mas-Hesse}, {Pagano}, {Paproth}, {Pertenais}, {Pollacco}, {Ragazzoni}, {Ramsay}, {Rauer}, \& {Udry}}]{Montalto2021}
{Montalto}, M., {Piotto}, G., {Marrese}, P.~M., {et~al.} 2021, \aap, 653, A98

\bibitem[{{Nascimbeni} {et~al.}(2022){Nascimbeni}, {Piotto}, {B{\"o}rner}, {Montalto}, {Marrese}, {Cabrera}, {Marinoni}, {Aerts}, {Altavilla}, {Benatti}, {Claudi}, {Deleuil}, {Desidera}, {Fabrizio}, {Gizon}, {Goupil}, {Granata}, {Heras}, {Magrin}, {Malavolta}, {Mas-Hesse}, {Ortolani}, {Pagano}, {Pollacco}, {Prisinzano}, {Ragazzoni}, {Ramsay}, {Rauer}, \& {Udry}}]{Nascimbeni2022}
{Nascimbeni}, V., {Piotto}, G., {B{\"o}rner}, A., {et~al.} 2022, \aap, 658, A31

\bibitem[{{Nielsen} {et~al.}(2013){Nielsen}, {Gizon}, {Schunker}, \& {Karoff}}]{Nielsen2013}
{Nielsen}, M.~B., {Gizon}, L., {Schunker}, H., \& {Karoff}, C. 2013, \aap, 557, L10

\bibitem[{{Noraz} {et~al.}(2022){Noraz}, {Breton}, {Brun}, {Garc{\'\i}a}, {Strugarek}, {Santos}, {Mathur}, \& {Amard}}]{Noraz2022}
{Noraz}, Q., {Breton}, S.~N., {Brun}, A.~S., {et~al.} 2022, \aap, 667, A50

\bibitem[{{Notsu} {et~al.}(2019){Notsu}, {Maehara}, {Honda}, {Hawley}, {Davenport}, {Namekata}, {Notsu}, {Ikuta}, {Nogami}, \& {Shibata}}]{Notsu2019}
{Notsu}, Y., {Maehara}, H., {Honda}, S., {et~al.} 2019, \apj, 876, 58

\bibitem[{{Noyes} {et~al.}(1984){Noyes}, {Hartmann}, {Baliunas}, {Duncan}, \& {Vaughan}}]{Noyes1984}
{Noyes}, R.~W., {Hartmann}, L.~W., {Baliunas}, S.~L., {Duncan}, D.~K., \& {Vaughan}, A.~H. 1984, \apj, 279, 763

\bibitem[{{{\"O}zavc{\i}} {et~al.}(2018){{\"O}zavc{\i}}, {{\c{S}}enavc{\i}}, {I{\c{s}}{\i}k}, {Hussain}, {O'Neal}, {Y{\i}lmaz}, \& {Selam}}]{Ozavci2018}
{{\"O}zavc{\i}}, I., {{\c{S}}enavc{\i}}, H.~V., {I{\c{s}}{\i}k}, E., {et~al.} 2018, \mnras, 474, 5534

\bibitem[{pandas~development team(2020)}]{reback2020pandas}
pandas~development team, T. 2020, pandas-dev/pandas: Pandas

\bibitem[{Pedregosa {et~al.}(2011)Pedregosa, Varoquaux, Gramfort, Michel, Thirion, Grisel, Blondel, Prettenhofer, Weiss, Dubourg, Vanderplas, Passos, Cournapeau, Brucher, Perrot, \& Duchesnay}]{scikit-learn}
Pedregosa, F., Varoquaux, G., Gramfort, A., {et~al.} 2011, Journal of Machine Learning Research, 12, 2825

\bibitem[{{Raetz} {et~al.}(2020){Raetz}, {Stelzer}, {Damasso}, \& {Scholz}}]{Raetz2020}
{Raetz}, S., {Stelzer}, B., {Damasso}, M., \& {Scholz}, A. 2020, \aap, 637, A22

\bibitem[{{Rauer} {et~al.}(2014){Rauer}, {Catala}, {Aerts}, {Appourchaux}, {Benz}, {Brandeker}, {Christensen-Dalsgaard}, {Deleuil}, {Gizon}, {Goupil}, {G{\"u}del}, {Janot-Pacheco}, {Mas-Hesse}, {Pagano}, {Piotto}, {Pollacco}, {Santos}, {Smith}, {Su{\'a}rez}, {Szab{\'o}}, {Udry}, {Adibekyan}, {Alibert}, {Almenara}, {Amaro-Seoane}, {Eiff}, {Asplund}, {Antonello}, {Barnes}, {Baudin}, {Belkacem}, {Bergemann}, {Bihain}, {Birch}, {Bonfils}, {Boisse}, {Bonomo}, {Borsa}, {Brand {\~a}o}, {Brocato}, {Brun}, {Burleigh}, {Burston}, {Cabrera}, {Cassisi}, {Chaplin}, {Charpinet}, {Chiappini}, {Church}, {Csizmadia}, {Cunha}, {Damasso}, {Davies}, {Deeg}, {D{\'\i}az}, {Dreizler}, {Dreyer}, {Eggenberger}, {Ehrenreich}, {Eigm{\"u}ller}, {Erikson}, {Farmer}, {Feltzing}, {de Oliveira Fialho}, {Figueira}, {Forveille}, {Fridlund}, {Garc{\'\i}a}, {Giommi}, {Giuffrida}, {Godolt}, {Gomes da Silva}, {Granzer}, {Grenfell}, {Grotsch-Noels}, {G{\"u}nther}, {Haswell}, {Hatzes}, {H{\'e}brard}, {Hekker}, {Helled}, {Heng}, {Jenkins},
  {Johansen}, {Khodachenko}, {Kislyakova}, {Kley}, {Kolb}, {Krivova}, {Kupka}, {Lammer}, {Lanza}, {Lebreton}, {Magrin}, {Marcos-Arenal}, {Marrese}, {Marques}, {Martins}, {Mathis}, {Mathur}, {Messina}, {Miglio}, {Montalban}, {Montalto}, {Monteiro}, {Moradi}, {Moravveji}, {Mordasini}, {Morel}, {Mortier}, {Nascimbeni}, {Nelson}, {Nielsen}, {Noack}, {Norton}, {Ofir}, {Oshagh}, {Ouazzani}, {P{\'a}pics}, {Parro}, {Petit}, {Plez}, {Poretti}, {Quirrenbach}, {Ragazzoni}, {Raimondo}, {Rainer}, {Reese}, {Redmer}, {Reffert}, {Rojas-Ayala}, {Roxburgh}, {Salmon}, {Santerne}, {Schneider}, {Schou}, {Schuh}, {Schunker}, {Silva-Valio}, {Silvotti}, {Skillen}, {Snellen}, {Sohl}, {Sousa}, {Sozzetti}, {Stello}, {Strassmeier}, {{\v{S}}vanda}, {Szab{\'o}}, {Tkachenko}, {Valencia}, {Van Grootel}, {Vauclair}, {Ventura}, {Wagner}, {Walton}, {Weingrill}, {Werner}, {Wheatley}, \& {Zwintz}}]{Rauer2014}
{Rauer}, H., {Catala}, C., {Aerts}, C., {et~al.} 2014, Experimental Astronomy, 38, 249

\bibitem[{{Reinhold} \& {Hekker}(2020)}]{Reinhold2020}
{Reinhold}, T. \& {Hekker}, S. 2020, \aap, 635, A43

\bibitem[{Reinhold {et~al.}(2013)Reinhold, Reiners, \& Basri}]{Reinhold2013a}
Reinhold, T., Reiners, A., \& Basri, G. 2013, A\&A, 560, A4

\bibitem[{{Reinhold} {et~al.}(2023){Reinhold}, {Shapiro}, {Solanki}, \& {Basri}}]{Reinhold2023}
{Reinhold}, T., {Shapiro}, A.~I., {Solanki}, S.~K., \& {Basri}, G. 2023, \aap, 678, A24

\bibitem[{{Ricker} {et~al.}(2015){Ricker}, {Winn}, {Vanderspek}, {Latham}, {Bakos}, {Bean}, {Berta-Thompson}, {Brown}, {Buchhave}, {Butler}, {Butler}, {Chaplin}, {Charbonneau}, {Christensen-Dalsgaard}, {Clampin}, {Deming}, {Doty}, {De Lee}, {Dressing}, {Dunham}, {Endl}, {Fressin}, {Ge}, {Henning}, {Holman}, {Howard}, {Ida}, {Jenkins}, {Jernigan}, {Johnson}, {Kaltenegger}, {Kawai}, {Kjeldsen}, {Laughlin}, {Levine}, {Lin}, {Lissauer}, {MacQueen}, {Marcy}, {McCullough}, {Morton}, {Narita}, {Paegert}, {Palle}, {Pepe}, {Pepper}, {Quirrenbach}, {Rinehart}, {Sasselov}, {Sato}, {Seager}, {Sozzetti}, {Stassun}, {Sullivan}, {Szentgyorgyi}, {Torres}, {Udry}, \& {Villasenor}}]{Ricker2015}
{Ricker}, G.~R., {Winn}, J.~N., {Vanderspek}, R., {et~al.} 2015, Journal of Astronomical Telescopes, Instruments, and Systems, 1, 014003

\bibitem[{{Rieger} {et~al.}(1984){Rieger}, {Share}, {Forrest}, {Kanbach}, {Reppin}, \& {Chupp}}]{Rieger1984}
{Rieger}, E., {Share}, G.~H., {Forrest}, D.~J., {et~al.} 1984, \nat, 312, 623

\bibitem[{{Salabert} {et~al.}(2016){Salabert}, {Garc{\'\i}a}, {Beck}, {Egeland}, {Pall{\'e}}, {Mathur}, {Metcalfe}, {do Nascimento}, {Ceillier}, {Andersen}, \& {Trivi{\~n}o Hage}}]{Salabert2016}
{Salabert}, D., {Garc{\'\i}a}, R.~A., {Beck}, P.~G., {et~al.} 2016, \aap, 596, A31

\bibitem[{{Salabert} {et~al.}(2017){Salabert}, {Garc{\'\i}a}, {Jim{\'e}nez}, {Bertello}, {Corsaro}, \& {Pall{\'e}}}]{Salabert2017}
{Salabert}, D., {Garc{\'\i}a}, R.~A., {Jim{\'e}nez}, A., {et~al.} 2017, \aap, 608, A87

\bibitem[{{Samadi} {et~al.}(2019){Samadi}, {Deru}, {Reese}, {Marchiori}, {Grolleau}, {Green}, {Pertenais}, {Lebreton}, {Deheuvels}, {Mosser}, {Belkacem}, {B{\"o}rner}, \& {Smith}}]{Samadi2019}
{Samadi}, R., {Deru}, A., {Reese}, D., {et~al.} 2019, \aap, 624, A117

\bibitem[{{Santos} {et~al.}(2021{\natexlab{a}}){Santos}, {Breton}, {Mathur}, \& {Garc{\'\i}a}}]{Santos2021}
{Santos}, A.~R.~G., {Breton}, S.~N., {Mathur}, S., \& {Garc{\'\i}a}, R.~A. 2021{\natexlab{a}}, \apjs, 255, 17

\bibitem[{{Santos} {et~al.}(2017){Santos}, {Cunha}, {Avelino}, {Garc{\'\i}a}, \& {Mathur}}]{Santos2017}
{Santos}, A.~R.~G., {Cunha}, M.~S., {Avelino}, P.~P., {Garc{\'\i}a}, R.~A., \& {Mathur}, S. 2017, \aap, 599, A1

\bibitem[{{Santos} {et~al.}(2019){Santos}, {Garc{\'\i}a}, {Mathur}, {Bugnet}, {van Saders}, {Metcalfe}, {Simonian}, \& {Pinsonneault}}]{Santos2019}
{Santos}, A.~R.~G., {Garc{\'\i}a}, R.~A., {Mathur}, S., {et~al.} 2019, \apjs, 244, 21

\bibitem[{{Santos} {et~al.}(2023){Santos}, {Mathur}, {Garc{\'\i}a}, {Broomhall}, {Egeland}, {Jim{\'e}nez}, {Godoy-Rivera}, {Breton}, {Claytor}, {Metcalfe}, {Cunha}, \& {Amard}}]{Santos2023}
{Santos}, A.~R.~G., {Mathur}, S., {Garc{\'\i}a}, R.~A., {et~al.} 2023, \aap, 672, A56

\bibitem[{{Santos} {et~al.}(2021{\natexlab{b}}){Santos}, {Mathur}, {Garc{\'\i}a}, {Cunha}, \& {Avelino}}]{Santos2021_ACF}
{Santos}, A.~R.~G., {Mathur}, S., {Garc{\'\i}a}, R.~A., {Cunha}, M.~S., \& {Avelino}, P.~P. 2021{\natexlab{b}}, \mnras, 508, 267

\bibitem[{{Scargle}(1982)}]{Scargle1982}
{Scargle}, J.~D. 1982, \apj, 263, 835

\bibitem[{{See} {et~al.}(2023){See}, {Roquette}, {Amard}, \& {Matt}}]{See2023}
{See}, V., {Roquette}, J., {Amard}, L., \& {Matt}, S. 2023, \mnras, 524, 5781

\bibitem[{{See} {et~al.}(2021){See}, {Roquette}, {Amard}, \& {Matt}}]{See2021}
{See}, V., {Roquette}, J., {Amard}, L., \& {Matt}, S.~P. 2021, \apj, 912, 127

\bibitem[{{Skumanich}(1972)}]{Skumanich1972}
{Skumanich}, A. 1972, \apj, 171, 565

\bibitem[{{Spada} \& {Lanzafame}(2020)}]{Spada2020}
{Spada}, F. \& {Lanzafame}, A.~C. 2020, \aap, 636, A76

\bibitem[{{van Saders} {et~al.}(2016){van Saders}, {Ceillier}, {Metcalfe}, {Silva Aguirre}, {Pinsonneault}, {Garc{\'\i}a}, {Mathur}, \& {Davies}}]{vanSaders2016}
{van Saders}, J.~L., {Ceillier}, T., {Metcalfe}, T.~S., {et~al.} 2016, \nat, 529, 181

\bibitem[{{van Saders} {et~al.}(2019){van Saders}, {Pinsonneault}, \& {Barbieri}}]{vanSaders2019}
{van Saders}, J.~L., {Pinsonneault}, M.~H., \& {Barbieri}, M. 2019, \apj, 872, 128

\bibitem[{Virtanen {et~al.}(2020)Virtanen, Gommers, Oliphant, Haberland, Reddy, Cournapeau, Burovski, Peterson, Weckesser, Bright, {van der Walt}, Brett, Wilson, Millman, Mayorov, Nelson, Jones, Kern, Larson, Carey, Polat, Feng, Moore, {VanderPlas}, Laxalde, Perktold, Cimrman, Henriksen, Quintero, Harris, Archibald, Ribeiro, Pedregosa, {van Mulbregt}, \& {SciPy 1.0 Contributors}}]{2020SciPy-NMeth}
Virtanen, P., Gommers, R., Oliphant, T.~E., {et~al.} 2020, Nature Methods, 17, 261

\bibitem[{{W}es {M}c{K}inney(2010)}]{mckinney-proc-scipy-2010}
{W}es {M}c{K}inney. 2010, in {P}roceedings of the 9th {P}ython in {S}cience {C}onference, ed. {S}t\'efan van~der {W}alt \& {J}arrod {M}illman, 56 -- 61

\bibitem[{{Wilson} {et~al.}(1988){Wilson}, {Altrocki}, {Harvey}, {Martin}, \& {Snodgrass}}]{Wilson1988}
{Wilson}, P.~R., {Altrocki}, R.~C., {Harvey}, K.~L., {Martin}, S.~F., \& {Snodgrass}, H.~B. 1988, \nat, 333, 748

\bibitem[{{Woodard}(1984)}]{Woodard1984}
{Woodard}, M.~F. 1984, PhD thesis, University of California, San Diego.

\bibitem[{{Yang} \& {Liu}(2019)}]{Yang2019}
{Yang}, H. \& {Liu}, J. 2019, \apjs, 241, 29

\bibitem[{{Zechmeister} \& {K{\"u}rster}(2009)}]{Zechmeister09}
{Zechmeister}, M. \& {K{\"u}rster}, M. 2009, \aap, 496, 577

\end{thebibliography}

\appendix

\section{Generation of rotational modulation parameters \label{appendix:draw_rot}}


The prescriptions from \citet{Meunier2019} are used to generate rotational modulation parameters from the input $T_\mathrm{eff}$ selected from the PIC.
In what follows, we denote as $\mathbb{X}$ a random variable with uniform distribution between 0 and 1. First, $B-V$ is interpolated from $T_\mathrm{eff}$ using Table~1 from \citet{Meunier2019}. Using $B-V$, an activity proxy $\log R'_{\rm HK}$ is randomly drawn with the following procedure.
\begin{equation}
    \log R'_{\rm HK} = \log R'_{\rm HK, min} 
    + (\log R'_{\rm HK, max} - \log R'_{\rm HK, min}) \mathbb{X} \; ,
\end{equation}
where 
\begin{equation}
\log R'_{\rm HK, max} = - 0.375 (B-V) - 4.4 \; ,
\end{equation}
and $\log R'_{\rm \rm HK, min}$ has been obtained through a minimum value of the activity index $S$, $S_\mathrm{min}$
\begin{equation}
     S_\mathrm{min} = \left\{
    \begin{aligned}
     &0.144 \text{ if } B-V < 0.94 \\
     &0.0269231 (B-V) + 0.118892 \text{ otherwise} 
    \end{aligned}
    \right . \; ,
\end{equation}
and then
\begin{equation}
    \log R'_{\rm \rm HK, min} = 
     - 4 + \log [1.34 (B-V)] + \log C_{cf} \; ,
\end{equation}
where, writing $\mu = 0.63 - (B-V)$
\begin{equation}
    \log C_{cf} = 
    \left \{
    \begin{aligned}
     & \log C_{cf,0} \text{ if } B-V \geq 0.63 \\
     & \log C_{cf,0} + 0.135 \mu - 0.814 \mu^2 + 6.03 \mu^3 \text{ otherwise} 
    \end{aligned}
    \right . \; ,
\end{equation}
with 
\begin{equation}
    \log C_{cf,0} = 1.13 (B-V)^3 - 3.91 (B-V)^2 + 2.84 (B-V) - 0.47 \; .
\end{equation}

At this stage, $B-V$ and $\log R'_{\rm HK}$ are combined to draw the average rotation period $P_\mathrm{rot}$, using the Rossby number, $Ro_{\rm Noyes}$, and the convective turnover time, $\tau_c$, estimates from \citet{Noyes1984}. We have,  writing $x = 1 - (B - V)$
\begin{equation}
    \log \tau_c = \left\{
    \begin{aligned}
     &1.362 - 0.166x + 0.025x^2 - 5.323x^3 \text{ if } x > 0 \\
     &1.362 - 0.14x \text{ if } x \leq 0 
    \end{aligned}
    \right . \; ,
\end{equation}
and 
\begin{equation}
    Ro_{\rm Noyes} = 0.808 - 2.966 (\log R'_{\rm HK} + 4.52) \; .
\end{equation}
Subsequently we draw $P_\mathrm{rot}$ as 
\begin{equation}
    P_\mathrm{rot} = (Ro_{\rm Noyes} + 0.4 \mathbb{X} - 0.2) \tau_c \; .
\end{equation}

Once $P_\mathrm{rot}$ has been drawn, the amplitude of latitudinal differential rotation, given by the parameters $P_\mathrm{min}$ and $P_\mathrm{max}$, is deterministic and depends on $T_\mathrm{eff}$ and $P_\mathrm{rot}$
\begin{equation}
\label{eq:differential_rotation}
    \left\{
    \begin{aligned}
    &p_0 = - 3.485 + 2.4781 \, \frac{T_\mathrm{eff}}{10^4 K}  \; , \\
    &p_1 = 1.597 - 1.351 \, \frac{T_\mathrm{eff}}{10^4 K} \; , \\
    &\log \alpha = p_0 + p_1 \log \left (\frac{P_\mathrm{rot}}{1 \text{ day}} \right) \; , \\
    &P_\mathrm{max} = 2 P_\mathrm{rot} / (2 - \alpha) \; , \\
    &P_\mathrm{min} = (1 - \alpha) P_\mathrm{max} \; ,
    \end{aligned}
    \right .
\end{equation}
where $p_0$, $p_1$ and $\alpha$ are dimensionless parameters.
The latitudinal extent of the active region formation range between $\lambda = 0$ and $\lambda_\mathrm{max}$ drawn as 
\begin{equation}
   \lambda_\mathrm{max} = 32^{\rm o} + 20^{\rm o} \times \mathbb{X} \; .
\end{equation}

The activity cycle length, $P_\mathrm{cycle}$ is drawn as 
\begin{equation}
\label{eq:prot_pcycle}
    \log \left ( \frac{P_\mathrm{cycle}}{1 \, \mathrm{year}} \right) = 0.16 \log \left (\frac{P_\mathrm{rot}}{1 \text{ day}} \right) + 3.14 + 0.6\mathbb{X} - 0.3  \; ,
\end{equation}
and the overlap between consecutive cycles, $\delta_\mathrm{cycle}$, as
\begin{equation}
    \delta_{\rm cycle} = 0.1 \mathbb{X} P_{\rm cycle} \; .
\end{equation}



{
\section{Light curves properties and preprocessing}

In this appendix, we present and discuss some additional aspects related to the PSLS light curves and their preprocessing. 

\subsection{Photon noise level \label{appendix:photon_noise}}

\begin{figure}[ht!]
    \centering
    \includegraphics[width=0.49\textwidth]{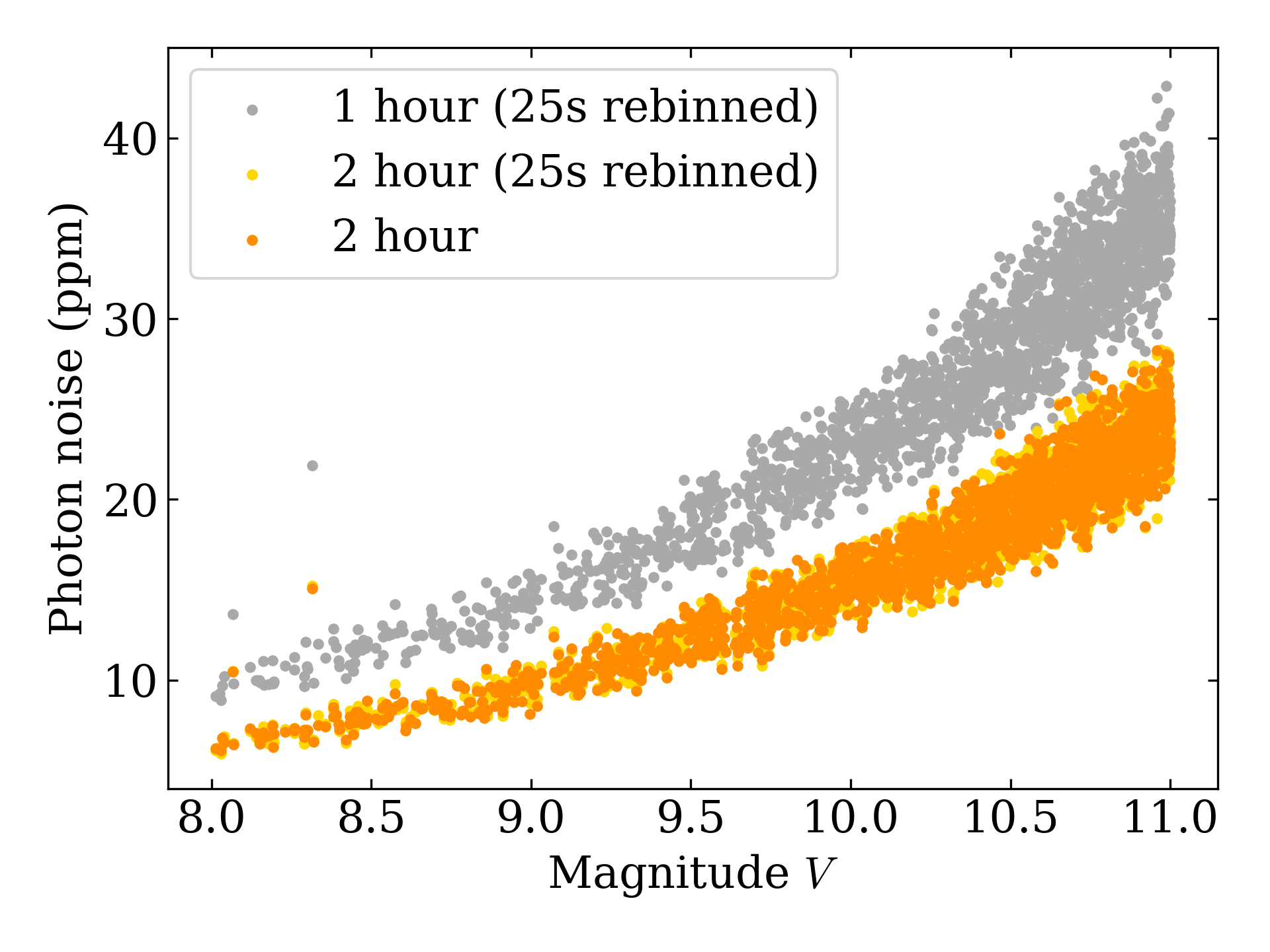}
    \caption{Photon noise level as a function of magnitude $V$. The orange distribution corresponds to photon noise level computed for pure noise time series generated directly with a 2-hour sampling. The yellow distribution corresponds to photon noise level computed for pure noise time series generated with a 25s sampling, rebinned at 2-hour. The grey distribution corresponds to photon noise level computed for pure noise time series generated with a 25s sampling, rebinned at 1-hour.}
    \label{fig:noise_level}
\end{figure}

In order to isolate the contribution of photon noise level in the simulations presented in this work, we generated a set of 90-day (one quarter) light curves with only instrumental signal, using as input the same magnitude $V$ and PSLS random seed as for the 8-year simulations with rotational modulations. 
In order to validate that using the PSLS to directly generate two-hour light curves does not affect the photon noise, we generate these 90 day light curves with both a two-hour and a 25 second sampling. We rebin the 25 second sampled time series at one hour and two hours. 
After filtering out the low-frequency drift, we compute the photon noise level as the standard deviation of the time series. We show the magnitude $V$ versus photon noise diagram in Fig.~\ref{fig:noise_level}.

The orange and yellow distributions correspond to the photon noise of the simulations analysed in this work, and to what is currently expected for two-hour rebinned PLATO light curves in this range of magnitude. Comparing these two distributions also allows us to validate that using the PSLS to directly generate light curves with two-hour sampling does not introduce a bias in the resulting photon noise level.
We show for comparison (in grey) the noise level obtained for the one-hour rebinning of the 25 second time series. This timescale is the reference used to define PLATO stellar sample in terms of noise: P1 and P2 sample will have, by construction, a photon noise below 50 ppm in one hour \citep[e.g.][]{Montalto2021}. We remind that, for all of our simulations, we considered the case of stars observed by the 4 groups of 6 cameras together. For a given magnitude $V$, the noise level of a star observed by a smaller number of cameras is therefore expected to be slightly larger (but still under 50 ppm in one hour for P1 and P2). 

{
\subsection{Granulation level \label{appendix:granulation_level}}

\begin{figure}[ht!]
    \centering
    \includegraphics[width=0.49\textwidth]{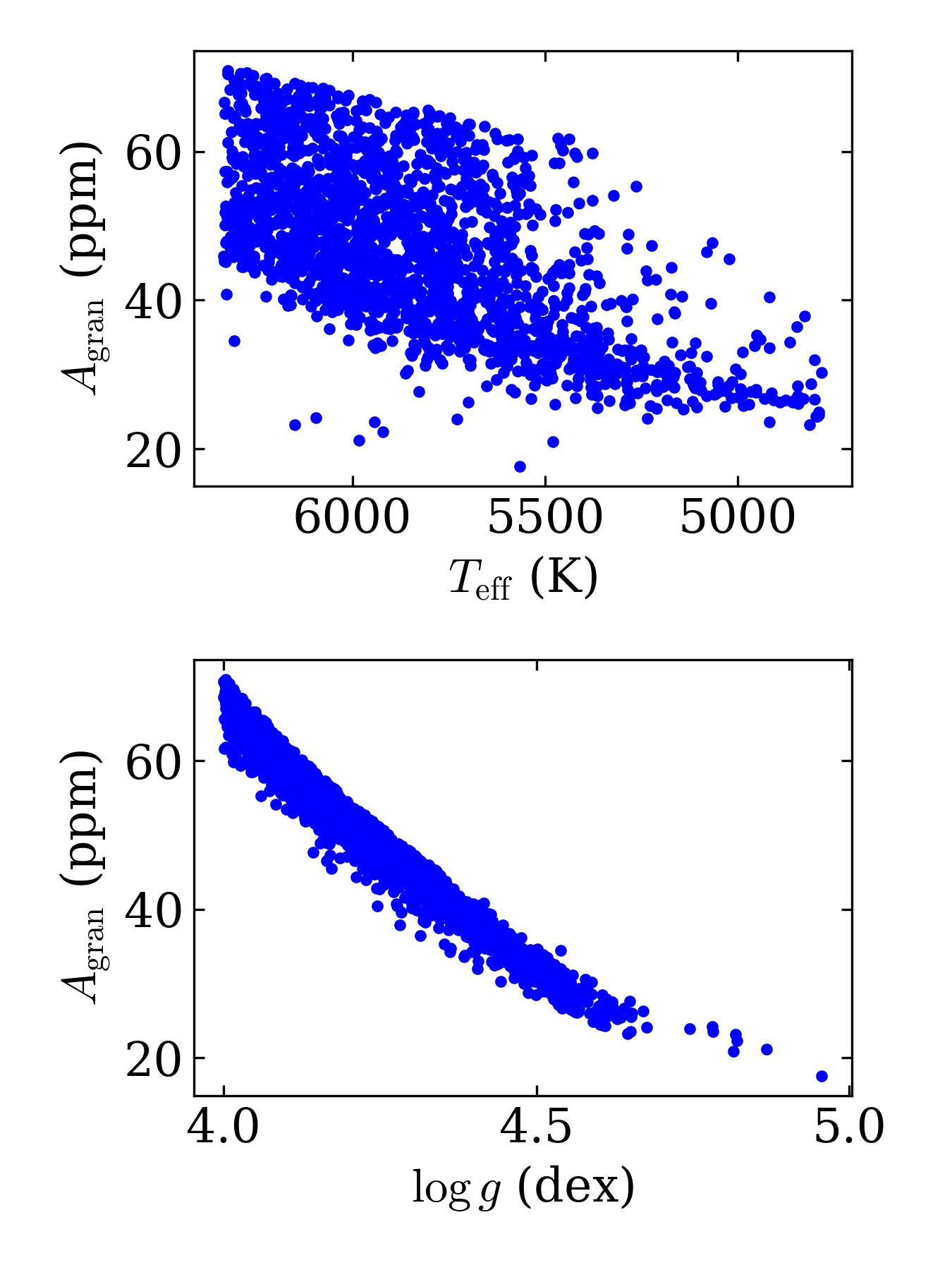}
    \caption{Granulation power level as a function of $T_{\rm eff}$ (\textit{top}) and $\log g$ (\textit{bottom}).}
    \label{fig:granulation_level}
\end{figure}

As mentioned in Sect.~\ref{sec:simulation_procedure}, granulation is included in the simulated light curves using the scaling laws derived by \citet{Kallinger2014}. The granulation is modelled by two pseudo-Lorentzian functions (also referred as Harvey models) with distinct characteristic timescales. The amplitudes $a_1$ and $a_2$ of theses functions, in ppm, is given by 
\begin{align}
    a_1 &= 3382 \times \nu_{\rm max}^{-0.609} \; , \\
    a_2 &= 3710 \times \nu_{\rm max}^{-0.613} M^{-0.26} \; .
\end{align}
where $M$ is the stellar mass and $\nu_{\rm max}$ is the frequency of maximal power of the stellar oscillations \citep[see e.g.][]{Garcia2019}. The total intensity fluctuation related to granulation $A_{\rm gran}$ in the background will therefore be 
\begin{equation}
    A_{\rm gran} = C_{\rm bol} \sqrt{a_1^2 + a_2^2} \; ,
\end{equation}
where the bolometric correction $C_{\rm bol}$ is given by $C_{\rm bol} = (T_{\rm eff} /  T_0)^{\alpha}$ with $T_0 = 5934$~K and $\alpha=0.8$. In Fig~\ref{fig:granulation_level}, we represent how $A_{\rm gran}$ scales with the $T_\mathrm{eff}$ and $\log g$ parameters of the star. The strong correlation between $A_{\rm gran}$ and $\log g$ is explained by the fact that 
the $\nu_{\rm max}$ values strongly depends on $\log g$ \citep[e.g.][]{Kjeldsen1995,Kallinger2016}  
}

\subsection{Light curve filtering \label{appendix:filtering}}

We illustrate here the effect that the 60 day high pass filtering applied on LC1 to obtain LC2 has on stellar variability. We show the comparison between the observed signal for a noise-free light curve before and after filtering in Fig.~\ref{fig:example_filtering}. We also compare the periodograms of the two light curves, where the impact of the 60 day cutoff ($\sim 0.19$~$\mu$Hz) on the LC2 periodogram is clearly visible. 

\begin{figure}[ht!]
    \centering
    \includegraphics[width=0.49\textwidth]{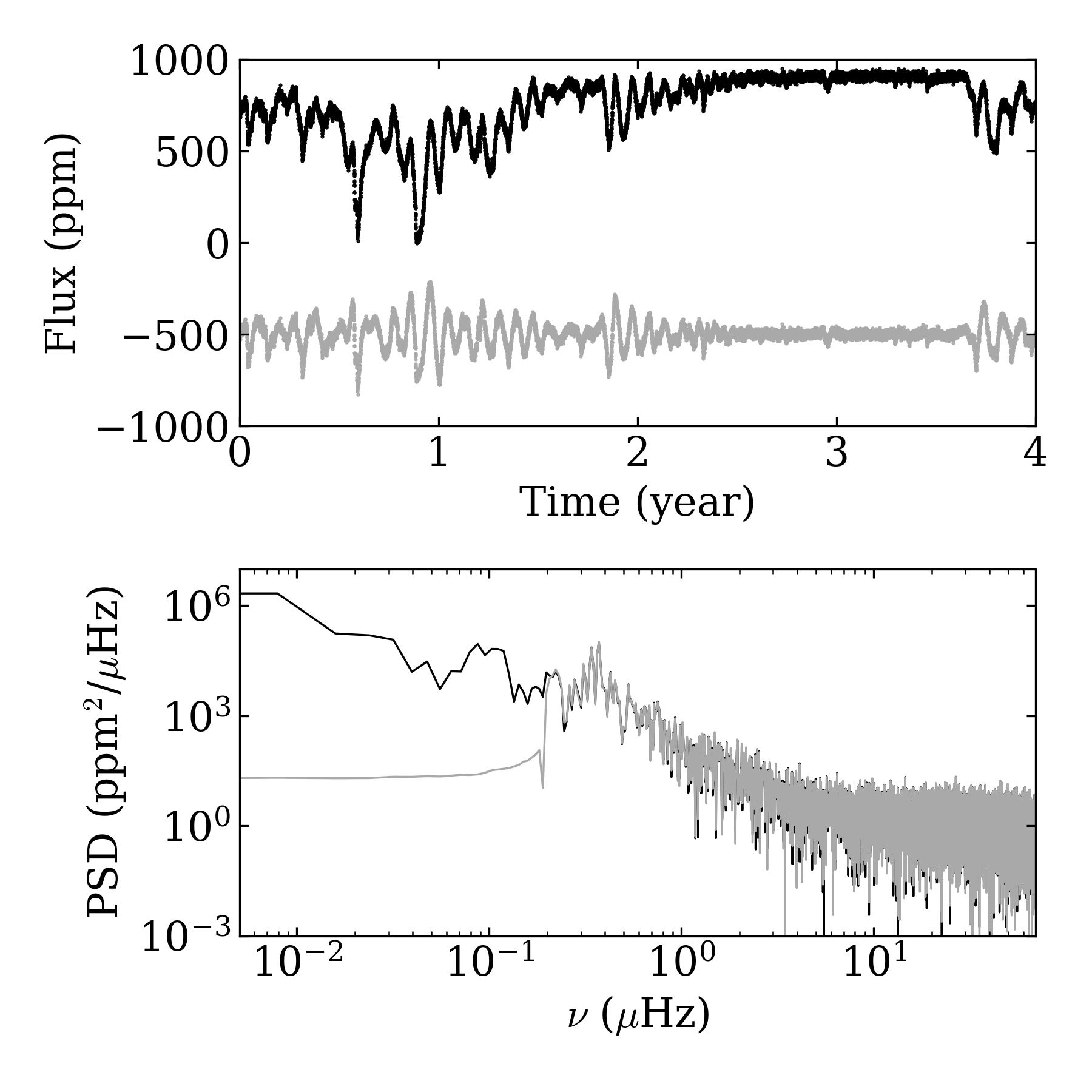}
    \caption{\textit{Top:} comparison between LC1 (black) and LC2 (grey) for a noise-free light curve. The light curves have been shifted to improve readibility. \textit{Bottom}: corresponding LC1 and LC2 periodograms.}
    \label{fig:example_filtering}
\end{figure}

\section{Additional rotation recovery diagnostics \label{appendix:additional_recovery_diagnostics}}

We provide in this appendix diagnostics complementary to those presented in Sect.~\ref{sec:rotation_period_recovery}. We show in Fig.~\ref{fig:additional_recovery_diagnostics} and Fig.~\ref{fig:additional_recovery_diagnostics_corr} the comparison between the true rotation periods $P_{\rm rot, true}$, and the recovered rotation periods $P_{\rm rot, recovered}$, for both noise-free light curves and noisy light curves.
Uncertainties on $P_{\rm rot, recovered}$ are represented when available (we remind that the case where uncertainties are not estimated means that ROOSTER selected $P_{\rm ACF}$ as $P_{\rm rot}$, see Sect.~\ref{sec:average_rotation_period}).
The uncertainties increases with $P_{\rm rot, recovered}$. 
For the considered range ($8 < V < 11$), we do not not find a dependence of the uncertainties on the magnitude.  
We note that for short temporal baselines, long periods are more generally selected by ROOSTER from $P_{\rm ACF}$, while, for longer baselines, ROOSTER favours $P_{\rm GLS}$ or $P_{\rm CS}$.
We finally note that, for a very small fraction of the sample, our methodology selects the second harmonic of $P_{\rm rot}$ rather than $P_{\rm rot}$ itself, an issue that was already discussed in \citet{Breton2021}. Nevertheless, the vast majority of the recovered periods are located in the $\pm$10\% interval around $P_{\rm rot,true}$: considering that the second harmonic was chosen if $0.49 P_{\rm rot, true} < P_{\rm rot, recovered} < 0.51 P_{\rm rot, true}$, we never have more than 1.2~\% of the recovered periods laying in this area.

\begin{figure*}[ht!]
    \centering
    \includegraphics[width=0.42\textwidth]{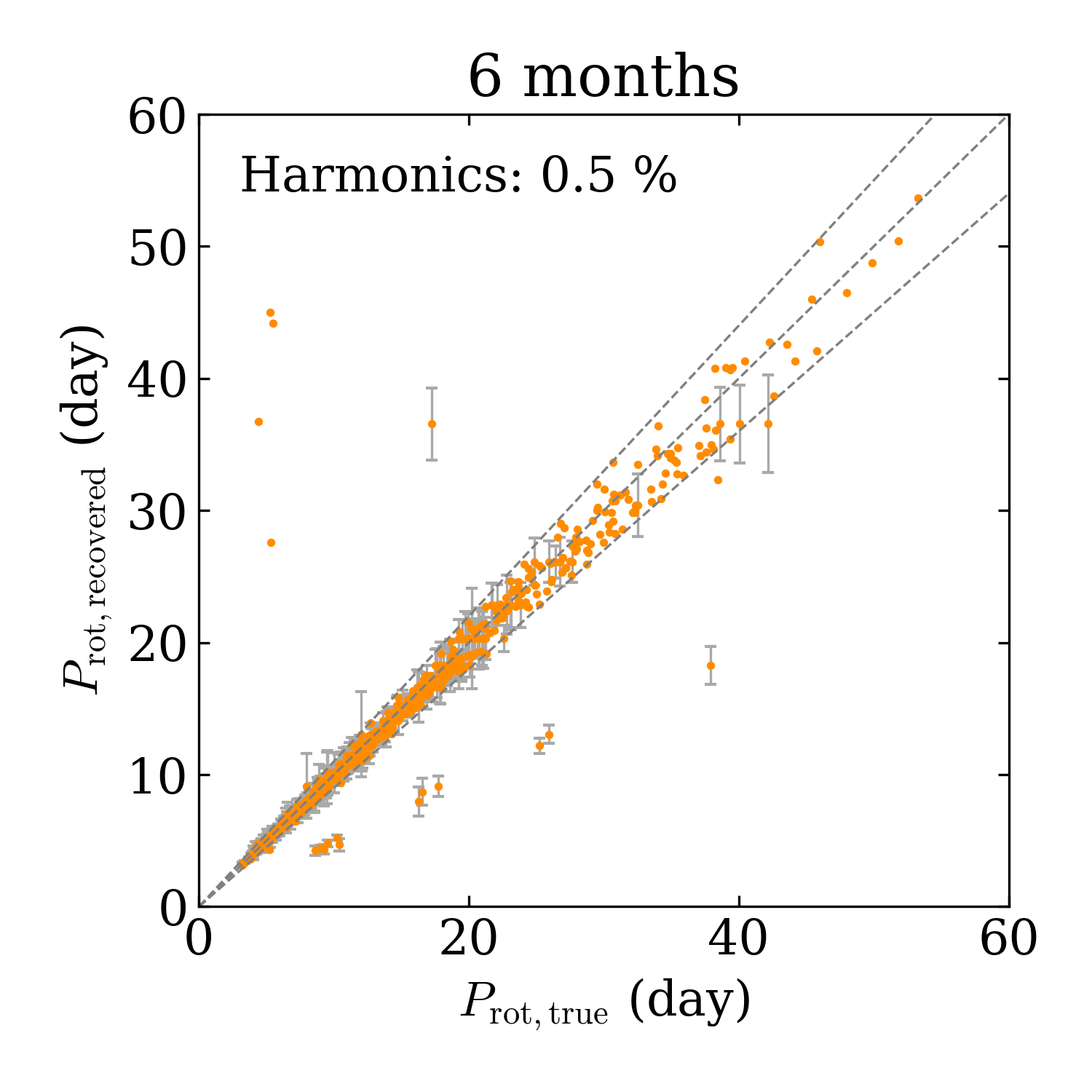}
    \includegraphics[width=0.42\textwidth]{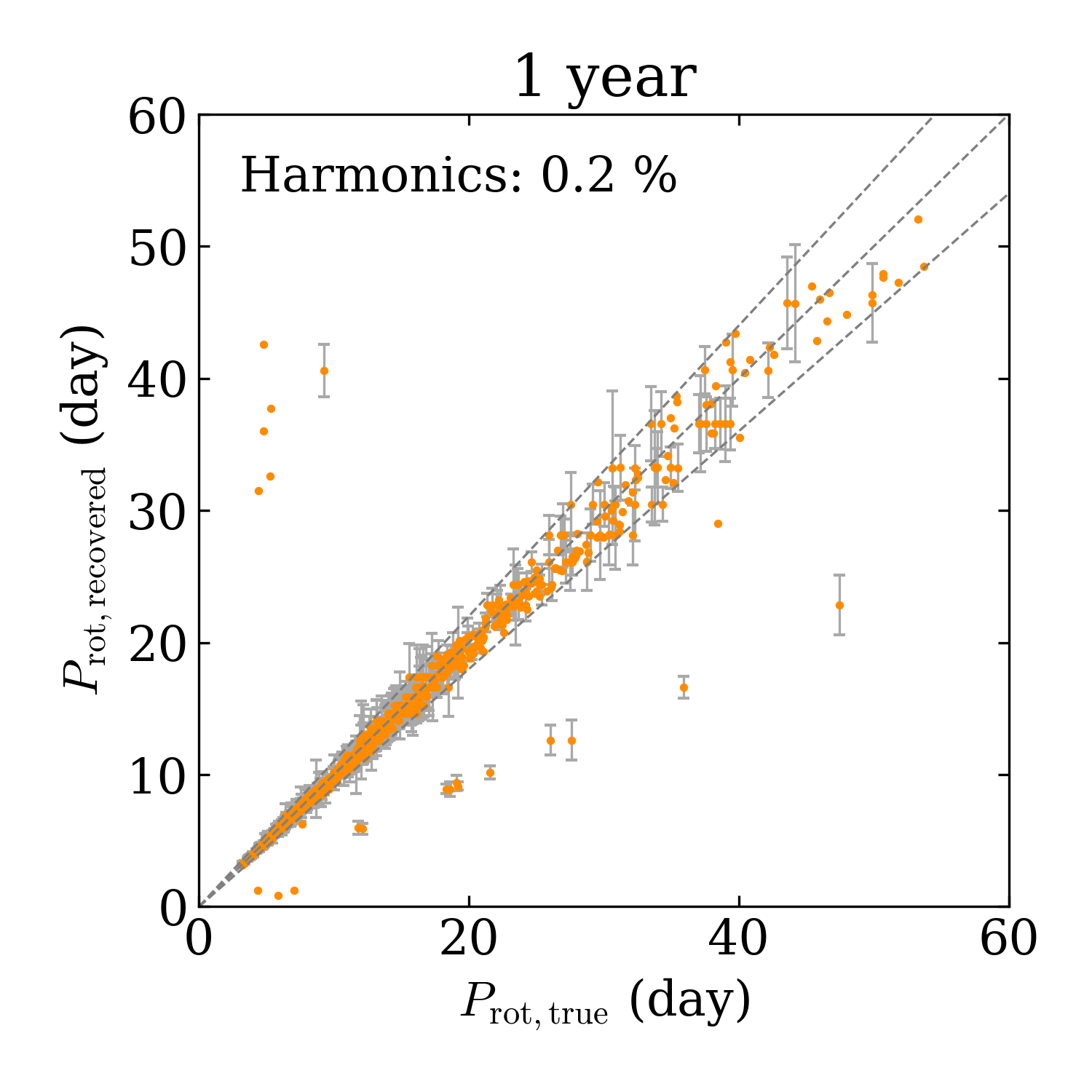}
    \includegraphics[width=0.42\textwidth]{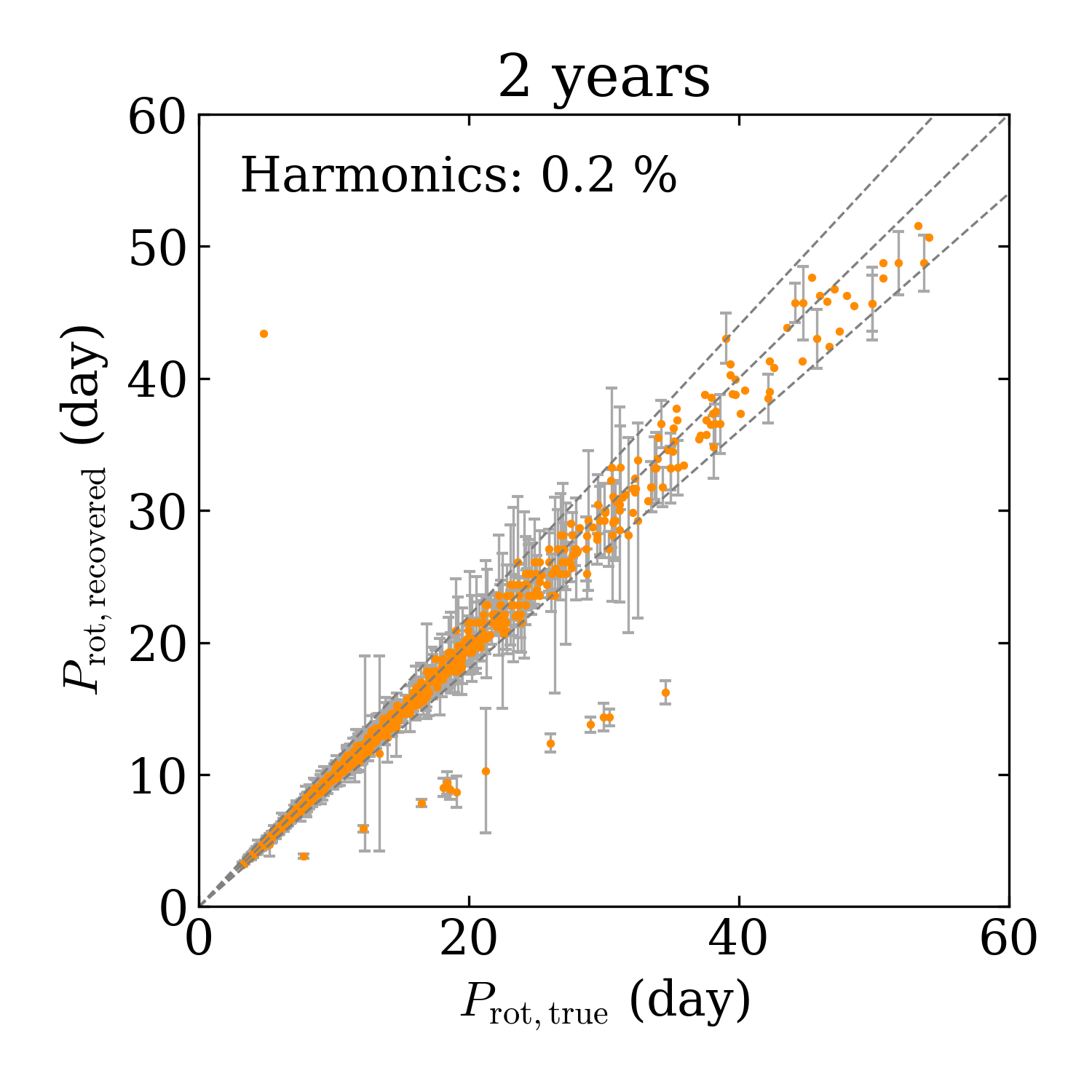}
    \includegraphics[width=0.42\textwidth]{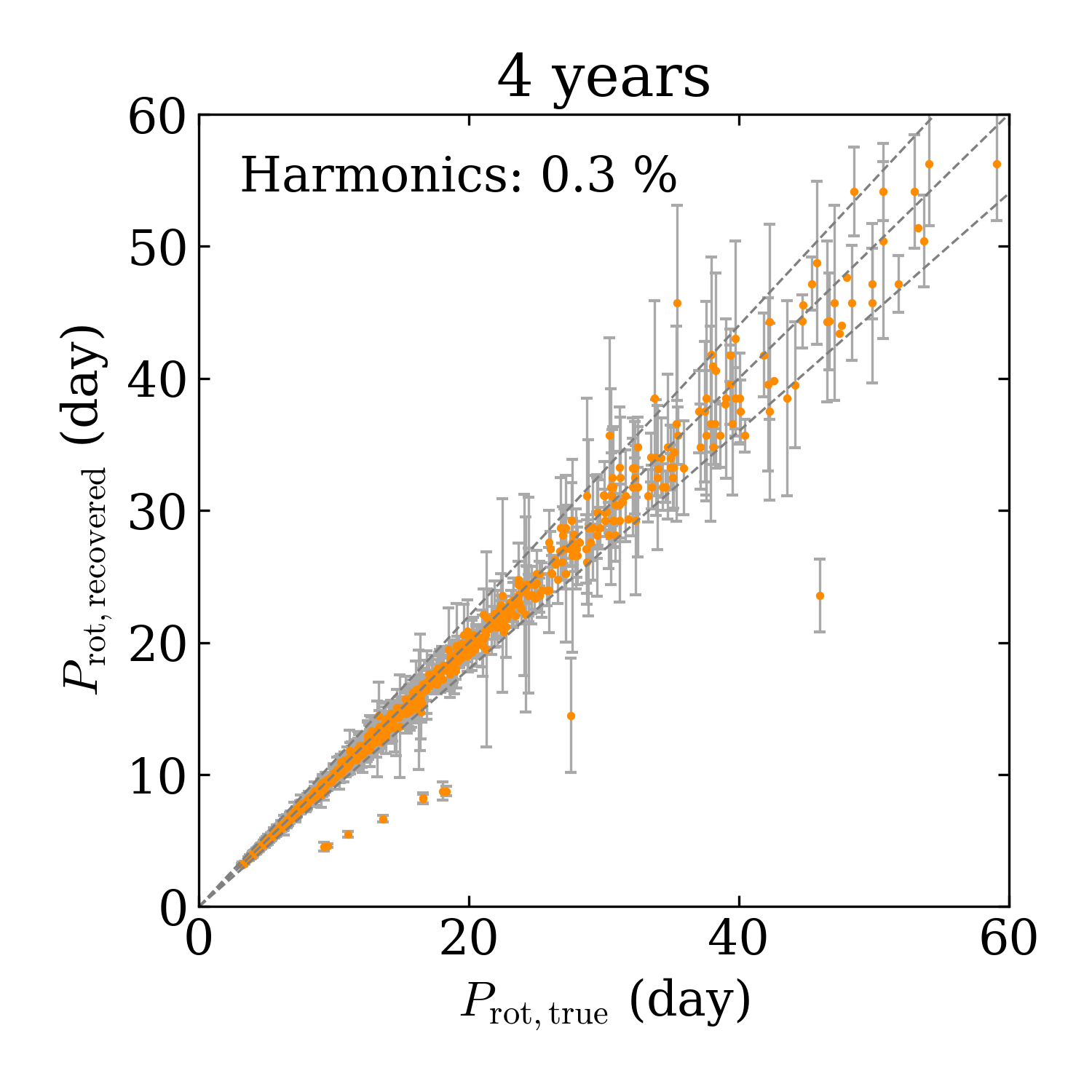}
    \includegraphics[width=0.42\textwidth]{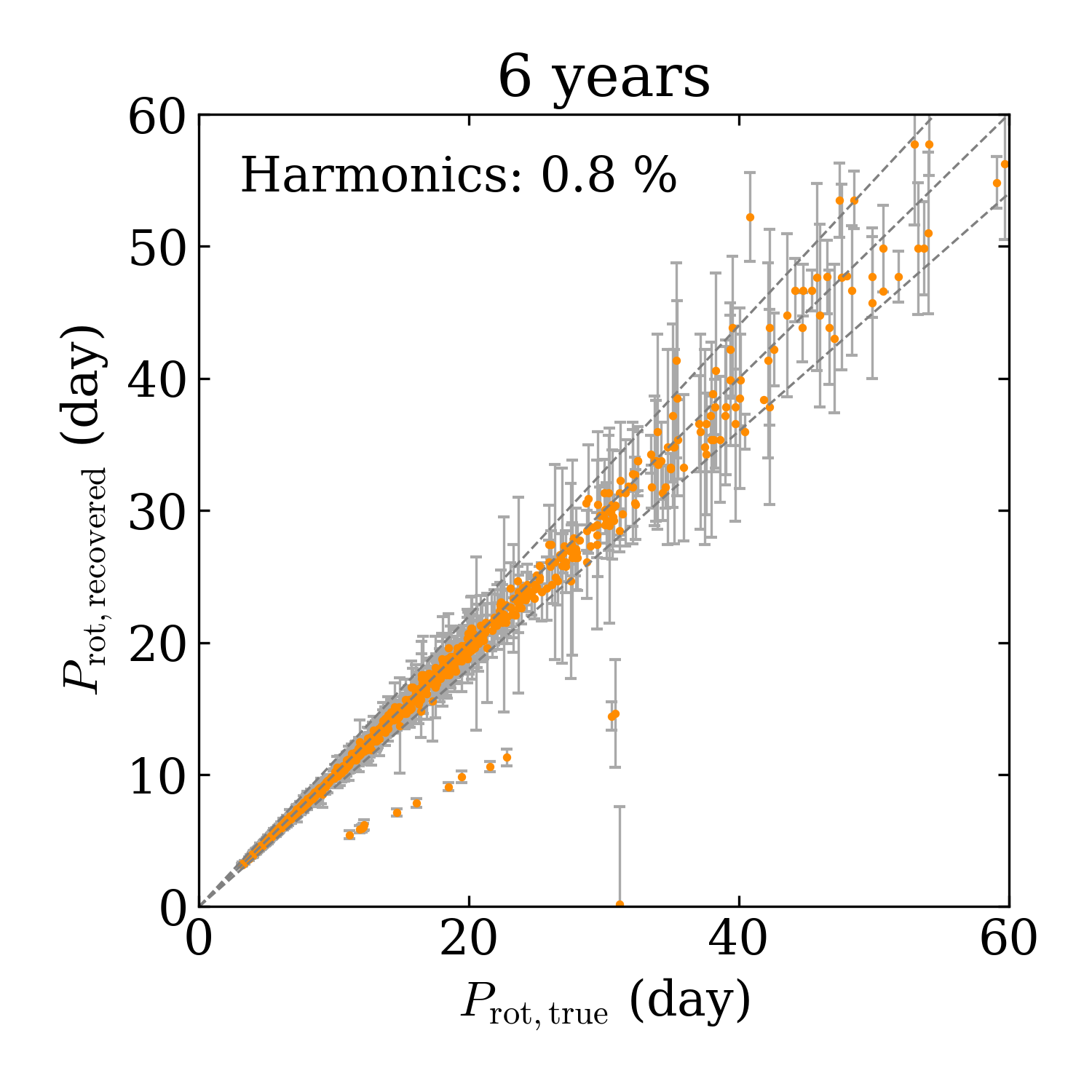}
    \includegraphics[width=0.42\textwidth]{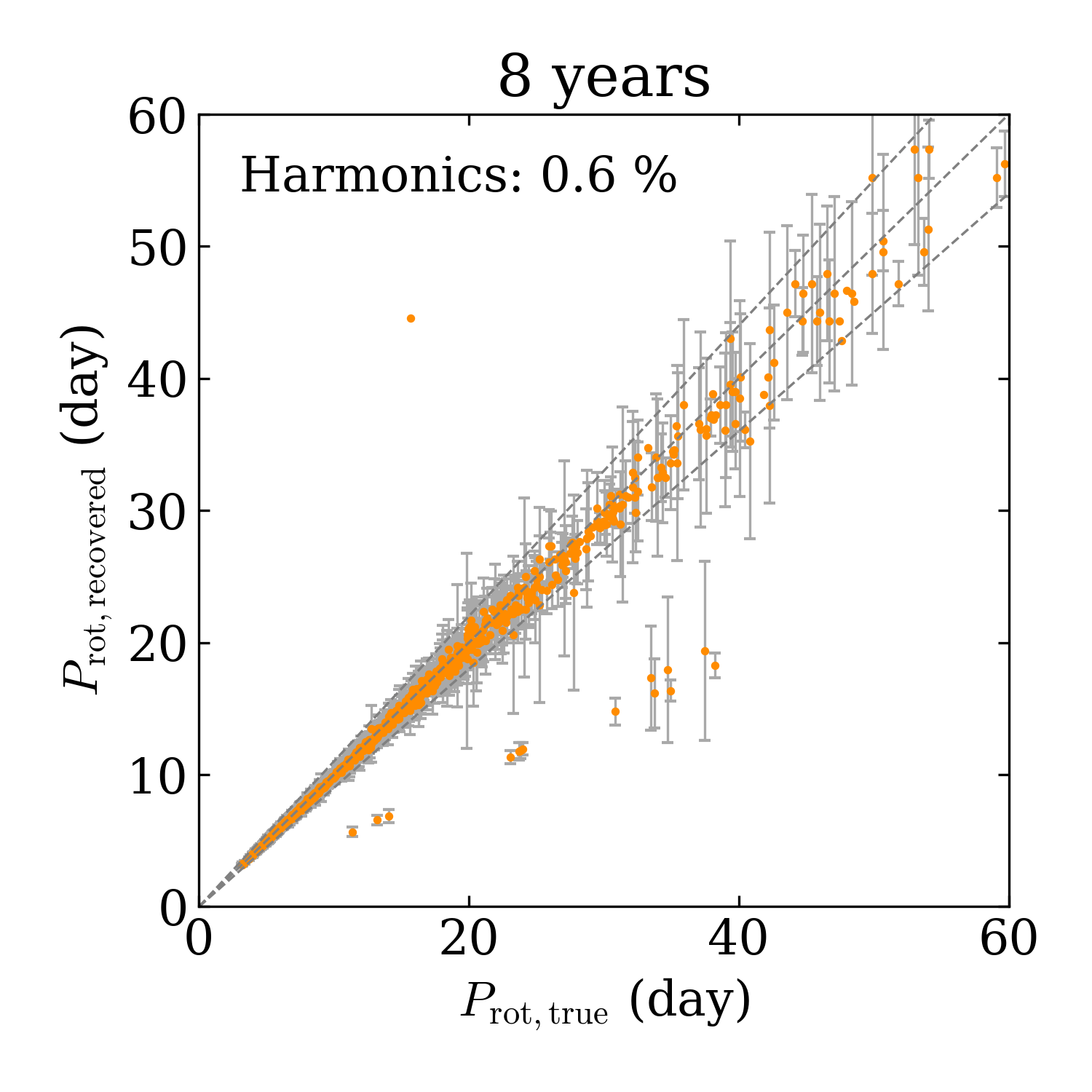}
    \caption{Comparison between the true rotation periods $P_{\rm rot, true}$, and the recovered rotation periods $P_{\rm rot, recovered}$ for the different temporal baselines considered in this work, in the case of noise free light curves: 6 months (\textit{top left}), 1 year (\textit{top right}), 2 years (\textit{middle left}), 4 years (\textit{middle right}), 6 years (\textit{bottom left}), and 8 years (\textit{bottom right}). Uncertainties on $P_{\rm rot, recovered}$ are shown when available. The dashed grey lines correspond to the 1:1 ratio and the $\pm$10\% interval. The percentage of recovered periods laying in the interval $0.49 P_{\rm rot, true} < P_{\rm rot, recovered} < 0.51 P_{\rm rot, true}$ is indicated.}
    \label{fig:additional_recovery_diagnostics}
\end{figure*}

\begin{figure*}[ht!]
    \centering
    \includegraphics[width=0.42\textwidth]{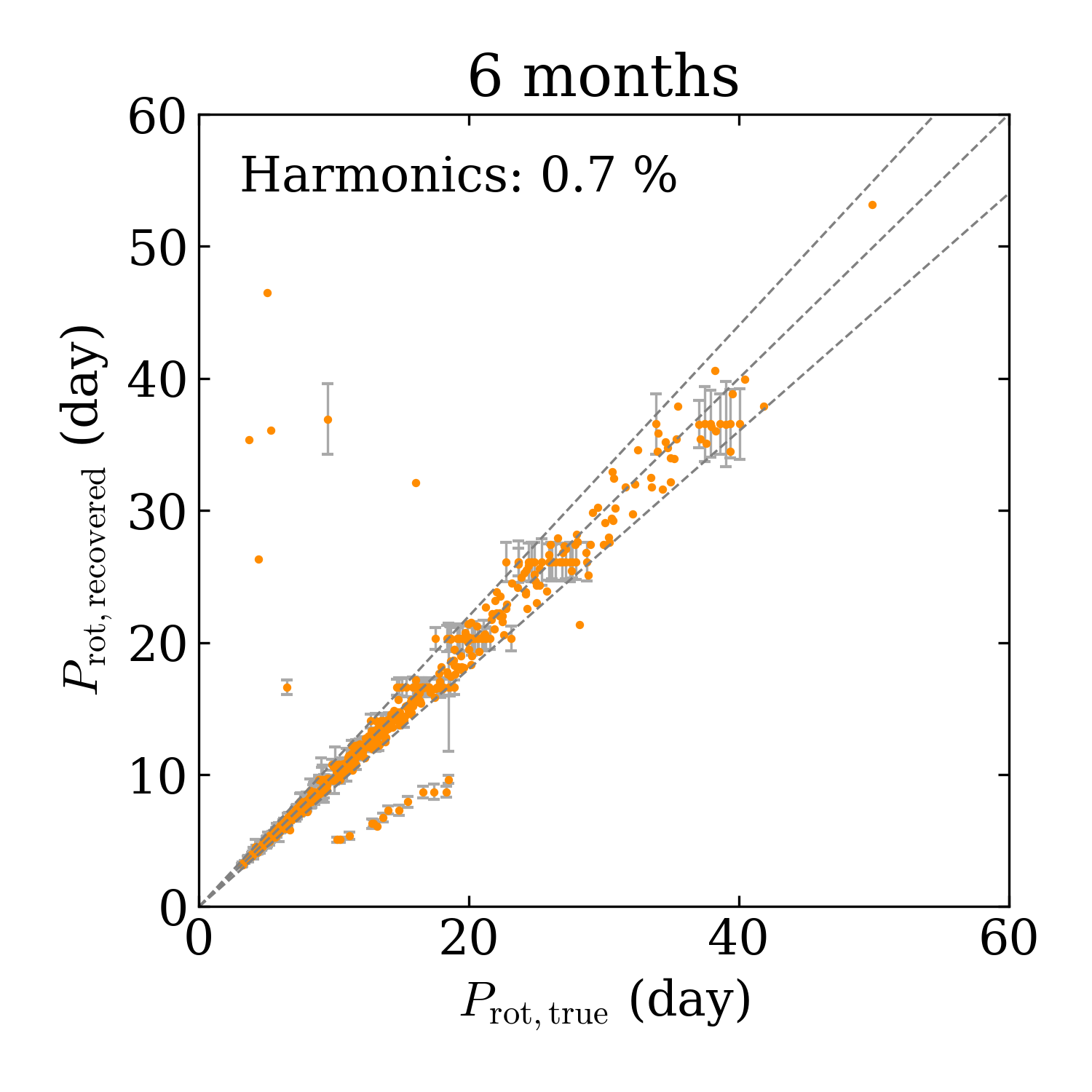}
    \includegraphics[width=0.42\textwidth]{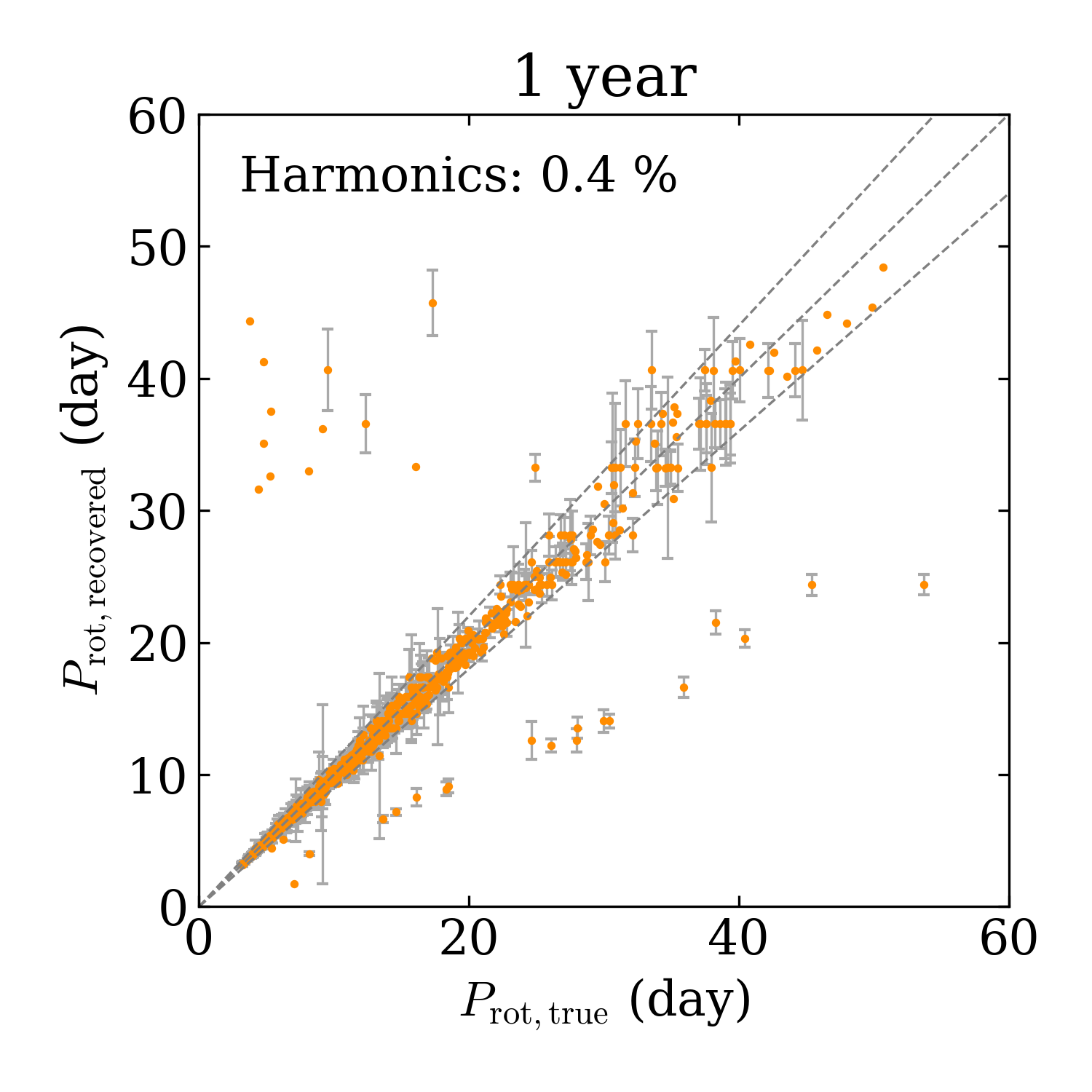}
    \includegraphics[width=0.42\textwidth]{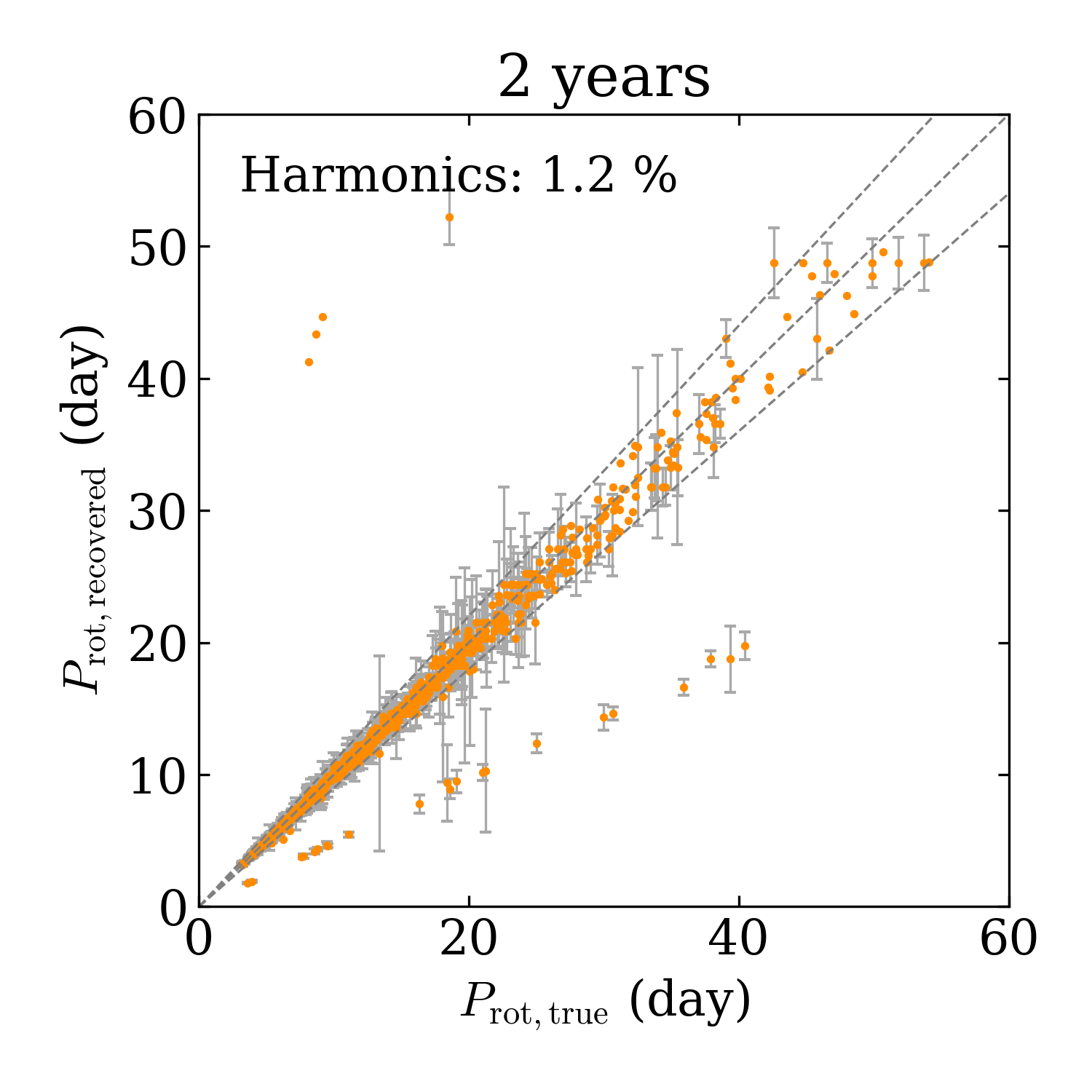}
    \includegraphics[width=0.42\textwidth]{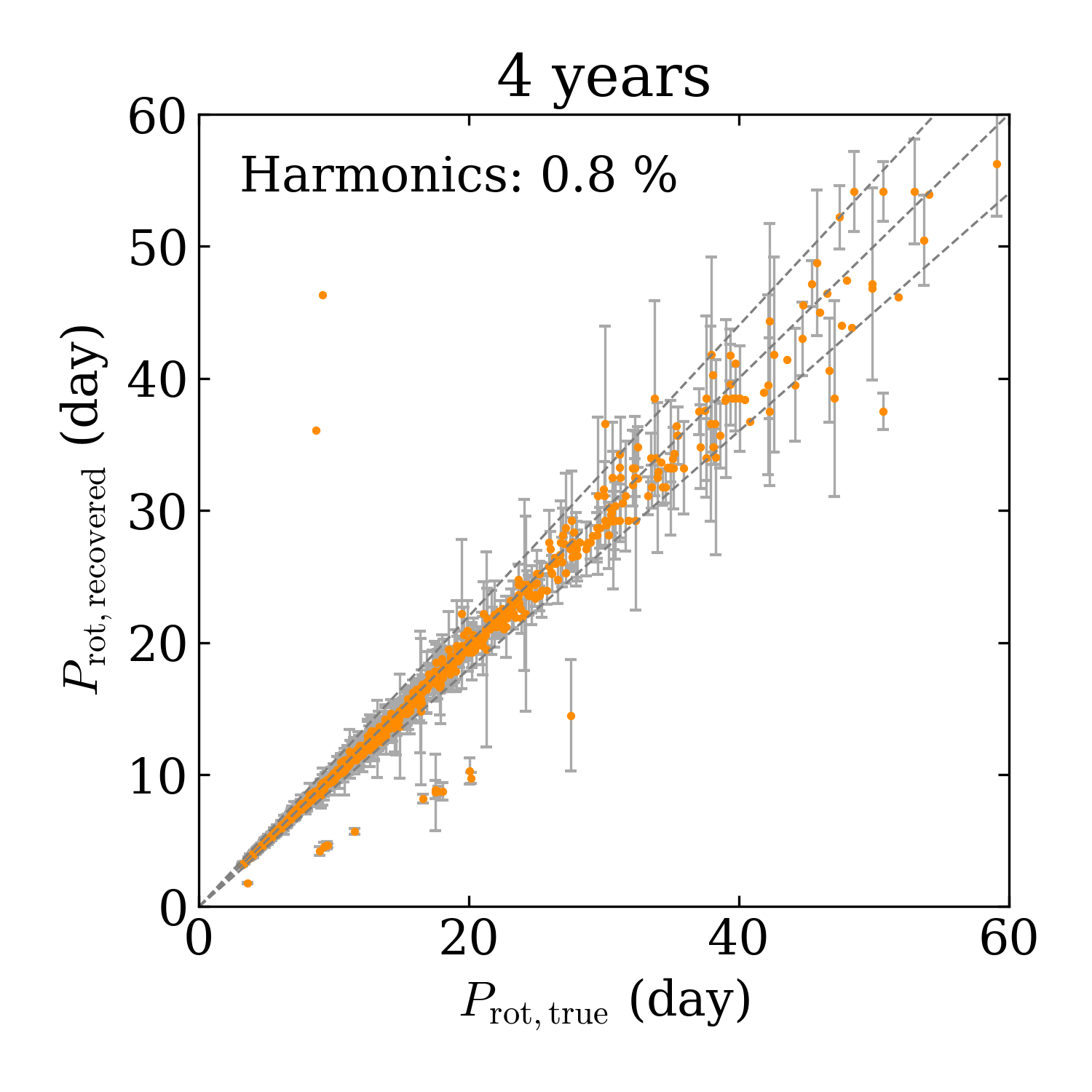}
    \includegraphics[width=0.42\textwidth]{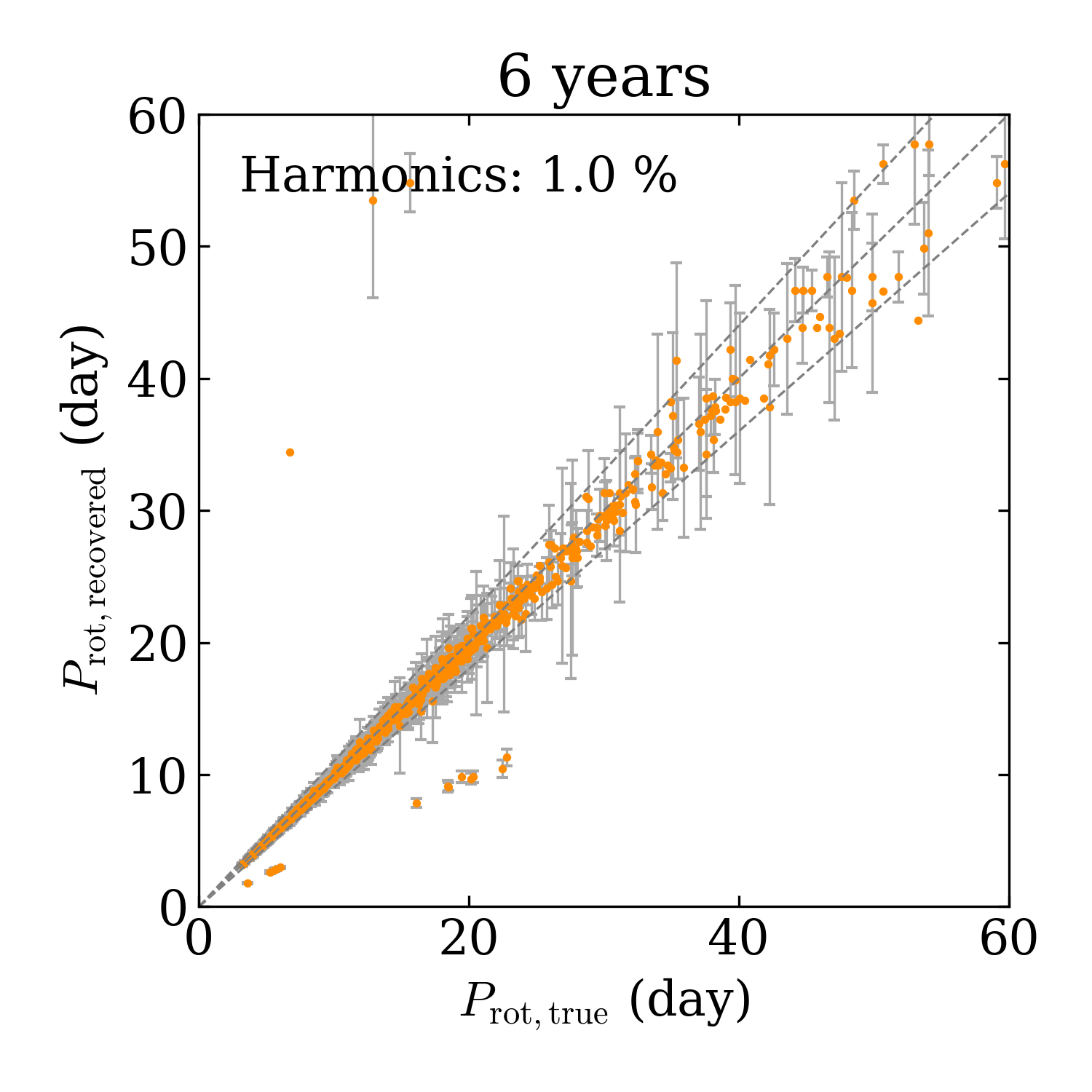}
    \includegraphics[width=0.42\textwidth]{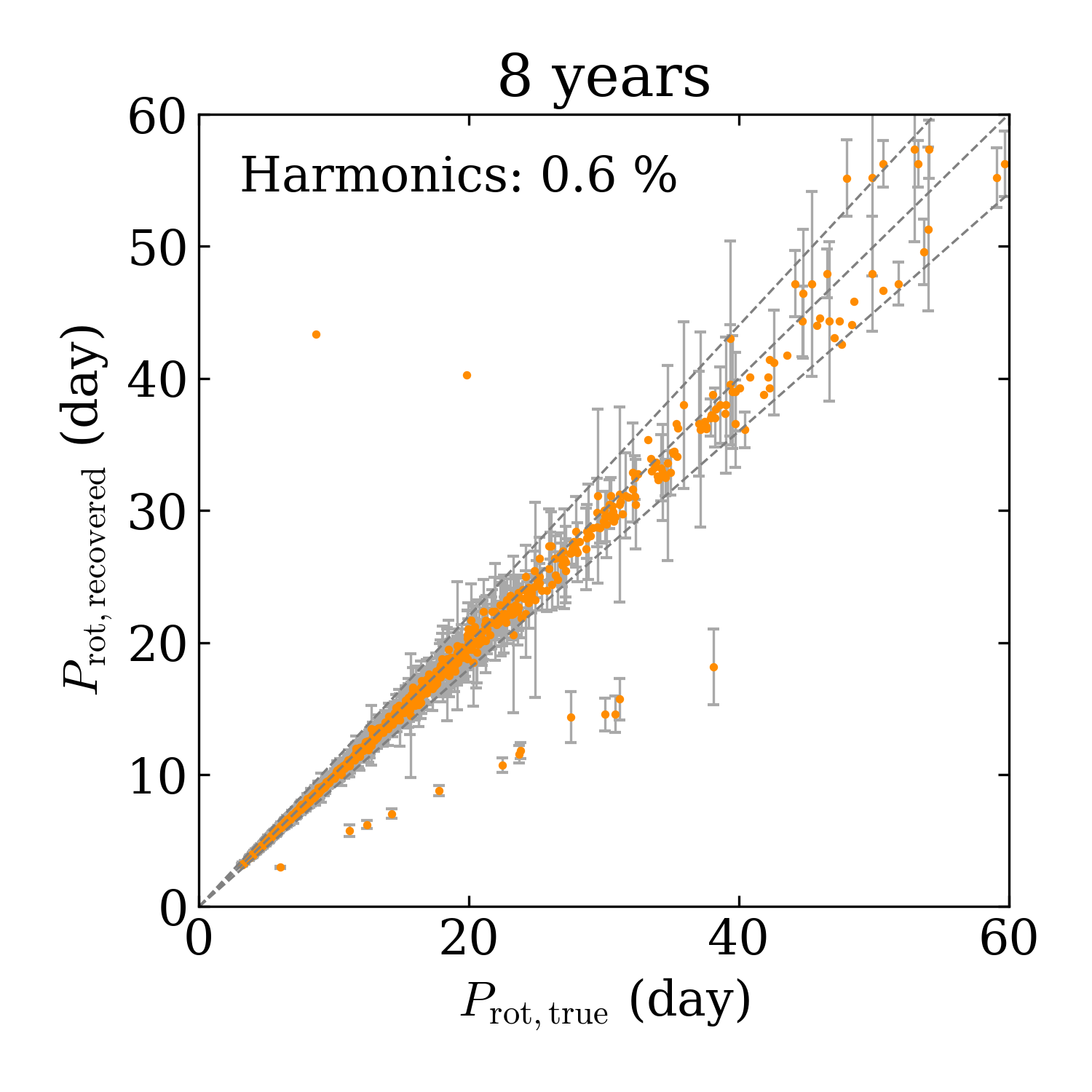}
    \caption{Same as Fig.~\ref{fig:additional_recovery_diagnostics} in the case of corrected noisy light curves.}
    \label{fig:additional_recovery_diagnostics_corr}
\end{figure*}

}

\section{Nomenclature}

For readability considerations, we made the choice to reduce as much as possible the use of PLATO technical acronyms and denominations.  
We provide in this appendix the correspondance between the simple nomenclature used in this work to describe the module structure and the current PLATO denominations. The rotation and activity module is a component of the PLATO Stellar Analysis System (SAS), under the name of Module for Stellar Astrophysics number 4 (MSAP4). Each submodule presented in this work is referred to using an additional index, as listed in Table~\ref{tab:nomenclature}.

\begin{table}[h!]
    \centering
    \caption{Correspondance of denominations between this work and PLATO nomenclature.}
    \begin{tabular}{cc}
    \hline\hline
    This work  &  PLATO nomenclature \\
    \hline
    Submodule 1   & MSAP4-01 \\
    Submodule 2   & MSAP4-02 \\
    Submodule 3   & MSAP4-03 \\
    Submodule 4   & MSAP4-04 \\
    Submodule 5   & MSAP4-06 \\
    \end{tabular}
\label{tab:nomenclature}
\end{table}

\end{document}